\documentclass[11pt]{article}

\usepackage{geometry}
\usepackage{enumitem}
\usepackage{amssymb,amsmath,amsfonts,eurosym,graphicx,caption,color,setspace,sectsty,comment,footmisc,caption,pdflscape,subfigure,array,bm,listings,multirow,  xcolor}
\usepackage[colorlinks,linkcolor=purple, urlcolor  = purple, anchorcolor=brown, citecolor=violet]{hyperref}
\usepackage{algorithm}
\usepackage{algorithmicx}
\usepackage{algpseudocode}
\usepackage{amsthm}
\usepackage{amsmath}
\usepackage[normalem]{ulem}
\usepackage{apacite}
\usepackage{mathrsfs}
\usepackage{booktabs}
\usepackage{adjustbox}
\usepackage{xr}
\usepackage{cleveref}
\usepackage{etoolbox} 
\usepackage{multirow} 
\usepackage{centernot}
\usepackage{cancel}

\usepackage{booktabs} 
\usepackage{array} 

\usepackage{rotating}

\usepackage[title]{appendix}

\newcommand\iid{\stackrel{\rm i.i.d.}{\sim}}

\def\qed{\rule{2mm}{2mm}}

\newtheoremstyle{myremark}
  {\topsep}   
  {\topsep}   
  {\normalfont}  
  {}          
  {\bfseries} 
  {.}         
  { }         
  {}          

\usepackage[pagewise,mathlines]{lineno}
\mathchardef\dash="2D

\newtheorem{theorem}{Theorem}[section]
\newtheorem{lemma}{Lemma}[section]

\newtheorem{corollary}{Corollary}[section]
\newtheorem{proposition}{Proposition}[section]
\newtheorem{assumption}{Assumption}[section]
\usepackage{bbm} 

\Crefname{assumption}{Assumption}{Assumptions}

\theoremstyle{myremark}

\newtheorem{remark}{Remark}[section]

\AtEndEnvironment{remark}{~\qed}
\AtEndEnvironment{example}{~\qed}

\DeclareMathOperator*{\var}{Var}
\DeclareMathOperator*{\cov}{Cov}

\begin{document}
\lstset{
basicstyle=\ttfamily\small,
numbers=left,
keywordstyle= \color{ blue!70},commentstyle=\color{red!50!green!50!blue!50}
}

\title{Degrees of Freedom  and Information Criteria for \\the Synthetic Control
Method\thanks{First draft: December 16, 2019. We thank Alberto Abadie, Dmitry Arkhangelsky, Guanglei Hong,  Kaspar Wuthrich, as well as the editor and anonymous referees for insightful comments and questions.  The first two authors are in alphabetical order and made equal contributions.}}

\author{%
Guillaume A.~Pouliot%
\thanks{Department of Economics, Rice University, 
6100 Main St, Houston, TX 77005, United States. Email: \href{mailto:guillaume.pouliot@rice.edu}{guillaume.pouliot@rice.edu}.}%
\and Zhen Xie%
\thanks{Department of Economics, Northwestern University, 
2211 Campus Drive, Evanston, IL 60208. Email: \href{mailto:zhenxie@u.northwestern.edu}{zhenxie@u.northwestern.edu}.}%
\and Ziyi Liu%
\thanks{Haas School of Business, University of California, Berkeley, 
2220 Piedmont Ave, Berkeley, CA 94720. Email: \href{mailto:zyliu2023@berkeley.edu}{zyliu2023@berkeley.edu}.}
}

\date{\today}

\maketitle

\vspace{-0.2in}

\begin{abstract}


We provide an analytical characterization of the model flexibility of the synthetic control method (SCM) in the familiar form of degrees of freedom. We obtain estimable information criteria, which may be used to circumvent cross-validation when selecting either the tuning parameter in 
penalized variants of SCM or the weighting matrix in the SCM with covariates.  We assess the impact of car license rationing in Tianjin; while a natural match is available, both it and other donors are noisy, inviting the use of SCM to average over approximately matching donors. The very large number of candidate donors calls for
penalized variants of SCM and we observe that model selection using information criteria outperforms that based on cross-validation.
\end{abstract}

\noindent\textbf{Keywords}: Synthetic Controls, Model Selection, Information Criteria, Degrees
of Freedom, Lagrange Multiplier Theory, Chinese Automotive Industry.\\[4pt]


\newpage

\section{Introduction}

The synthetic control method has become a standard regression tool
in economics, political science, and a handful of other fields. See
\citeA{abadie2021using} for a recent
survey and pedagogical introduction. As such, methodological research
has endeavored to append the synthetic control estimator with standard
regression output. This includes quality of fit criteria, confidence intervals, $p$-values,
etc \cite{abadie2010synthetic,chernozhukov2018t, cattaneo2021prediction}.
This paper endeavors to do just that and delivers the degrees of freedom and information criteria for the synthetic control method.  As such, our motivation, albeit somewhat jejune, is commensurately uncontroversial.

Our first contribution is to produce
the degrees of freedom. We find that the degrees
of freedom, or the effective number of estimated parameters, of the synthetic
control method without covariates is \emph{one less than the expected
number of donors having nonzero estimated coefficients}. We
obtain a more general result that covers the case with covariates.

Our motivation for producing this as-of-yet-unavailable statistic
is two-fold. First, the question ``does the synthetic control method
overfit?'' is, as we argue below, non-trivial and important.
The degrees of freedom offer a clear and intuitive answer to that
question. 
We find that SCM does \emph{not} overfit in most of the seminal
applications we revisited.  It does however overfit in ``high-dimensional" applications such as the one we investigate in this paper.

Second, we produce information criteria for synthetic control methods.
We find that, in applications where the number of donors is large relative to the number of pre-treatment observations,
overfitting
(through model selection flexibility) does arise and regularization
is required.
The short pre-treatment series, relative to the number of donors, makes it such that cross-validation
strategies
can perform poorly. An information
criterion, which assesses the out-of-sample performance of the estimator
evaluated on the entire pre-treatment data, is expected to do better. 
Equipped with closed-form expressions for the degrees of freedom
that have sample analogs, we produce just such information criteria
and indeed observe (in simulation and placebo cases) that they outperform
cross-validation in terms of producing accurate counterfactual and treatment effect estimates.

Importantly, the produced information criteria can also be used to select the weighting matrix in the synthetic control problem with
covariates, see problem formulation (\ref{eq:SCwithCOVbeginning})-(\ref{eq:SCwithCOVend}).

\subsubsection*{Degrees of freedom}

\begin{figure}
  \centering
\begin{center}
\begin{minipage}[t]{0.30\textwidth}
  \centering
  \resizebox{1.45\textwidth}{!}{%
    \begin{tabular}{@{} l r  l r @{}}
      \toprule
      \hline
      State & Weight & State & Weight \\
      \midrule
      Alabama & 0     & Montana         & 0     \\
      Alaska  & --    & Nebraska        & 0     \\
      Arizona & --    & Nevada          & 0.062 \\
      Arkansas & 0    & New Hampshire   & 0.001 \\
      Colorado & 0    & New Jersey      & --    \\
      Connecticut & 0 & New Mexico      & 0     \\
      Delaware & 0    & New York        & --    \\
      District of Columbia & -- & North Carolina & 0 \\
      Florida  & --   & North Dakota    & 0     \\
      Georgia  & 0    & Ohio            & 0     \\
      Hawaii   & --   & Oklahoma        & 0     \\
      Idaho    & 0    & Oregon          & --    \\
      Illinois & 0    & Pennsylvania    & 0     \\
      Indiana  & 0.175& Rhode Island    & 0     \\
      Iowa     & 0    & South Carolina  & 0.343 \\
      Kansas   & 0    & South Dakota    & 0     \\
      Kentucky & 0    & Tennessee       & 0     \\
      Louisiana& 0    & Texas           & 0.182 \\
      Maine    & 0.236& Utah            & 0     \\
      Maryland & --   & Vermont         & 0     \\
      Massachusetts & -- & Virginia    & 0     \\
      Michigan & --   & Washington      & --    \\
      Minnesota& 0    & West Virginia   & 0     \\
      Mississippi & 0 & Wisconsin       & 0     \\
      Missouri & 0    & Wyoming         & 0     \\
      \bottomrule
    \end{tabular}
  }
\end{minipage}
  \hfill
   \begin{minipage}[t]{0.53\textwidth}
   \vspace{-15em}
    \centering
\includegraphics[scale=0.64]{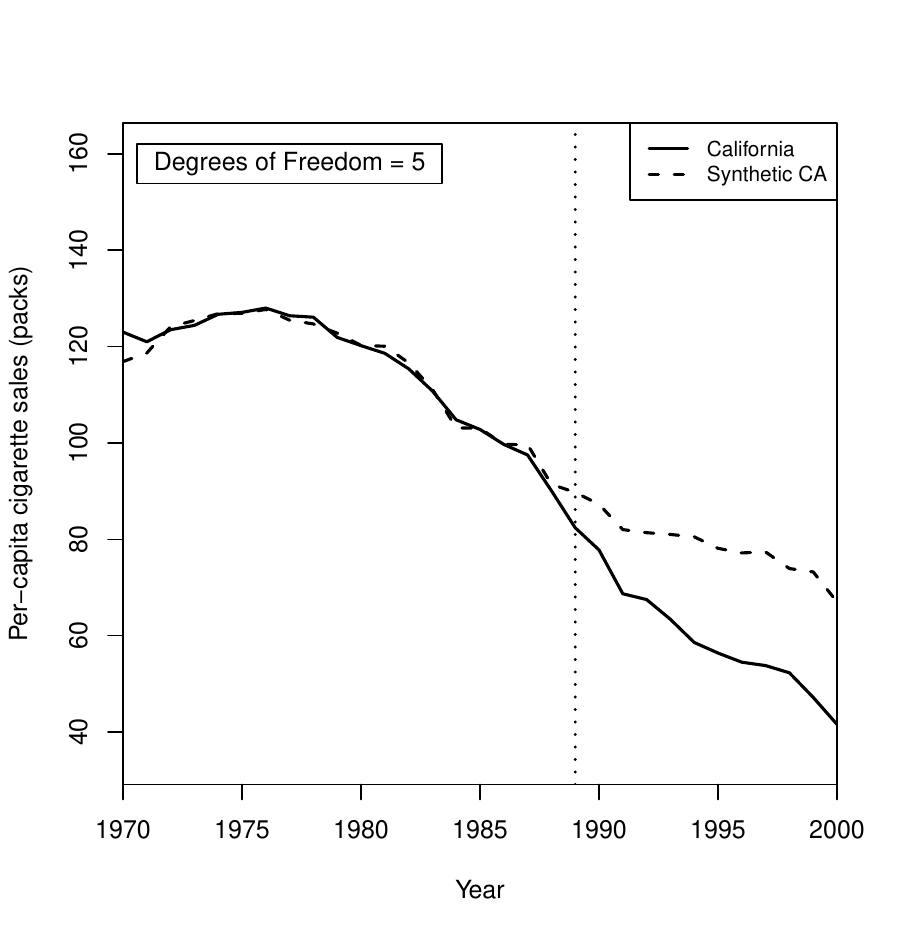}
\end{minipage}
\end{center}
\caption{\emph{Synthetic control output of California Proposition 99 investigation. The degrees of freedom estimate is 5. The left-hand side and right-hand side panels are, respectively, our replications of Table 2 and Figure 2 of \protect\citeA{abadie2010synthetic}.}}
\label{figure1}
\end{figure}

Degrees of freedom expressions are inherently interesting for the
synthetic control method. Indeed, consider two of the more striking
--and attractive-- features of the synthetic control method, typically displayed in its standard output. We reproduce in \Cref{figure1}
the output of \citeA{abadie2010synthetic} as an example.
First, the fitted regression coefficient estimates typically exhibit
substantial sparsity; the synthetic control is a linear combination
of a few donors, with many donors getting an estimated weight of zero.
Second, the observed and fitted paths, before treatment, often suggest
a high in-sample fit, as is the case in \Cref{figure1}.

These two features capture our motivation for inquiring into the degrees
of freedom of the synthetic control method. The high degree of sparsity in the estimated regression coefficients suggests the possibility
of extensive implicit model selection. Given that the series being fitted are often short, one may worry that this additional model flexibility brings about overfitting, and that this is in fact what explains the surprising quality of the in-sample fit, hence casting doubt on the
quality of the counterfactual and treatment effect estimates.

The question \emph{``does synthetic control overfit?''} thus stands open. On the one hand, the best subset selection method with sample
size and number of active independent variables typical of synthetic
control applications would be expected to overfit (see \Cref{figure2}).
On the other hand, placebo exercises in which one forecasts non-treated
series and compares the forecasted to the realized series often suggest
a reasonable quality of forecast (e.g., \citeNP{abadie2010synthetic}).
This motivates the development of an analytical measure of the model
flexibility of the method, the most compelling of which is, in our
opinion, its degrees of freedom.

The key idea is to produce an estimable expression for the degrees of freedom by applying Stein's Lemma, see Section \ref{section3}.  To the best of our knowledge, this was first done by \citeA{meyer2000degrees}.  The closest paper to ours is \citeA{zou2007degrees}, which uses Stein's Lemma to compute the degrees of freedom of the lasso (see Figure \ref{figure2}).  Their result and the similarities between the lasso and SCM estimation problems inspired this analysis.

\subsubsection*{Information criteria}

The wide applicability of the synthetic control method has brought
applied analysts to consider ``high-dimensional'' applications --i.e.,
with many donors-- even though the training, or pre-treatment, period may be short
relative to the number of donors.
\footnote{For instance, \citeA{cavallo2013catastrophic}
study the causal impact of catastrophic natural disasters on economic
growth, they have 196 donors and fewer than 40 training periods in
each regression. \citeA{bifulco2017using} study the
impact of an education program on district enrollments and graduation
rates, they have 275 donors and 10 training periods. \citeA{bohn2014did}
study the effect of Arizona\textquoteright s 2007 Legal
Arizona Workers Act (LAWA) on the proportion of the state\textquoteright
s population, they have 46 donors and 9 training periods. \citeA{pieters2016effect}
study the effect of democratization on child mortality, they have
24 untreated units and 10 training periods. \citeA{heersink2017disasters}
study the effect of natural disasters on politicians\textquoteright{}
electoral fortunes, they have 100 donors and 29 training periods.
\citeA{peri2019labor} study the effects of the
Mariel Boatlift on wages and employment, they have 43 donors and six
training periods. \citeA{billmeier2013assessing}
study the impact of economic liberalization on the real GDP per capita,
they have 180 donors and 38 training periods.} Such applications
of course raise \emph{a priori} concerns about overfitting,
and these have brought about the development of penalized 
methods,
all of which require the specification of a tuning parameter
(\citeNP{abadie2021penalized, athey2021matrix, doudchenko2016balancing}).
Typically, such tuning parameters are estimated by cross-validation,
with different specific algorithms preferred by different authors,
see \Cref{table1}.

\begin{table}[H]\centering


\begin{tabular}{
    >{\centering\arraybackslash}p{0.28\linewidth}
    >{\centering\arraybackslash}p{0.67\linewidth}
}
\toprule
CV method & Reference\\
\midrule
Rolling window & \citeA{kellogg2020combining}
\\[8pt]
\multirow{3}{*}{\parbox{\linewidth}{\centering Pre-Intervention Holdout Validation on the Treated}} & \citeA{abadie2015comparative}, \citeA{xu2017generalized},\\
 & \citeA{abadie2021penalized},\\
 & \citeA{ben2021augmented}
\\[8pt]
\multirow{2}{*}{\parbox{\linewidth}{\centering Leave-one-out Validation on the Untreated}} & \citeA{abadie2021penalized}\\
\\
\bottomrule
\end{tabular}

\caption{\emph{Survey of cross-validation procedures suggested in the synthetic
controls literature.}}
\label{table1}
\end{table}

Cross-validation is also called upon in synthetic control applications
to estimate the weighting matrix of the inner problem in the synthetic
control method with covariates. See (\ref{eq:SCwithCOVbeginning})-(\ref{eq:SCwithCOVend})
below for details about the construction of synthetic controls using
covariates.

However, cross-validation is often ill-suited for model selection
with the synthetic control method. 
On the one hand, ``pre-intervention holdout validation on the treated''
is a data-splitting procedure; the estimator is trained on the first
half of the pre-treatment data, and the second half is used as a test
set. As such, we may expect it to be severely biased with short pre-treatment series; conceptually, the synthetic control method trained on, say, half of the pre-treatment data may behave very differently from one trained on the whole pre-treatment data. On the other hand, the ``leave-one-out validation on the untreated'' approach treats a donor as a placebo to-be-treated unit, and uses the post-treatment data as a test set.
It ignores the true to-be-treated unit, and averages over the post-treatment
forecast error of a selection of such placebo to-be-treated units.
It relies on the assumption that, in the absence of a treatment effect,
the conditional distribution of any donor --perhaps after some preselection--
given the other donors is the same as that of the to-be-treated unit
given the donors.
This is a strong, additional assumption to make about the underlying data generating process.

As investigated in \citeA{kellogg2020combining}, the pre-intervention holdout approach may produce a misleading estimate
of the prediction mean-squared error. \citeA{kellogg2020combining} instead propose to use rolling-window cross-validation which they find to perform best, at least for the estimator they propose.

As detailed below, these procedures remain data hungry, and we are well motivated to consider out-of-sample error estimation methods that 
use information from the entire pre-treatment data without implicitly assuming symmetry between the donor and the to-be-treated unit.

A classical alternative to cross-validation is to compare models using
an information criterion (\citeNP{claeskens2008model}).
Information criteria append to the in-sample loss a penalty for model
flexibility, thus allowing for the comparison of models of different
flexibility while training them on the entire pre-treatment data.

The key point is that while both information criteria (may they be AIC, BIC, or other) and cross-validation techniques estimate out-of-sample fit, information criteria do so using all available data, while cross-validation requires splitting the data.  This brings about a tradeoff between the stronger assumptions required for the validity of information criteria and the larger sample required for the accuracy of cross-validation.  The two approaches thus being complementary, a complete toolkit for a regression method should include both.

It is worthwhile to note that the aforementioned cross-validation procedures are not only
data-hungry, they also require
tuning from the user. The rolling window and pre-intervention holdout
cross-validation approaches require careful tuning of the size of the
training and test sets. Likewise, for leave-one-out validation on the
untreated, the pool of placebo units (those ``similar enough'' to
the treated unit) must be selected; this can be delicate, as illustrated
in \citeA{abadie2010synthetic}. Meanwhile, the information
criteria approach requires no tuning and our publicly available code
can be used directly.

The challenge in producing information criteria is to give a computable
expression for a penalty term capturing model flexibility. As detailed
in \Cref{section3} below, this exercise is intimately
tied to the estimation of degrees of freedom.


The intellectual history of this approach to producing information criteria can be summarized as follows. \citeA{stein1981estimation} proved the now famous Stein's Lemma, displayed below in (\ref{eq:SteinLemma}), and used it to produce what is now called Stein's unbiased risk estimate (SURE), displayed below in (\ref{eq:homoskedasticSURE}). This implicitly defines a general formulation for degrees of freedom as part of the penalty term in SURE, when the latter is thought of as an information criterion. The literature attributes to \citeA{efron2004estimation} and \citeA{hastie1990generalized} the definition of degrees of freedom in terms of covariance between observed and fitted values, which matches that deduced from the penalty term of SURE, before applying Stein's lemma. \citeA{meyer2000degrees} used Stein's lemma to compute the degrees of freedom for shape restricted regression, setting the stage for other such applications.  Amongst those, we find the degrees of freedom of the lasso, a result first given in \citeA{zou2007degrees}, further generalized in \citeA{tibshirani2012degrees}, and which has inspired the analysis herein. 
General results producing degrees of freedom estimates for classes of models are given in \citeA{kato2009degrees} and \citeA{chen2020degrees}.

\subsubsection*{Impact of rationing on the demand for cars}

The aforementioned methodological developments are of general interest
but were, for the authors, motivated by the analysis of the market
for new automobiles in China after the introduction of rationing.
We investigate model-specific changes in sales in Tianjin following the introduction of a lottery-auction hybrid for the distribution
of car licenses. In order to build counterfactual --as if there
had been no rationing-- time series for each individual car model,
the synthetic control method turns out to be a natural choice but falls prey to overfitting, and the penalized synthetic control method (\citeNP{abadie2021penalized})
is more reliable. However, cross-validation performs poorly for tuning
parameter selection and we are able to carry out a more robust analysis
using our information criteria instead. We present model-specific
treatment effects of the policy, thus allowing for a detailed study
of the market for new cars in Tianjin.

This is a somewhat novel, or at least uncommon, application of the synthetic
control method. While a natural match is available --the same car
model in an untreated city-- it is very noisy. By instead averaging
over many approximate matches, we can produce a synthetic control
having a much smaller variance than the natural donor does, at a relatively
small cost in bias. %

\subsubsection*{Reduced form framework, target of inference, and notation}

The typical notation in the synthetic control and in the degrees of
freedom literature is somewhat at odds. We clarify the connection.
The synthetic control literature uses notation
appealing to the potential outcomes framework with, say, $(Y_{j,t}(0),Y_{j,t}(1))$,
$j=0,1,...,p$, being the potential outcomes under control and treatment
of unit $j$ at time $t$. Some units will remain untreated over time
while some will be treated after some time period $T^{*}$, i.e., $Y_{j,t}(1)$
is observed instead of $Y_{j,t}(0)$ for such $j$'s for $t>T^{*}$.\footnote{This is the canonical framework, scattered treatment times may also
be considered.}

The object of interest is some treatment effect $\tau=Y_{j,t}(1)-Y_{j,t}(0)$ for the --here, unique-- treated unit $j=0$ at some time $t>T^{*}$.
The synthetic control estimator being a constrained least-squares
estimator, its implicit target is the constrained best linear predictor
and, under correct specification of the regression function, the conditional
expectation. Specifically, under correct specification, the population
estimand is $Y_{j,t}(1)-E\left[\left.Y_{j,t}(0)\right|Y_{1,t}(0),\dots,Y_{p,t}(0)\right]$.
Since only a single $Y_{j,t}(1)$ is observed, quality of inference
about $\tau$ is assessed conditionally on $Y_{j,t}(1)$ and is thus
commensurate to quality of inference about $E\left[\left.Y_{j,t}(0)\right|Y_{1,t}(0),\dots,Y_{p,t}(0)\right]$.
Furthermore, since only pre-treatment data is involved in the study
of this object, the notation more typical of the degrees of freedom
literature suffices, and in fact will prove quite handy and natural.

Bridging to the notation we will use, we define the vectors $\mathbf{Y}=\left(Y_{0,1}(0),\dots,Y_{0,T^{*}}(0)\right)^{T}$
, $\mathbf{X}_{j}=\left(Y_{j,1}(0),\dots,Y_{j,T^{*}}(0)\right)^{T}$, $j=1,...,p$, and $\mathbf{X}=\left(\mathbf{X}_{1},\dots,\mathbf{X}_{p}\right)$.  
Letting $n=T^{*}$, we recover the conventional $\mathbf{X}\in\mathbb{R}^{n \times p}$.
We use $\mathbf{X}_{J}$, $J\subset\{1,\dots,p\}$, to designate the submatrix
made of the columns designated by $J$. We use $\mathbf{1}_{p}$, $p\in\mathbb{N}$,
to designate the vertical vector of ones of length $p$. For any vector
$\mathbf{w}\in\mathbb{R}^{n}$, we use $\left\Vert \mathbf{w}\right\Vert _{2}$
to indicate its $\ell_{2}$ norm and for any set $\mathcal{B}$, we
use $\left|\mathcal{B}\right|$ to indicate its cardinality.  


\bigskip{}

Our methodological contributions can be succinctly summarized as follows.
We produce information criteria for  synthetic control methods.
In the case of the penalized synthetic control method, this allows for the selection of the tuning parameter controlling model selection. 
In the classical case of the synthetic control method with covariates, this allows for the selection of the weighting matrix. 
Importantly, model
selection based on the information criteria circumvents cross-validation,
which we argue and demonstrate can be misleading in practice.
We furthermore produce closed form expressions for the degrees of freedom for the synthetic control method, with or without covariates, as well as for penalized versions. 
These have sample analogs and may be reported alongside standard output.
See Figures \ref{figure1} and \ref{fig:one model fits} for a suggestion of modified output.

The remainder of the article is divided as follows. \Cref{section2}
defines the synthetic control method and several penalized 
extensions. \Cref{section3} produces the information
criteria and degrees of freedom estimates for the different methods.
\Cref{section4} uses some of the herein developed
methodology to study the impact of license rationing on the sales
of individual car models in Tianjin. \Cref{section5}
discusses and concludes.

\section{The Synthetic Control Method}
\label{section2}

We are interested in the classical synthetic control method, and particularly
interested in its penalized extensions.

\subsection{The Synthetic Control Method Without Covariates}

Given an observed series $\mathbf{Y}\in\mathbb{R}^{n}$ and $p$ observed
series from ``donors'' collected as a matrix $\mathbf{X}\in\mathbb{R}^{n\times p}$,
the vector of optimal donor weights $\hat{\beta} \in \mathbb{R}^{p}$ is the solution
of the optimization problem
\begin{equation}
\min_{\beta \in \mathbb{R}^{p}}\left\Vert \mathbf{Y}-\mathbf{X}\beta\right\Vert^{2} _{2}\label{eq:SCwithoutCOVbeginning}
\end{equation}
subject to
\begin{equation}
\mathbf{1}_{p}^{T}\beta=1,\ \beta\ge0,\label{SCwithoutCOVend}
\end{equation}
where the inequality applies pointwise to vectors and $\mathbf{1}_{p} \in \mathbb{R}^{p}$ is the $p$-tuple whose entries are all 1.

The standard regression output of the synthetic control method contains
a table of regression coefficient estimates, as well as a plot of
the observed series before and after treatment, overlaid with the
treated unit's series fitted by the synthetic control method before treatment,
and forecasted after treatment. Because the forecasted series is a
function of untreated units --called ``donors''-- estimated on
pre-treatment data, it is interpreted as forecasting a non-treated
counterfactual --a ``synthetic control''-- for the treated unit.
The difference between the realized and forecasted series provides
an estimate of the treatment effect.

\subsection{The Synthetic Control Method With Covariates}

In some cases, $n_{\mathrm{cov}}$ ``covariate'' variables are believed to
satisfy --at least approximately-- the same linear relationship
as the aforementioned series, and the coefficient $\beta$ is constrained
to be a synthetic control solution for these covariates. Let the matrix
$\mathbf{D}\in\mathbb{R}^{n_{\mathrm{cov}}\times p}$ and the vector
$\mathbf{Z}\in\mathbb{R}^{n_{\mathrm{cov}}}$ collect the independent
and dependent covariate variables, respectively. The optimal solution
$\hat{\beta}$ is then obtained by solving the two-level
program
\begin{equation}
\min_{\beta}\left\Vert \mathbf{Y}-\mathbf{X}\beta\right\Vert_{2}^2
\label{eq:SCwithCOVbeginning}
\end{equation}
subject to
\begin{equation}
\beta\in \underset{\beta'\in\mathbb{S}}{\arg\min}\ \left\Vert \mathbf{Z}-\mathbf{D}\beta'\right\Vert_{V}^2,
\label{eq:SCwithCOVend}
\end{equation}
where $\mathbb{S}=\left\{ \beta\in\mathbb{R}^p: \mathbf{1}_p^T\beta=1, \beta\geq 0 \right\}$, $\left\Vert \bm{a}\right\Vert _{V}=(\bm{a}^{T}V\bm{a})^{1/2}$ for $V\succ 0$,
and $\left\Vert {\bm a}\right\Vert _{2}=({\bm a}^T{\bm a})^{1/2}$ for conformable vectors ${\bm a}$.

The diagonal matrix $V$ effectively weighs the ``observations''
of the inner regression problem. As such, it may be considered as a hyperparameter in the above formulation and may be selected using an out-of-sample criterion. Indeed, we may pick $V$ by cross-validation, solving at
each iteration the problem (\ref{eq:SCwithCOVbeginning})-(\ref{eq:SCwithCOVend})
with $V$ fixed, and searching for a ``good" $V$ matrix. This is a suggestion of \citeA{abadie2015comparative}. As detailed below, our proposed information criteria will be an attractive alternative to cross-validation.

\subsection{The Penalized Synthetic Control Method}

In settings where there are many donors relative to the number of
pre-treatment time periods, one may be concerned about overfitting.
Indeed, such a relatively large number of donors makes it more likely
to find a linear combination of donors that closely matches the to-be-treated
unit, even though they in fact have little predictive power. 
In particular,
a collection of very ``far away'' donors may give a tight in-sample
fit, even though we may have \emph{a priori} knowledge or belief that such
``far away'' donors are more likely to be poor individual matches and are expected
as such to make for a poor synthetic control. 
We may therefore want
to limit the choice of donors by relying on the conventional prior that donors more ``similar'' to the to-be-treated unit are more reliable in the specific sense that they offer a
better out-of-sample forecast.

A natural way to implement such a prior belief for frequentist estimation
is to add to the objective function a shrinkage term penalizing coefficients that put more weight on \emph{a priori} bad matches. This was suggested
in \citeA{abadie2010synthetic} and implemented in \citeA{abadie2021penalized}.
The estimator $\hat{\beta}_{\mathrm{pen}}$ is defined as the minimizer
of the penalized synthetic control method (PSCM) problem
\begin{equation}
\min_{\beta}\left\Vert \mathbf{Y}-\mathbf{X}\beta\right\Vert _{2}^{2}+\lambda\sum_{j=1}^{p}\beta_{j}\left\Vert \mathbf{Y}-\mathbf{X}_{j}\right\Vert _{2}^{2}\label{eq:PenalizedCSbeginning}
\end{equation}
subject to
\begin{equation}
\mathbf{1}_{p}^{T}\beta=1,\ \beta\ge0,\label{eq:PenalizedSCend}
\end{equation}
where $\mathbf{X}_{j}$ is the $j^{\mathrm{th}}$ column of $\mathbf{X}$.
Note that $\lambda>0$ must be selected by the user.

\subsection{Constrained Ridge SCM}

A different and more flexible approach to regularizing the synthetic control method is proposed in \citeA{arkhangelsky2021synthetic}. The optimization problem is
\begin{equation}
\min_{\beta_{0},\beta} \ \left\Vert \mathbf{Y}-\beta_{0}\mathbf{1}_{n}-\mathbf{X}\beta\right\Vert _{2}^{2}+\lambda\left\Vert \beta\right\Vert _{2}^{2}
\label{eq:CR objective}
\end{equation}
subject to
\begin{equation}
\mathbf{1}_p^{T}\beta=1,\ \beta\ge0,
\label{eq:CR constraint}
\end{equation}
where $\beta_0\in\mathbb{R}$ and $\lambda > 0$ is a regularization parameter that must be selected by the user.

First, this modified synthetic control method includes an intercept term. 
As pointed out in \citeA{doudchenko2016balancing}, it allows for a systematic additive difference between the treated unit and control units, which is an important feature of the standard difference-in-difference strategy. \citeA{ferman2021synthetic} also show that including an intercept term can improve the estimator in terms of bias and variance when the pre-treatment fit is not perfect.

Second, following \citeA{doudchenko2016balancing}, adding an $\ell_2$ regularization penalty can increase the dispersion and ensure the uniqueness of the synthetic control weights.

\subsection{Elastic Net SCM}

Another popular cousin of the SCM is the elastic net variant of \citeA{doudchenko2016balancing}. Their modified problem is
\begin{equation}
\min_{\beta_{0},\beta} \ \left\Vert \mathbf{Y}-\beta_{0}\mathbf{1}_n-\mathbf{X}\beta\right\Vert _{2}^{2}+\lambda_{1}\left\Vert \beta\right\Vert _{1}+\lambda_{2}\left\Vert \beta\right\Vert _{2}^{2},
\label{eq:EN}
\end{equation}
with $\lambda_1, \lambda_2>0$.  Remark that the optimization is not subject to the simplex constraint.

\section{Information Criteria and Degrees of Freedom}
\label{section3}

Because the selection of the tuning parameter $\lambda$ in (\ref{eq:PenalizedCSbeginning})-(\ref{eq:PenalizedSCend}), (\ref{eq:CR objective})-(\ref{eq:CR constraint}),
(\ref{eq:EN}), or of the weighting matrix $V$ in (\ref{eq:SCwithCOVbeginning})-(\ref{eq:SCwithCOVend}),
boils down to comparing models of different flexibility, we want to
use a measure of out-of-sample fit as our criterion.

Our preferred such notion of fit is the error between the model's fitted values and
the target regression function, the conditional expectation.
Specifically, we want to select a model, or tuning parameter, which minimizes the risk
\begin{equation}
\mathcal{R}:=E\left\Vert \hat{\mathbf{Y}}-E\left[\mathbf{Y}\left|\mathbf{X}\right.\right]\right\Vert _{2}^{2},
\end{equation}
where $\hat{\mathbf{Y}}\in\mathbb{R}^{n}$ are the fitted values produced
by the candidate model.



We begin by noting that, since the goal is model selection, we will care about the risk only up to a constant. Note that
\begin{equation}
\mathcal{R} =E\left\Vert \mathbf{Y}-\hat{\mathbf{Y}}\right\Vert _{2}^{2}+2E\left[\sum_{i=1}^{n}\cov\left(\left.\hat{Y}_{i},Y_{i}\right|\mathbf{X}\right)\right]+\mathrm{const} ,
\label{eq:proprisk}
\end{equation}
where ``$\mathrm{const}$" always stands for some constant that does not depend on the fitted values, and is thus irrelevant for assessment of the quality of fit.


We speak of an \emph{information criterion} to refer to estimates of $\mathcal{R}$ based on the decomposition  (\ref{eq:proprisk}), i.e., a measure of in-sample fit plus a penalty term for model flexibility.

The information criteria approach requires an estimable penalty term.
To accomplish this, we rely on the modern theory for degrees of freedom (\citeNP{meyer2000degrees}, \citeNP{zou2007degrees}) based on Stein's lemma and the computation of divergences.

Stein's
lemma --and thus the entire degrees of freedom literature-- requires the Gaussian assumption
\begin{equation}
\mathbf{Y}|\mathbf{X}\sim N\left(E\left[\mathbf{Y}|\mathbf{X}\right],\Sigma_{Y|X}\right).\label{GaussianAssumption}
\end{equation}
This assumption is discussed in detail in \Cref{sec:discussion-Gaussian}.
Assuming (\ref{GaussianAssumption}) with diagonal covariance matrix $\Sigma_{Y|X}$ and almost differentiability
(see \Cref{subsectionA4}), Stein's Lemma (\citeNP{stein1981estimation}) states that
\begin{equation}
\cov\left(\left.Y_{i},\hat{Y}_{i} \ \right| \ \mathbf{X}\right)=\sigma_{i}^{2}E_{Y|X}\left[\frac{\partial\hat{Y}_{i}}{\partial Y_{i}}\right],\label{eq:SteinLemma}
\end{equation}
where $\sigma_{i}^{2}=V\left(\left.Y_{i}\right|\mathbf{X}\right)$. 
This is a simple yet powerful result. 
While the left-hand side of (\ref{eq:SteinLemma}) is intractable,\footnote{This is immediate. Consider the favorable case of unbiased forecasts; computation of the conditional covariance requires knowledge of the conditional expectation $E\left[\mathbf{Y}|\mathbf{X}\right]$,
which is the object we are trying to estimate in the first place.} 
the right-hand side may be estimated in closed form for the regression models covered in this article.

Under the Gaussian assumption (\ref{GaussianAssumption}), equation
(\ref{eq:proprisk}) is equivalently expressed as
\begin{equation}
\mathcal{R} = E\left\Vert \mathbf{Y}-\hat{\mathbf{Y}}\right\Vert _{2}^{2}+2E\left[\sum_{i=1}^{n}\sigma_{i}^{2}E_{Y|X}\left[\frac{\partial\hat{Y}_{i}}{\partial Y_{i}}\right]\right]+\mathrm{const.}\label{genSURE}
\end{equation}

Remarkably, this closed form has a sample analog as long as we can plug in estimates for the divergence and the variance.




To produce a simple plug-in estimate of (\ref{genSURE}), we assume the data is independently and identically distributed and conditionally homoskedastic such that
\begin{equation}
\sigma_{i}^{2}=\sigma^{2},\  \forall \ i, \label{eq:Ahomoskedastic}
\end{equation}
for some $\sigma^{2}$.

Specifically, supposing (\ref{eq:Ahomoskedastic}) holds and omitting the additive constant, expression (\ref{genSURE})
rewrites as
\begin{equation}
\mathrm{IC} := E\left\Vert \mathbf{Y}-\hat{\mathbf{Y}}\right\Vert _{2}^{2}+2\sigma^{2} E \left[\mathrm{df}\left(\hat{\mathbf{Y}}\right)\right],\label{eq:homoskedasticSURE}
\end{equation}
where
\begin{equation}
\label{eq:homoskedastic_df}
\mathrm{df}\left(\hat{\mathbf{Y}}\right)=\frac{1}{\sigma^{2}}\sum_{i=1}^{n}\cov\left(\left.Y_{i},\hat{Y}_{i}\right| \mathbf{X} \right)=\mathrm{Tr}\left(E\left[\left.\nabla\hat{\mathbf{Y}}\right|\mathbf{X}\right]\right),
\end{equation}
and the divergence $\nabla\hat{\mathbf{Y}}$ is the Jacobian of $\hat{\mathbf{Y}}$ with respect to $\mathbf{Y}$.

The information criterion $\mathrm{IC}$ is our main estimand of --proportional--
risk. \emph{The key is to develop theory that produces computable
estimates of the divergence $\nabla\hat{\mathbf{Y}}$}. We do so 
in \Cref{subsectionA4}
and produce sample analogs of the degrees of freedom \eqref{eq:homoskedastic_df}
in \Cref{sec:info-criteria}.


Recognizing similarities between the lasso and SCM estimation problems, we recuperate the insight of \citeA{zou2007degrees} and use Stein's Lemma to produce an expression for the degrees of freedom and a risk estimate for the SCM that have sample analogs.


We obtained analytical and computational simplifications by assuming that the data was normally distributed, which is not exact in typical applications.  However, the resulting expression appears to be a good estimate of model flexibility in non-Gaussian regimes. 
Indeed,
both theory and simulations substantiate the claim of robustness against departure from Gaussianity.  See \Cref{sec:simulation}.
The conditional homoskedasticity invoked in (\ref{eq:Ahomoskedastic}) can, however, be more challenging.
Under heavy heteroskedasticity, (\ref{eq:homoskedasticSURE}) may not be a good approximation of (\ref{genSURE}).  As detailed below, in such regimes, we recommend the less simple but more robust alternative estimator (\ref{eq:robustIC}).

We remark that under homoskedasticity, the risk expression (\ref{eq:homoskedasticSURE})
relates in an immediate fashion to degrees of freedom as well as to
more familiar expressions for information criteria, such as AIC, BIC
or Mallow's $C_{p}$.

The other classical approach for comparing the quality of fit of different models without falling prey to overfitting is to estimate the fit on held-out data; this is referred to as \emph{cross-validation}.  
The cross-validation approach is understood to estimate
$\mathcal{CV}:=E\left\Vert \mathbf{Y}_{\mathrm{new}}-\hat{\mathbf{Y}}_{\mathrm{new}}\right\Vert _{2}^{2},$
which is equal, up to a constant, to
\begin{equation}
E\left\Vert \hat{\mathbf{Y}}_{\mathrm{new}}-E\left[\mathbf{Y}\left|\mathbf{X}_{\mathrm{new}}\right.\right]\right\Vert _{2}^{2},
\end{equation}
where $\left(\mathbf{X}_{\mathrm{new}},\mathbf{Y}_{\mathrm{new}}\right)$
is an observation drawn from the same data generating process as,
but not included in, the data set which the model forecasting $\mathbf{Y}_{\mathrm{new}}$
with $\hat{\mathbf{Y}}_{\mathrm{new}}$ was trained on.\footnote{Explicitly, $E\left\Vert \mathbf{Y}_{\mathrm{new}}-\hat{\mathbf{Y}}_{\mathrm{new}}\right\Vert _{2}^{2}=E\left\Vert E\left[\mathbf{Y}\left|\mathbf{X}_{\mathrm{new}}\right.\right]+\varepsilon_{\mathrm{new}}-\hat{\mathbf{Y}}_{\mathrm{new}}\right\Vert _{2}^{2} = E\left\Vert \hat{\mathbf{Y}}_{\mathrm{new}}-E\left[\mathbf{Y}\left|\mathbf{X}_{\mathrm{new}}\right.\right]\right\Vert _{2}^{2}+\mathrm{const},$
because $\varepsilon_{\mathrm{new}}=\mathbf{Y}_{\mathrm{new}}-E\left[\mathbf{Y}\left|\mathbf{X}_{\mathrm{new}}\right.\right]$
is independent of $\hat{\mathbf{Y}}_{\mathrm{new}}$.}

Of course, both information criteria and cross-validation, as they have their own advantages and disadvantages, are standard model selection tools in modern statistics and econometrics.
Information criteria typically require strong assumptions in order to deliver closed-form expressions.  Cross-validation typically involves some form of sample-splitting, and as such may be biased or data hungry.

Finally, remark that using Stein's lemma and computable expressions
for the divergence, we obtain an analytical and estimable expression
for $\mathrm{df}(\hat{\mathbf{Y}})$ in order to quantify analytically
the model flexibility of the synthetic control method in typical applications.  In \Cref{sec:df-sc}, we carry out such an inquiry and answer our motivating question ``does the synthetic control method overfit?''.


\subsection{The Degrees of Freedom of the Synthetic Control Methods}\label{sec:df-sc}

Once a tractable analytic expression obtains for the divergence, the
risk estimate and degrees of freedom obtain upon verifying regularity
conditions. However, computing the divergence in specific cases can
be non-trivial. Correspondingly, as \citeA{tibshirani2015stein} point out, a small industry of computing divergences has blossomed.\footnote{For instance, \citeA{meyer2000degrees} compute divergences for convex
constrained regression estimators, \citeA{mukherjee2015degrees} compute divergences for reduced rank regressions, \citeA{candes2013unbiased} compute the degrees of freedom of singular value thresholding and
spectral estimators, \citeA{deledalle2012risk} compute divergences for
singular value thresholding, \citeA{mazumder2020computing} compute divergences
for estimators with rank penalty, \citeA{minami2020degrees} gives the degrees
of freedom of estimators with submodular penalties, and \citeA{chen2020degrees} study least-squares estimators with linear penalties and constraints.} 
\Cref{subsectionA4} collects the technical derivations of the divergences.  With those in hand, the desired degrees of freedom expressions, and thus the information criteria, readily obtain as corollaries.
The degrees of freedom sometimes obtain in rather elegant closed-form expressions.


Recall that our definition of degrees of freedom for a model and its
fitted values $\hat{\mathbf{Y}}$ is
\[
\mathrm{df}\left(\hat{\mathbf{Y}}\right)=\frac{1}{\sigma^{2}}\sum_{i=1}^{n}\cov\left(\left.Y_{i},\hat{Y}_{i}\right|\mathbf{X}\right),
\]
assuming (\ref{eq:Ahomoskedastic}) holds. This is an attractive measurement of model flexibility.
Indeed, define the noise perturbations $\varepsilon_{i}=Y_{i}-E\left[\left.Y_i\right|\mathbf{X}\right]$,
$i=1,...,n$, then the definitional equivalence $\cov(Y_{i},\hat{Y}_{i} \ | \ \mathbf{X})=\cov(\varepsilon_{i},\hat{Y}_{i} \ | \ \mathbf{X})$
makes explicit that $\mathrm{df}(\hat{\mathbf{Y}})$ is
measuring how much the fitted values are adapting to noise. A familiar
instance of degrees of freedom is the number of regression coefficients
in ordinary least-squares (OLS). 
We indeed obtain that statistic back in the Gaussian setting, as a special case of the more general definition. For instance, for OLS with full column rank design matrix $\mathbf{X}\in\mathbb{R}^{n\times p}$,
under assumptions (\ref{GaussianAssumption}) and  (\ref{eq:Ahomoskedastic}), application of Stein's lemma delivers
the familiar quantity
\[
\mathrm{df}\left(\hat{\mathbf{Y}}_{\mathrm{ols}}\right)=E\left[\mathrm{Tr}\left(\mathbf{X}\left(\mathbf{X}^{T}\mathbf{X}\right)^{-1}\mathbf{X}^{T}\right)\right]=p,
\]
where $\hat{\mathbf{Y}}_{\mathrm{ols}}=\mathbf{X}\left(\mathbf{X}^{T}\mathbf{X}\right)^{-1}\mathbf{X}^{T}\mathbf{Y}$.

We are able to quantify the model flexibility, or ``effective
number of coefficients'', of the synthetic control method. Consider
the case of the synthetic control method without covariates.
Quite remarkably, we find that in spite of the implicit-- and sometimes extensive-- model selection carried out by the synthetic control method,
the expected model flexibility remains that of a linear regression with sum-to-one constraint using only the selected donors.
In other words, the implicit model selection is carried out at no additional
cost in degrees of freedom.\footnote{There are different ways to develop intuition for that result. For the lasso, \citeA{tibshirani2015stein} informally speak of the penalization on the value of the coefficients perfectly offsetting the model selection flexibility,
thus producing the same degrees of freedom, in expectation, as if
carrying out ordinary least-squares on the selected model. Problem
(\ref{eq:SCwithoutCOVbeginning})-(\ref{SCwithoutCOVend}) has an
equivalent representation as a lasso problem for non-negative coefficients and for a specific tuning parameter, where the coefficients are furthermore constrained to sum to one, thus explaining the one fewer degrees of freedom.}
\begin{proposition} [Degrees of Freedom of SCM Without Covariates]
Let $\tilde{\mathbf{X}}=(\mathbf{X}^\top,\mathbf{1}_p)^\top\in\mathbb{R}^{(n+1)\times p}$, an augmented donor matrix.
Suppose that $\mathbf{Y}|\mathbf{X}$  follows the probability law stipulated in (\ref{GaussianAssumption}) and  (\ref{eq:Ahomoskedastic}). 
Then, 
\begin{equation*}
\mathrm{df}\left(\mathbf{X}\hat{\beta}_{\mathrm{sc}}\right)=E_{Y|X}\left[\mathrm{rank}(\tilde{\mathbf{X}}_{\mathcal{A}}) \right]-1,
\end{equation*}
where $\mathcal{A}=\mathcal{A}\left(\mathbf{Y}\right):=\{ j : \hat{\beta}_{\mathrm{sc},j}(\mathbf{Y}) > 0 \}$ is the active support corresponding to a solution $\hat{\beta}_{\mathrm{sc}}(\mathbf{Y})$ of the synthetic control problem (\ref{eq:SCwithoutCOVbeginning})-(\ref{SCwithoutCOVend}).
\label{proposition4}
\end{proposition}
With continuous data, the result is even more directly interpretable
as it can be formulated directly in terms of the number of nonzero
weight coefficients.\footnote{The rank deficient case can still be handled by selecting a ``canonical''
active set $\mathcal{A}^{*}$, for instance that selected by the penalized
estimator of \citeA{abadie2021penalized} for an arbitrarily small but
nonzero tuning parameter for the penalty term.}
Let $\mathcal{H}\left(\mathbf{X}\right)
$ designate the convex hull of the columns of the design matrix.\footnote{Specifically, $\mathcal{H}\left(\mathbf{X}\right)
=\left\{  \sum_{j=1}^p \beta_j\mathbf{X}_j: \beta_j\geq 0 \ \text{for all} \ j, \  \sum_{j=1}^p \beta_j=1 \right\}$.}
\begin{corollary}
Suppose that the conditions of \Cref{proposition4} hold and that the conditioned upon design matrix $\mathbf{X}$ follows a distribution that is absolutely continuous with respect to Lebesgue measure on $\mathbb{R}^{n\times p}$. If 
$\mathbf{Y}\notin \mathcal{H}\left(\mathbf{X}\right)$,
then the synthetic control solution is unique. Moreover, with probability one over $\mathbf{X}$, we have
\begin{equation*}
\mathrm{df}\left(\mathbf{X}\hat{\beta}_{\mathrm{sc}}\right)=E_{Y|X}\left|\mathcal{A}\right|-1.
\end{equation*}
\label{corollary:dof}
\end{corollary}
\vspace{-2.5em}

An immediate implication of \Cref{corollary:dof} is that one less than the number of non-zero coefficients is an unbiased estimate of the degrees of freedom of the synthetic control method without covariates.
The finite-sample unbiasedness is an attractive feature of the estimate, as emphasized in \citeA{shen2002adaptive,efron2004estimation,shen2006optimal}.  
Furthermore, the estimate is consistent, as is the analogous degrees of freedom estimate for the lasso
\citeA{zou2007degrees}.  See \Cref{subsectionA5}.

\begin{figure}
\begin{center}
\includegraphics[scale=0.45]{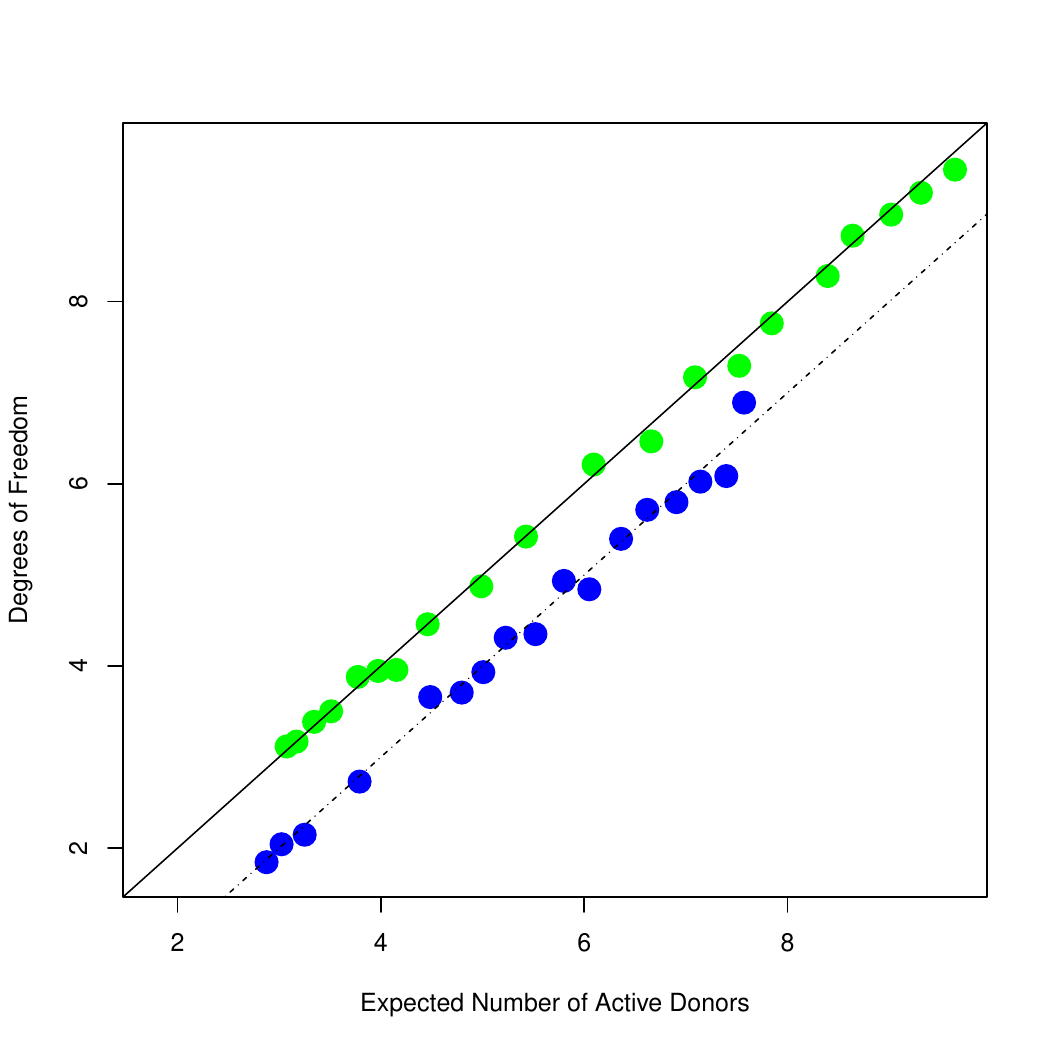}
\includegraphics[scale=0.45]{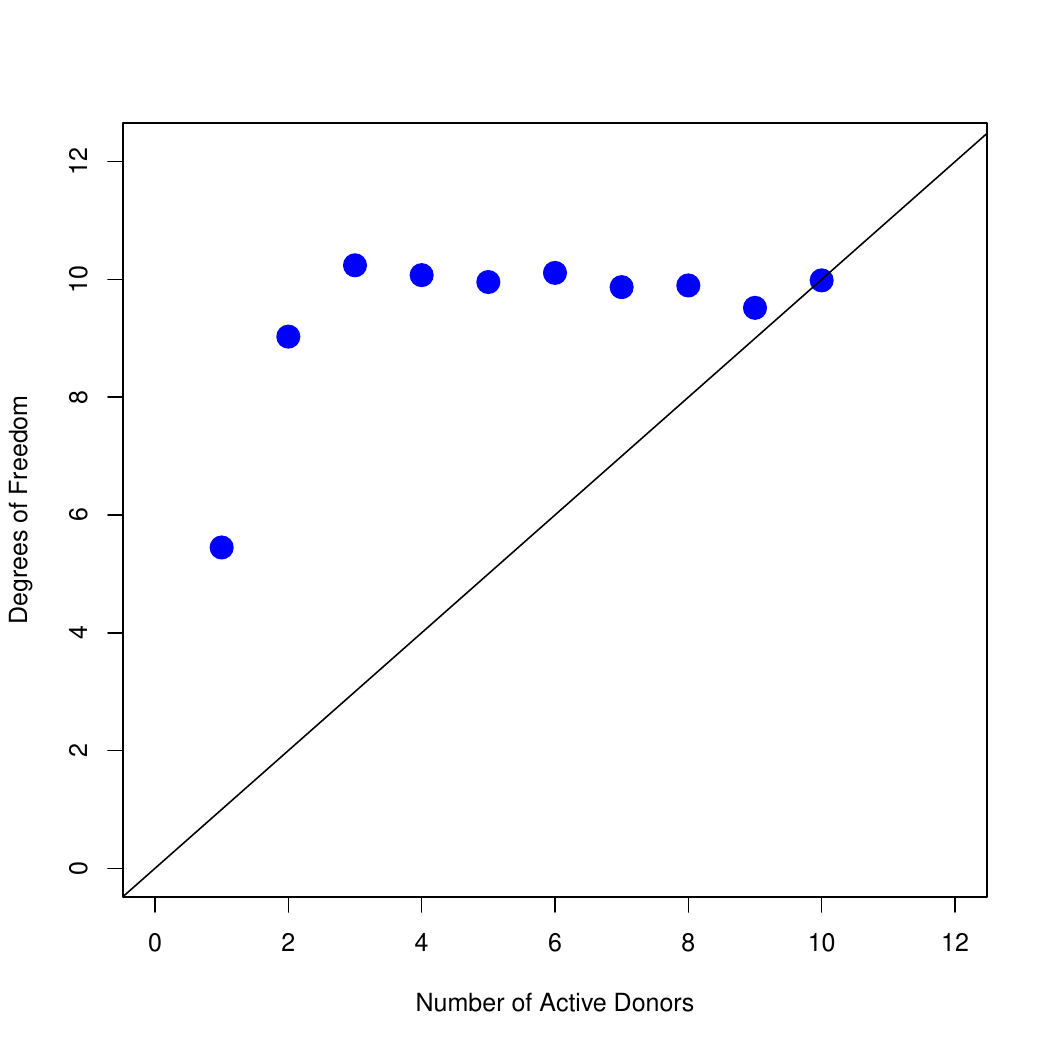}
\end{center}
\caption{\emph{Degrees of freedom of the synthetic control method without covariates
(blue) and of the lasso (green) on the left-hand side, and of the
best subset selection regression on the right-hand side.}}
\label{figure2}
\end{figure}

 \Cref{figure2} illustrates the theory on simulated data.\footnote{The simulation was done with $p=40$ donors, $n=200$ observations
for each data set, and each point is produced by averaging over 400
simulations. In order to get more variation in the degrees of freedom
of the synthetic control method, the equality constraint $\mathbf{1}_{p}^{T}\beta=1$
was replaced with $\mathbf{1}_{p}^{T}\beta=a$ for different values
of $a$.} We compare with the famous lasso result (\citeNP{zou2007degrees}), which
states that $\mathrm{df}(\mathbf{X}\hat{\beta}_{\mathrm{Lasso}})=E_{Y|X}\left|\mathcal{A}_{\mathrm{Lasso}}\right|$,
where $\mathcal{A}_{\mathrm{Lasso}}$ is the index set of the active independent variables in the lasso regression.

The right-hand side plot of \Cref{figure2} illustrates the ``real cost'' of unrestricted model selection in terms of degrees of freedom and potential overfitting. The degrees of freedom of the best subset selection method,
with respect to the size of the subset, are presented for comparison
with their synthetic control analog. While a synthetic control fitted
model has degrees of freedom one smaller than the expected number
of active donors, we can see that the explicit model search introduces
much more model flexibility.

Our main interest lies in penalized synthetic control methods.
We are therefore interested in their degrees of
freedom.
\begin{proposition}[Degrees of Freedom of Penalized Synthetic Controls]
Suppose that $\mathbf{Y}|\mathbf{X}$ follows the probability law stipulated in (\ref{GaussianAssumption}) and (\ref{eq:Ahomoskedastic}).
Suppose that the conditioned upon design matrix $\mathbf{X}$ follows a distribution that is absolutely continuous with respect to Lebesgue measure on $\mathbb{R}^{n\times p}$. 
Then, with probability one over $\mathbf{X}$, 
\begin{equation*}
\mathrm{df}\left(\mathbf{X}\hat{\beta}_{\mathrm{pen}}\right)=(1+\lambda)(E_{Y|X}\left[\mathrm{rank}(\mathbf{X}_{\mathcal{A}}) \right]-1),
\end{equation*}
where $\mathrm{rank}(\mathbf{X}_{\mathcal{A}})=\min\{|\mathcal{A}|,n\}$ and $\mathcal{A}=\mathcal{A}\left(\mathbf{Y}\right):=\{ j : \hat{\beta}_{\mathrm{pen},j}(\mathbf{Y}) > 0 \}$ is the active support corresponding to the solution $\hat{\beta}_{\mathrm{pen}}\left(\mathbf{Y}\right)$ of the penalized synthetic control problem (\ref{eq:PenalizedCSbeginning})--(\ref{eq:PenalizedSCend}).
\label{proposition5}
\end{proposition}

When interpreting \Cref{proposition5}, it is important to note that $E|\mathcal{A}|$ is a --typically decreasing-- function of $\lambda$.
It may also be instructive to observe that, if one takes $\lambda$ to be arbitrarily large, one obtains (abstracting away from ties) the one-nearest neighbor matching estimator with $\ell_2$ distance.

\begin{proposition}[Degrees of Freedom of Constrained Ridge SCM]
Suppose $\mathbf{Y}|\mathbf{X}$ follows the probability law stipulated in (\ref{GaussianAssumption}) and (\ref{eq:Ahomoskedastic}).
 Then, the constrained ridge synthetic control fit $\hat{\mathbf{Y}}_{\mathrm{crsc}}=\mathbf{X}\hat{\beta}_{\mathrm{crsc}}+\hat{\beta}_{\mathrm{crsc},0}\mathbf{1}_n$ has degrees of freedom
\begin{equation}
\mathrm{df}\left(\hat{\mathbf{Y}}_{\mathrm{crsc}}\right)=\operatorname{E}_{Y|X}\left[\sum_{j=1}^{|\mathcal{A}|}
\frac{s_j^2}{s_j^2+\lambda}+\lambda
\cfrac{\mathbf{1}_{|\mathcal{A}|}^T\left(  \mathbf{V}\left( \mathbf{S}^2+\lambda\mathbf{I}_{|\mathcal{A}|} \right)^2\mathbf{V}^T \right)^{-1}\mathbf{1}_{|\mathcal{A}|}}{\mathbf{1}_{|\mathcal{A}|}^T\left(  \mathbf{V}\left( \mathbf{S}^2+\lambda\mathbf{I}_{|\mathcal{A}|} \right)\mathbf{V}^T \right)^{-1}\mathbf{1}_{|\mathcal{A}|}}\right],
\label{dfcrsc}
\end{equation}
where $\mathcal{A}=\mathcal{A}\left(\mathbf{Y}\right):=\{ j: \hat{\beta}_{\mathrm{crsc},j}(\mathbf{Y}) > 0 \}$ is the active support, the singular value decomposition of $\tilde{\mathbf{X}}_{\mathcal{A}}=(\mathbf{I}_n-\frac{1}{n}\mathbf{1}_n\mathbf{1}_n^T)\mathbf{X}_{\mathcal{A}}\in\mathbb{R}^{n\times|\mathcal{A}|}$ has the form $\tilde{\mathbf{X}}_{\mathcal{A}}=\mathbf{U}\mathbf{S}\mathbf{V}^T$,  $\mathbf{U}\in\mathbb{R}^{n\times|\mathcal{A}|}$ and $ \mathbf{V}\in\mathbb{R}^{|\mathcal{A}|\times|\mathcal{A}|}$ are orthogonal matrices, and $\mathbf{S}\in\mathbb{R}^{|\mathcal{A}|\times|\mathcal{A}|}$ is a diagonal matrix with diagonal entries $s_1\geq s_2\geq...\geq s_{|\mathcal{A}|}\geq 0$, the singular values of $\tilde{\mathbf{X}}_{\mathcal{A}}$.
\label{prodfcrsc}
\end{proposition}
\begin{remark}
The change-of-variable from  $\mathbf{X}_{\mathcal{A}}$ to $\tilde{\mathbf{X}}_{\mathcal{A}}$ aligns with the observation that demeaning the data before applying the SCM estimator is equivalent to adding an intercept (Footnote 4, \citeNP{ferman2021synthetic}).
\end{remark}
\begin{remark}
We discuss two special cases. As $\lambda\downarrow0$, the constrained ridge SCM approaches the synthetic control estimator with an intercept. Its degrees of freedom is the expected number of nonzero coefficients $E|\mathcal{A}|$, implying that adding the intercept costs one degree of freedom. If we remove the sum-to-one and nonnegative constraints, then the estimator (after demeaning) reduces to the usual ridge regression, whose degrees of freedom is exactly the first term of (\ref{dfcrsc}). 
We thus recuperate, as a special case, the result of \citeA[~Equ. 3.50]{hastie2009elements}.
\end{remark}

The elastic net SCM can be expressed as a lasso problem and is treated as such by \citeA{tibshirani2012degrees}. 

\begin{proposition}[Degrees of Freedom of Elastic Net SCM, \cite{tibshirani2012degrees}]
Suppose that $\mathbf{Y}|\mathbf{X}$ follows the probability law stipulated in (\ref{GaussianAssumption}) and (\ref{eq:Ahomoskedastic}).
Then, the elastic net synthetic control fit $\hat{\mathbf{Y}}_{\mathrm{elast}}=\mathbf{X}\hat{\beta}_{\mathrm{elast}}+\hat{\beta}_{\mathrm{elast},0}\mathbf{1}_n$ has degrees of freedom
\begin{align*}
\mathrm{df}(\hat{\mathbf{Y}}_{\mathrm{elast}})&=E_{Y|X}\left[\mathrm{Tr}\left(\tilde{\mathbf{X}}_{\mathcal{A}}\left(\tilde{\mathbf{X}}_{\mathcal{A}}^{T}\tilde{\mathbf{X}}_{\mathcal{A}}+\lambda_{2}\mathbf{I}_{|\mathcal{A}|}\right)^{-1}\tilde{\mathbf{X}}_{\mathcal{A}}^{T}\right)\right]+1
=E_{Y|X}\left[\sum_{j=1}^{|\mathcal{A}|}
\frac{s_j^2}{s_j^2+\lambda_2}\right]+1,
\end{align*}
where 
$\mathcal{A}=\mathcal{A}\left(\mathbf{Y}\right):=\{ j: \hat{\beta}_{\mathrm{elast},j}(\mathbf{Y}) \neq 0 \}$ is the active support and $\tilde{\mathbf{X}}_{\mathcal{A}}$ as well as $\mathbf{S}=\text{diag}(s_1,...,s_{|\mathcal{A}|})$ are defined in \Cref{prodfcrsc}.
\label{propescm}
\end{proposition}



\begin{remark}
The degrees of freedom for elastic net SCM and ridge regression with an intercept share the same expression. However, their computed values generally differ because the active set $\mathcal{A}$ depends on the estimated coefficients, which vary by method.
\end{remark}


In the synthetic control problems with covariates (\ref{eq:SCwithCOVbeginning})-(\ref{eq:SCwithCOVend}), the ``covariates'' play the role of
special observations, or time periods, and the synthetic control coefficient $\beta$ is constrained
to produce the best fit possible on these special observations.

When the inner problem admits multiple solutions, it restrains the model flexibility
without imposing a unique solution, and that is reflected by a reduction in the degrees of
freedom equal to the number of covariates, $n_{\mathrm{cov}}$.

Define the augmented donor matrix $\tilde{\mathbf{X}}=(\mathbf{X}^\top,\mathbf{D}^\top,\mathbf{1}_p)^{\top}\in\mathbb{R}^{(n+n_{\mathrm{cov}}+1)\times p}$
and let $\mathrm{relint}\,\mathcal{H}(\mathbf{D})$ designate the relative (to its affine hull) interior of $\mathcal{H}(\mathbf{D})$. 

\begin{proposition}  [Degrees of Freedom of SCM With Covariates]
 
Suppose that, conditionally on $(\mathbf{X},\mathbf{D},\mathbf{Z})$, the outcome $\mathbf{Y}$ follows the probability law stipulated in (\ref{GaussianAssumption}) and (\ref{eq:Ahomoskedastic}).
Suppose that $\mathbf{D}\in\mathbb{R}^{n_{\mathrm{cov}}\times p}$ follows a distribution that is absolutely continuous with respect to Lebesgue measure and that $n_{\mathrm{cov}} \le p$.
 If $\mathbf{Z}\in \mathrm{relint}\, \mathcal{H}(\mathbf{D})$, then
\[
\mathrm{df}(\hat{\mathbf{Y}})
=
E_{Y\mid X,D,Z}\!\left[\mathrm{rank}\!\left(\tilde{\mathbf{X}}_{\mathcal{A}}\right)\right]
-
n_{\mathrm{cov}}-1,
\]
with probability one over $\mathbf{D}$, where $\hat{\mathbf{Y}}=\mathbf{X\hat{\beta}(\mathbf{Y})}$, and $\mathcal{A}(\mathbf{Y})=\{j:\hat{\beta}_j(\mathbf{Y})>0\}$ is the active set corresponding to a solution $\hat{\beta}(\mathbf{Y})$ of the synthetic control problem
(\ref{eq:SCwithCOVbeginning})--(\ref{eq:SCwithCOVend}).
\label{proposition7}
\end{proposition}

Analogously to the case without covariates, if the design matrix $\mathbf{X}$ is drawn from a continuous distribution,  the degrees of freedom expression can be further simplified. 


\begin{corollary}
Assume the conditions of \Cref{proposition7} hold and $(\mathbf{X},\mathbf{D})$ has a distribution that is absolutely continuous with respect to Lebesgue measure. If $\mathbf{Y}\notin \{ \mathbf{X}\beta: \mathbf{D}\beta=\mathbf{Z}, \mathbf{1}^T\beta=1, \beta \geq 0 \}$,
then the synthetic control solution is unique. Moreover, with probability one over $(\mathbf{X},\mathbf{D})$, then
\begin{equation*}
\mathrm{df}\left(\hat{\mathbf{Y}}\right)=E_{Y|X,D,Z}\left|\mathcal{A}\right|-n_{\mathrm{cov}}-1.
\end{equation*}
\label{cor:card_with_cov}
\end{corollary}
\vspace{-2.5em}

In some cases, the covariates uniquely determine the synthetic control coefficient $\beta$.
This happens when the independent variable covariate $\mathbf{Z}$ cannot be reproduced exactly
as a convex combination of the dependent covariate variables $\mathbf{D}$.
In other words, when $\mathbf{Z}$ is not in the convex hull of $\mathbf{D}$, which we denote
$\mathbf{Z} \notin \mathcal{H}(\mathbf{D})$.
When that is the case, the outcome variable $\mathbf{Y}$ does not influence the fit $\hat{\mathbf{Y}}$,
and no overfitting whatsoever arises.
With cross-validation for instance, when that is the case, the in-sample fit in the training
set is of the same magnitude---and equal in expectation---to the out-of-sample fit in the
test set.
This absence of fitting to $\mathbf{Y}$ is likewise captured by the degrees of freedom of the
synthetic control method with covariates, which are zero in that specific case.

\begin{proposition} [Degrees of Freedom of SCM With Covariates] 
Suppose all assumptions stated in \Cref{proposition7} hold.  If $\mathbf{Z}\notin \mathcal{H}(\mathbf{D})$, then the synthetic
control fitted values $\hat{\mathbf{Y}}$ have degrees of freedom $\mathrm{df}(\hat{\mathbf{Y}})=0$.
\label{prop:df_SCM_cov_notin}
\end{proposition}

\subsection{Information Criteria for the Synthetic Control Methods}\label{sec:info-criteria}

The information criteria estimate, the sample analog to \eqref{eq:homoskedasticSURE},
is
\begin{equation}
\widehat{\mathrm{IC}}:=\left\Vert \mathbf{Y}-\hat{\mathbf{Y}}\right\Vert _{2}^{2}+2\hat{\sigma}^{2}\hat{\mathrm{df}}\left(\hat{\mathbf{Y}}\right).
\label{eq:information-criteria-estimate}
\end{equation}
All the degrees of freedom expressions $\mathrm{df}(\hat{\mathbf{Y}})$
derived in \Cref{sec:df-sc} are expectations of observed quantities, the
latter are thus used as natural sample analogs $\hat{\mathrm{df}}(\hat{\mathbf{Y}})$.

The sample variance in (\ref{eq:information-criteria-estimate}) is estimated as
\[
\hat{\sigma}^{2}=\frac{1}{n-\hat{p}}\sum_{i=1}^{n}\hat{\varepsilon}_{i}^{2},
\]
where the fitted residuals are from the unpenalized synthetic control estimate
and $\hat{p} = \hat{\mathrm{df}}(\mathbf{X}\hat{\beta}_{\mathrm{sc}})$. 
This is the general approach suggested by \citeA{friedman2001elements}.
If one is concerned that the unpenalized synthetic control method is overfitting, our recommended, conservative approach is to use out-of-sample errors as $\hat{\varepsilon}_i$, for $i$ ranging over a test set in the pre-treatment period.
This will naturally tend to produce an inflated, rather than a deflated, variance estimate.
Because larger values of the tuning parameter in, say, the penalized synthetic control method, effectively shrink the estimate towards the matching estimate and tend to produce an estimate that is more stable, we consider this ``upward bias'' as erring on the side of caution.
See Section \ref{section4} for an example.

The estimated information criterion used to select the tuning parameter
$\lambda$ when implementing the penalized synthetic control method
is
\[
\widehat{\mathrm{IC}}_{\mathrm{pen}}(\lambda):=\left\Vert \mathbf{Y}-\hat{\mathbf{Y}}\right\Vert _{2}^{2}+2\hat{\sigma}^{2}\left(1+\lambda\right)\left(\left|\mathcal{A}\right|-1\right),
\]
where, importantly, both $\hat{\mathbf{Y}}$ and $\mathcal{A}$ are functions of $\lambda$.


Likewise, the tuning parameter $\lambda$ in the constrained ridge SCM, the pair $(\lambda_1, \lambda_2)$ in the elastic net SCM, and the weighting matrix $V$ in the SCM with covariates can be selected by minimizing the sample information criteria that append the in-sample loss $\|\mathbf{Y} - \hat{\mathbf{Y}}\|_2^2$ with the corresponding sample analog of the degrees of freedom, as provided in \Cref{prodfcrsc,propescm,proposition7}, respectively.




\subsection{Heteroskedasticity-Robust Information Criteria}\label{sec:discussion-Hete}



\begin{figure}
\begin{center}
\includegraphics[scale=0.25]{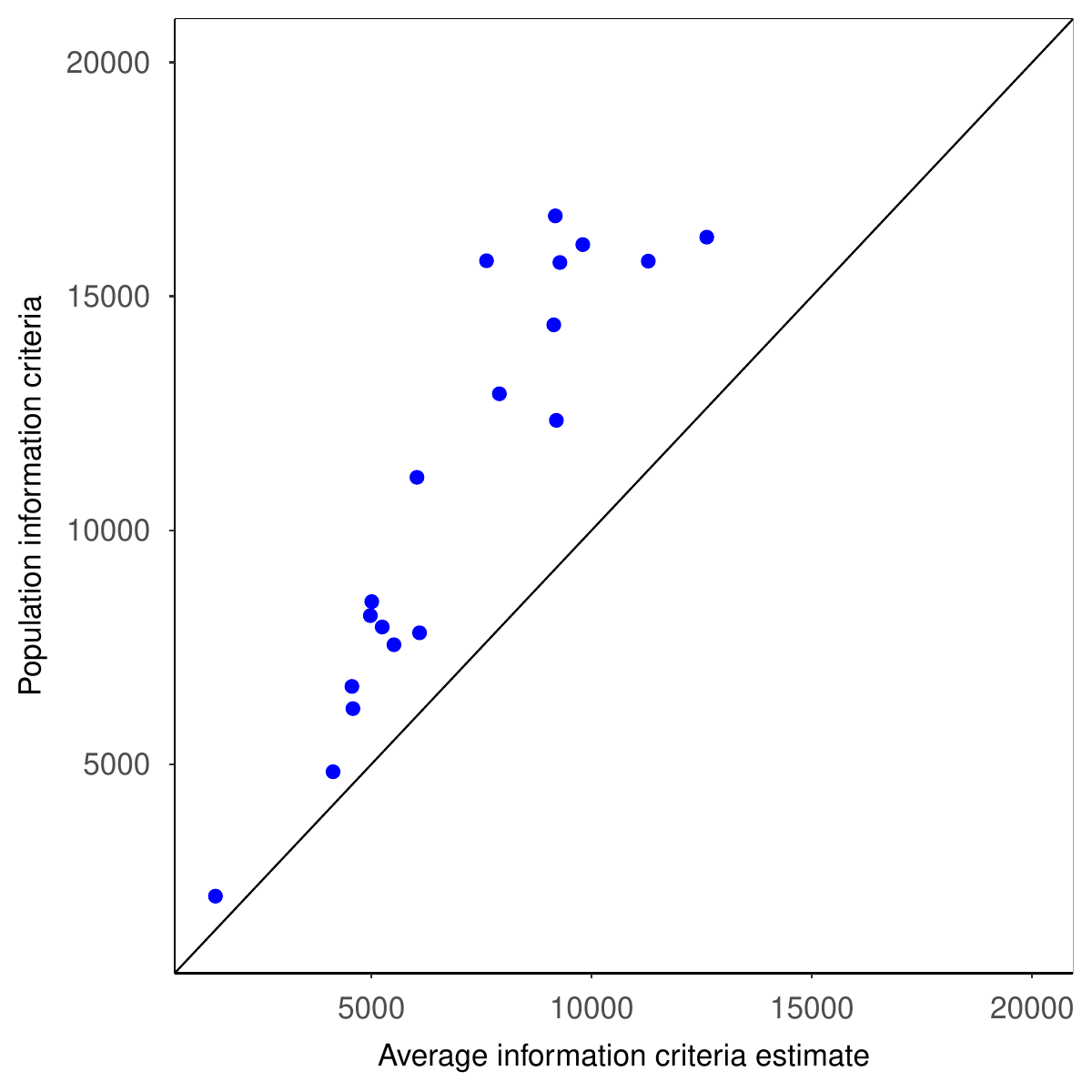} \
\includegraphics[scale=0.25]{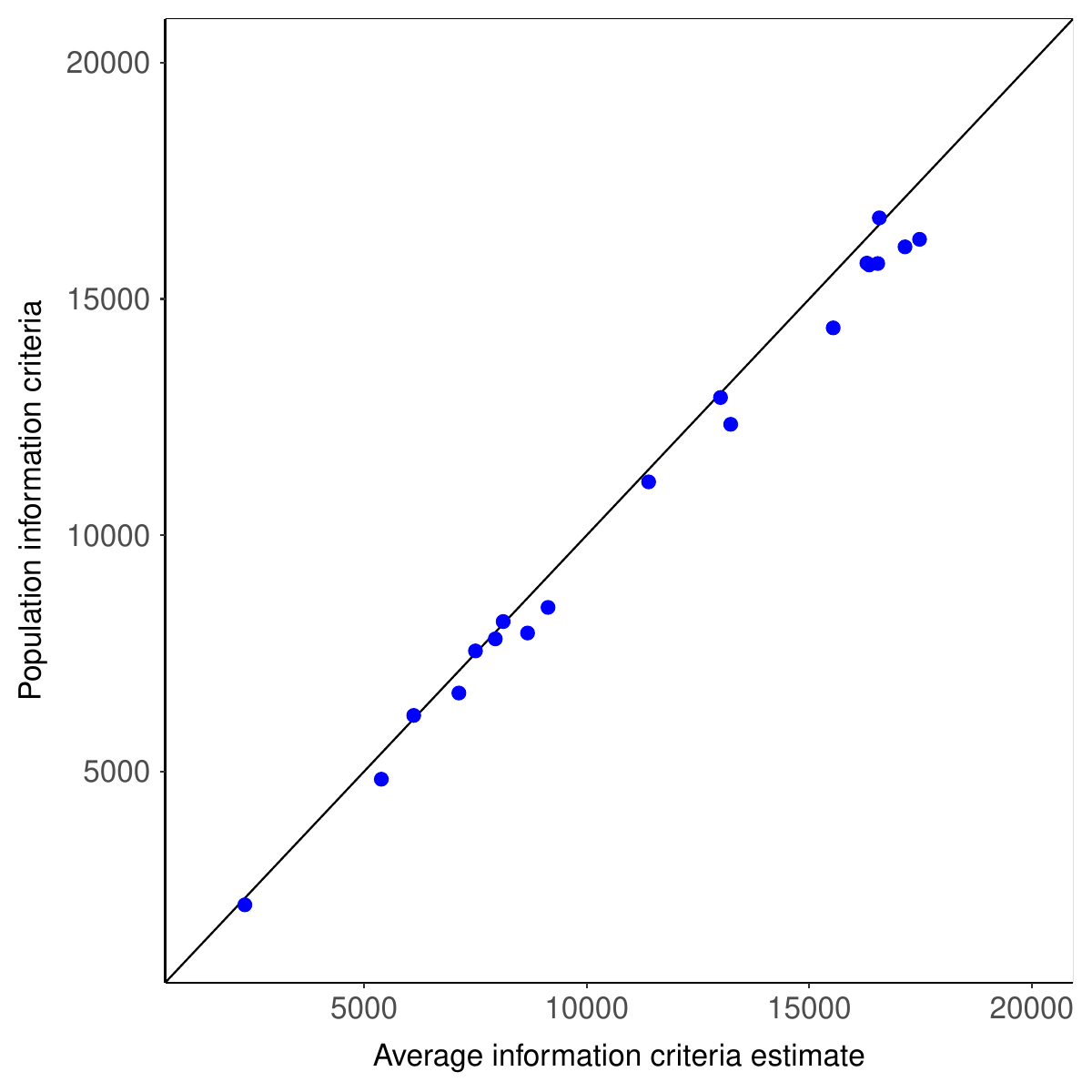} \
\includegraphics[scale=0.25]{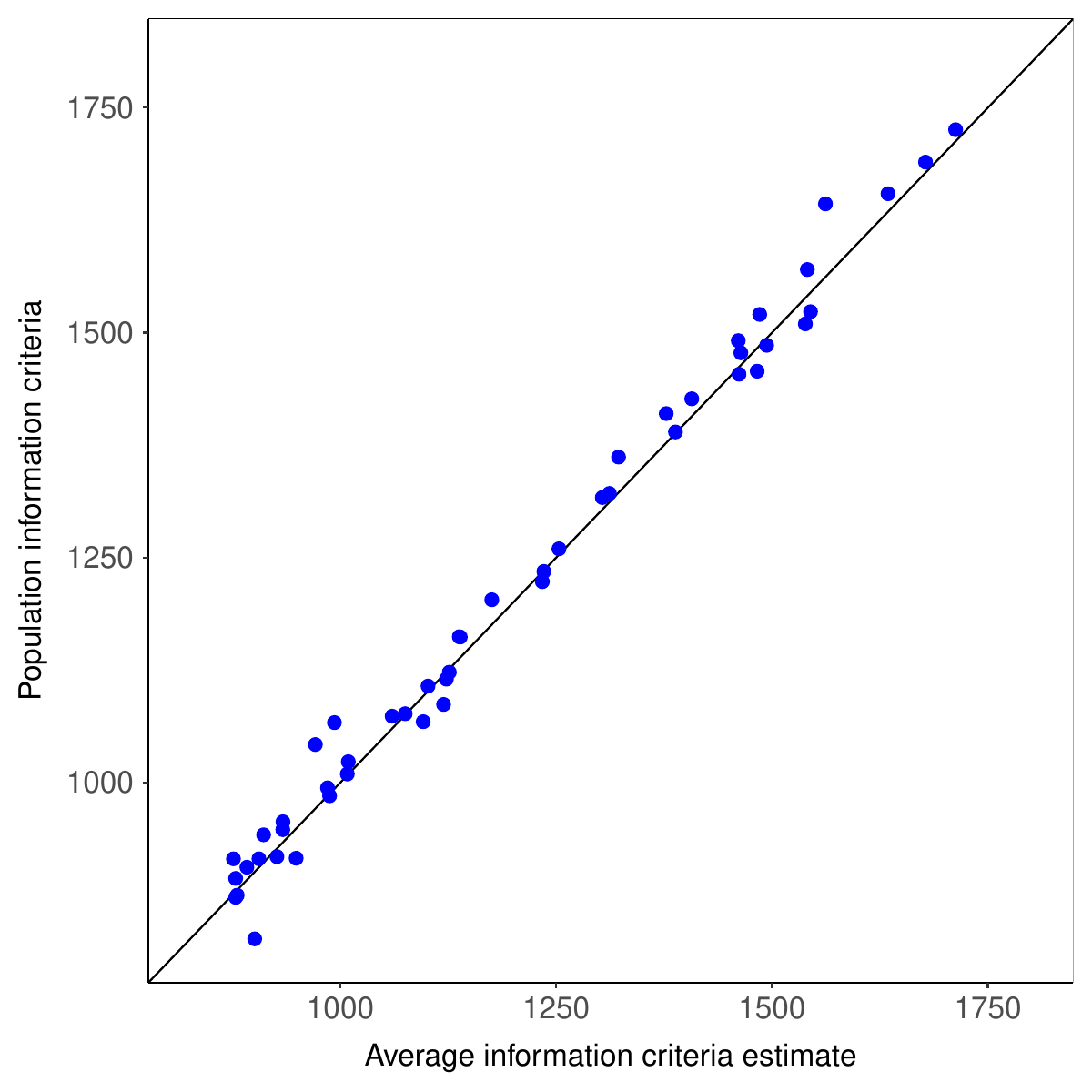}  
\end{center}

\vspace{1em}

\footnotesize\textit{} The left-hand side figure presents pairs of population and average estimated IC using (\ref{eq:homoskedasticSURE}) for different samples simulated from a high-heteroskedasticity data generating process.  The center and right-hand side figures present pairs of population and average estimated IC using (\ref{eq:robustIC}) for different samples simulated from a high-heteroskedasticity and homoskedastic  data generating process, respectively.  In the heteroskedastic DGP, the outcome variables are generated using a linear model with covariate $X_{t,i}$ and errors drawn from a normal distribution with mean zero and variance proportional to $\exp{(X_{t,1})}$.

\caption{\emph{Assessment of robustness to heteroskedasticity}}  

\label{fig:robustsim}
\end{figure}

The homoskedasticity assumption \eqref{eq:Ahomoskedastic} may appear restrictive to some users and this concern warrants discussion. 
We find that the proposed information criteria (\ref{eq:homoskedasticSURE}) performs well in the regime of moderate heteroskedasticity but exhibits substantial bias in simulation settings with high heteroskedasticity, see the left panel in \Cref{fig:robustsim}. 

We are thus motivated to produce an alternative information criterion that is more robust to heteroskedasticity.
As detailed in Appendix \ref{subsection:consistency_hc}, one such information criterion for high-heteroskedasticity regimes is
\begin{equation}
\widehat{\mathrm{IC}}_{\mathrm{HR}}=n^{-1}\left\Vert \mathbf{Y}-\hat{\mathbf{Y}}\right\Vert _{2}^{2}+\frac{2}{n-\hat{p}}\sum_{i=1}^{n}\hat{\varepsilon}_{i}^{2}\frac{\partial\hat{Y}_{i}}{\partial Y_{i}}
\label{eq:robustIC}
\end{equation}
where $\hat{p}:=\hat{\mathrm{df}}(\mathbf{X}\hat{\beta}_{\mathrm{sc}})$ is the degrees of freedom estimate of the unpenalized SCM estimator. The second term in (\ref{eq:robustIC}) serves as a consistent estimator for the covariance penalty term in (\ref{genSURE}).



The central and right panels of Figure \ref{fig:robustsim} display the performance of the heteroskedasticity robust information criteria estimate (\ref{eq:robustIC}) under both homoskedasticity and high heteroskedasticity regimes. We see that the heteroskedasticity robust information criteria remains unbiased in the homoskedastic case and approximates well, on average, its population analog even under severe heteroskedasticity. 

Our information criteria can also be extended to accommodate a non-diagonal conditional covariance matrix $\Sigma_{Y|X}$, as arises in the presence of serial dependence. 
By leveraging tools from heteroskedasticity-and-autocorrelation-robust (HAR) variance estimation, we produce a HAR information criteria that is robust to both conditional heteroskedasticity and serial correlation of unknown form. The formal procedure is detailed in Appendix \ref{section:hac_ic}.

The methodology can also accommodate structured conditional covariance matrices. 
 Consider \citeA{liu1994siegel}, who gives a multivariate
generalization of Stein's Lemma according to which the information
criteria become
\[
E\left\Vert \mathbf{Y}-\hat{\mathbf{Y}}\right\Vert _{2}^{2}+2E\left[\mathrm{Tr}\left(\Sigma_{Y|X}E\left[\left.\nabla\hat{\mathbf{Y}}\right|\mathbf{X}\right]\right)\right].
\]
The form is more general, but it brings about the challenge of estimating
$\Sigma_{Y|X}$. It may be estimated in different ways. 
If one is
willing to make the necessary model specification assumptions,
one may 
use a structured conditional covariance model.

\subsection{Discussion of Gaussian Assumption}\label{sec:discussion-Gaussian}

While developing an estimator under Gaussian assumptions is by no
means exceptional, it remains important to ask if such an assumption
can be relaxed, and if we are robust to its misspecification.

First,
recent developments have formalized the intuition that degrees of freedom as herein defined capture the flexibility of a method in a robust manner. Specifically, \citeA{fathi2022relaxing} provide theory formalizing how to assess the bias incurred when using standard forms of SURE even though the observations are not Gaussian.
They argue that the bias can be sufficiently small so as to yield estimates useful for the selection of tuning parameters. Formally, they produce bounds on the bias of the SURE estimate which go to zero as the distribution of the observations approaches a normal distribution.

Second, we can see from the simulations in \Cref{sec:simulation} that while the
semi-synthetic data is decidedly not Gaussian, our estimate of the
degrees of freedom, and thus of out-of-sample performance, is remarkably
robust.

\section{Forecasting Counterfactual Car Sales Under Rationing in Tianjin}
\label{section4}

In Tianjin, on December 16, 2013, the municipal government introduced
a hybrid half-lottery and half-auction system for the procurement
of license plates which heavily rationed the number of licenses
issued.\footnote{The measures to control vehicle purchase and restrict the traffic
in Tianjin were first announced in a press conference by the Tianjin
Municipal People\textquoteright s Government at 7 p.m. of December
15, 2013 (The State Council of the People\textquoteright s Republic
of China, 2013). The controls and restrictions were effective five
hours later on December 16, 2013 at midnight. The Tianjin Municipal
People\textquoteright s Government suspended any new vehicle registration
and vehicle transfer in Tianjin between December 16, 2013 and January
15, 2014 in order to ensure a smooth transition and preparation for
the new rules. For more detailed background information, see \citeA{daljord2021black}.} This was done in order to limit pollution from car emissions, which
had become a public health hazard.

Because the auction allowed wealthier individuals a better chance
of obtaining a license, the rationing induced a change in the population
of car buyers, and therefore a change in the demand for different
models (\citeNP{li2018better}).

This change in demand may be non-trivial. While \citeA{daljord2021black}
document the change in the population of car buyers in Beijing when there is a lottery without auction and the licenses are transacted solely
on a black market, they do not study the impact of rationing
on the demand for specific models. However, manufacturers and policymakers
would naturally be interested in the impact of such a policy on the sales
of car models of different prices and fuel consumption.

Specifically, we are interested in establishing model-specific counterfactual
demand, and thus the impact on sales, of individual car models.

The strategy is to use the city of Shijiazhuang to build counterfactuals.
Shijiazhuang is comparable to Tianjin in that, for instance, it is
geographically close and has a population on the same order of magnitude
(15 vs 11 million). Crucially, Shijiazhuang did not have rationing
and we observe sales data for that city.

We highlight two features of the empirical analysis.  The first feature is that we use the synthetic control method not because we are missing a good match, but to combine many good albeit noisy matches so as to attenuate variance while minimizing cost in bias.  Bias and variance are traded off according to the tuning parameter of the penalized synthetic control.

The second feature is that we consider the simulation and data analysis in conjunction,
and simulate from a data generating process meant to emulate that
of the observed data. We do this for two reasons. First, the simulation
should be thought of as part of the application; to best carry out model
selection (i.e., picking the tuning parameter), we must first elect a model selection method, and since we do not have enough observations to empirically compare model selection methods on observed data (a very data-hungry procedure), we resort to simulated data. Second, we are interested in the simulation for its own sake; in
order to properly assess how practical the theory is, we want to carry out robustness checks on simulated yet realistic data.

\subsection{Empirical Approach}

An apparently natural approach would be to use matching with a single
match. Indeed, all models investigated are sold in both Shijiazhuang
and Tianjin, and thus the Shijiazhuang sales for, say, a Toyota Highlander,
make a natural counterfactual for the Toyota Highlander sales in Tianjin,
post rationing. However, the relatively small number of sales for any
given model makes the time series relatively noisy and methodology
tackling this issue ought to be considered and compared.

On the one hand, filtering is a natural way to tackle the issue of
noisy donors data. Specifically, we want to average over similar time
series and thus reduce noise at little cost in bias. Intuitively,
the underlying demand patterns for analogous models from different
brands such as, say, a Toyota Highlander and a Honda CR-V, may be very
similar, thus allowing for variance reduction at little cost in bias
when averaging over both series to build a counterfactual series for
the Toyota Highlander. The synthetic control method is specifically aimed at
building a control unit by averaging over multiple possible controls.
We thus consider it as a natural alternative to matching.\footnote{Standard $k$-nearest-neighbor matching ($k>1$) would impose equal weights on potentially distant donors.  Kernel-weighted matching could relax this restriction, but in practice requires tuning bandwidth parameters.}

To the best of our knowledge, this is a novel use of the synthetic control method.
The SCM is typically used when an obvious match cannot be found; here,
a natural match can be found, but it is noisy and we prefer to combine
with slightly imperfect matches in order to attenuate the noise.

On the other hand, regularization is a natural way to tackle the issue of a noisy to-be-treated, or outcome variable, particularly when there are many independent variables. 
This suggests the use
of a regularized alternative to synthetic controls, such as
penalized synthetic controls (\citeNP{abadie2021penalized}). 
Conveniently, penalized methods can themselves provide evidence of overfitting when the unpenalized model is too flexible.  
This is explored in Figure \ref{Fig:single_model_control_city}, where we see that regularizing the synthetic control estimator improves its out-of-sample performance (left panel), and we find preliminary evidence that the SURE provides a reasonable estimate of the optimal tuning parameter.
Of course the probability simplex constraint, as well as the covariates if any are used, act as a form of regularization.
We find that the additional regularization induced by the above two methods is beneficial in our application.

Since our methodological premise is that the unpenalized synthetic control method overfits, we use out-of-sample errors to estimate the variance term in the information criteria.  Specifically, we fit an unpenalized synthetic control estimator on the first two-thirds of the pre-treatment data, and collect the out-of-sample errors $\hat{\varepsilon}_i$, for $i$ ranging over the last third of the pre-treatment data.

\subsection{Simulation}\label{sec:simulation}

We consider two complementary simulation designs. 
First, we consider a Gaussian factor model. In that design, the Gaussian assumption underlying the theory applies, and we can furthermore compute the true risk for comparison.
Second, we keep with the factor model but sample the outcome unit's residuals from their empirical distribution, thus assessing the robustness of the theory and the relative performance of different model selection methods in a more realistic design in which we can still compute the true risk.

\begin{figure}
\begin{center}
\includegraphics[scale=0.30]{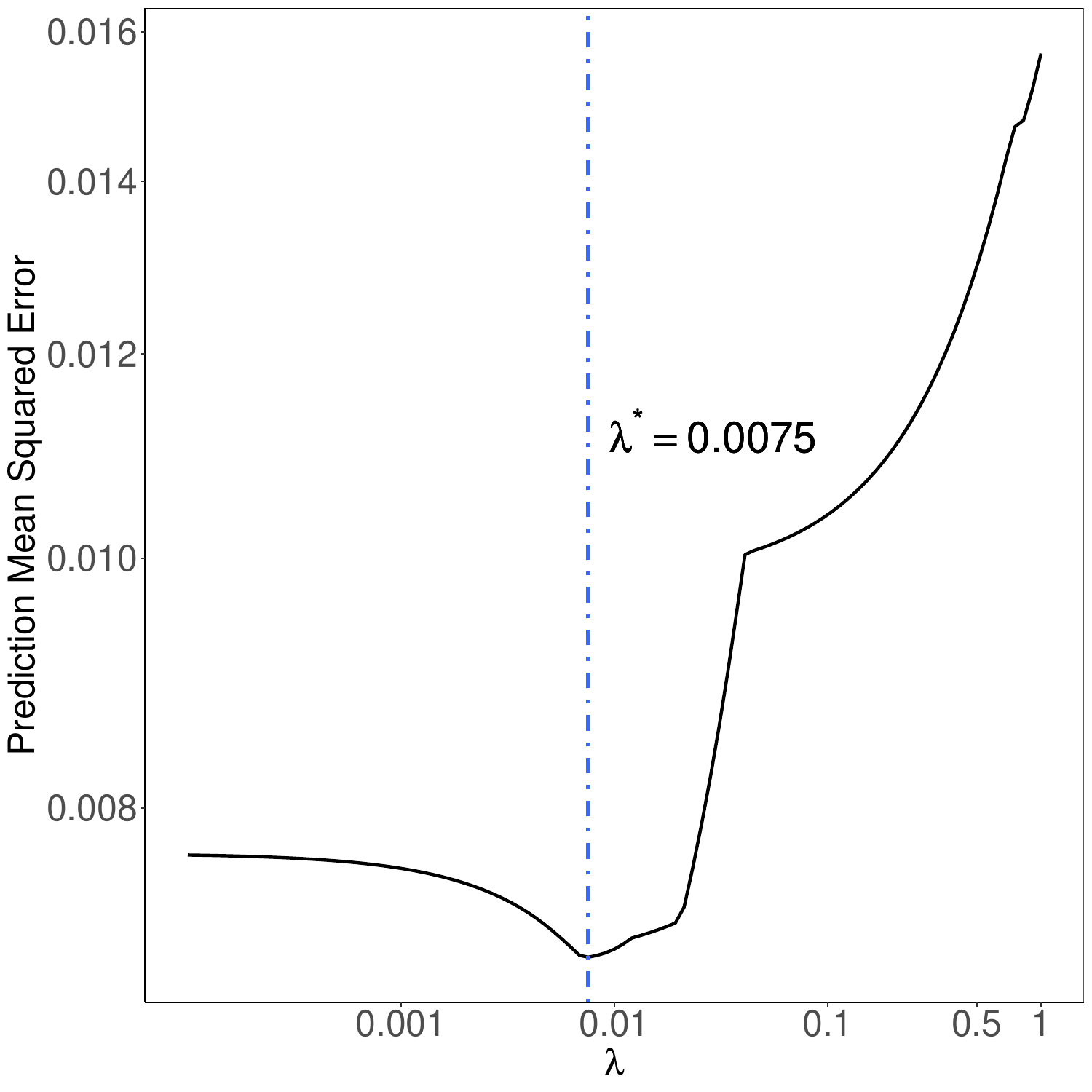}  \ \ \ \ \
\includegraphics[scale=0.30]{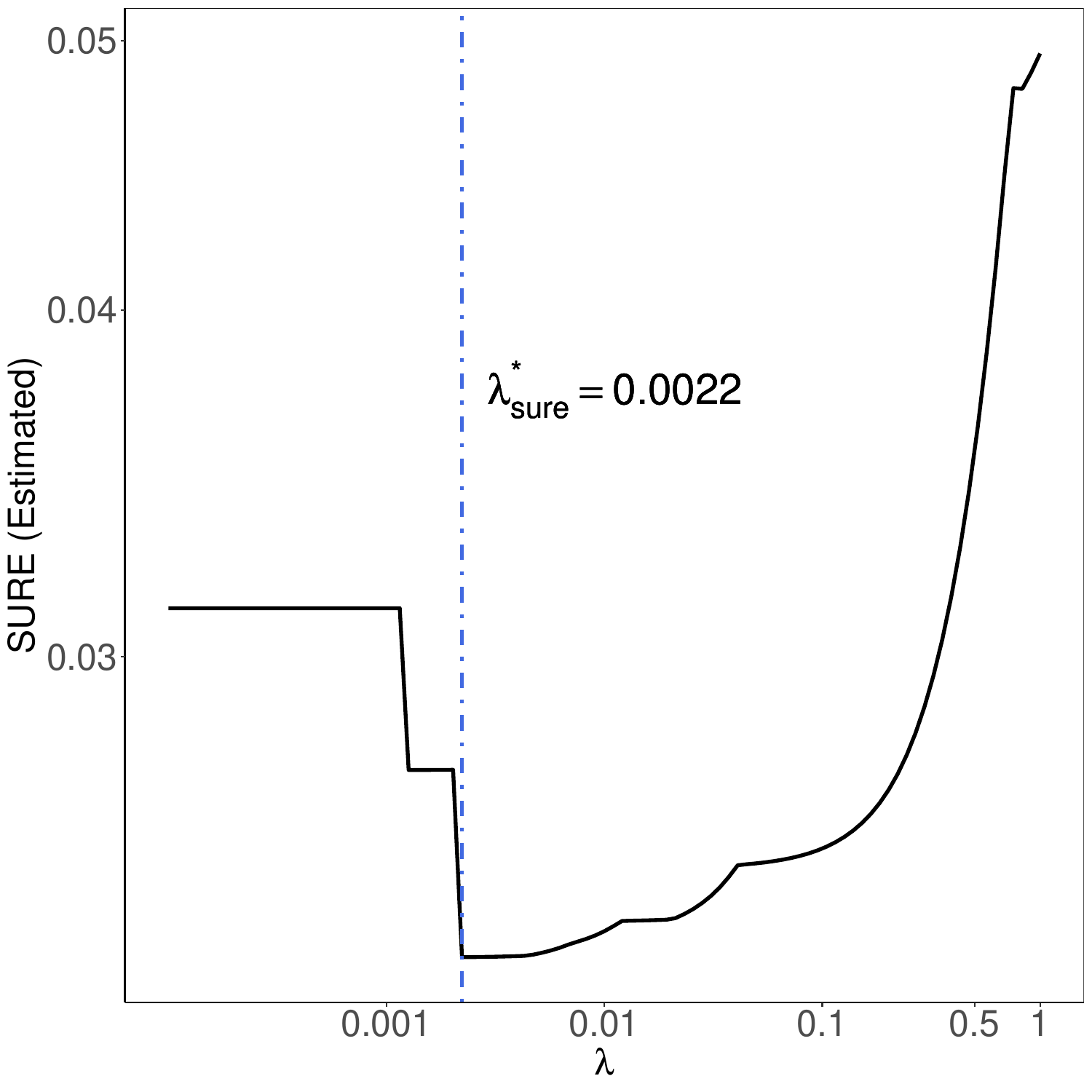}
\end{center}
\footnotesize\textit{} 
Left panel: out-of-sample prediction mean-squared error, computed by first estimating PSCM using pre-treatment Highlander market shares data from Shijiazhuang and then evaluating the mean-squared error in the post-treatment periods. Right panel:  estimated SURE computed using the same pre-treatment data. Both are plotted as functions of $\lambda$, with the vertical blue dashed line in each panel indicating the minimand.
\caption{\emph{\label{fig:bootstrap_risk}Prediction MSE and SURE Estimates for Different \(\lambda\) Values in PSCM (Shijiazhuang)}}
\label{Fig:single_model_control_city}
\end{figure}

While our interest is in the quality of fit of the synthetic control
method, we use and estimate a more involved but more flexible model to simulate from. Motivated by the theoretical insight of \citeA{abadie2010synthetic} and by the successful implementation on Current Population
Survey (\citeNP{ferman2021synthetic}), we fit a factor model.

We estimate a factor model on the Shijiazhuang data and simulate from the fitted model. Let $Y_{t}$ denote the sales of the treated unit in month $t$, and let $\mathbf X_{t}=(X_{1t},\dots ,X_{pt})^{T}$ collect sales for the $p$ donor units. We model sales as
\begin{equation}
Y_t=\psi_t^{T}L_0+U_{0t}\quad \mathrm{and}\quad X_{jt}=\psi_t^{T}L_j+U_{jt},
\qquad j=1,\dots,p,\; t=1,\dots,T,
\label{eq:fmy}
\end{equation}
where $\psi_t\in\mathbb{R}^r$ is a vector of common factors, $L_j\in\mathbb{R}^r$ is the factor loading for unit $j$, and $U_{jt}$ is an idiosyncratic error. Let $\mathbf{U}_t=(U_{0t},U_{1t},...,U_{pt})^T$, we normalize that
\begin{equation}
E\left[\psi_{t}\right]=\mathbf{0}_{r},\ E\left[\psi_{t}\psi_{t}^{T}\right]=\mathbf{I}_r,\ E[\mathbf{U}_t]=\mathbf{0}_{p+1}, \ E\left[\mathbf{U}_{t}\mathbf{U}_{t}^{T}\right]=\Sigma,
\label{equ:factor_restriction}
\end{equation}
and assume $E[\psi_t\mathbf{U}_t^T]=0$. We take $\Sigma \in \mathbb{R}^{(p+1)\times (p+1)}$ to be diagonal so the idiosyncratic shocks are uncorrelated across units. To connect the factor structure to synthetic controls, let $L_{-0}=[L_{1},\dots ,L_{p}]\in\mathbb R^{r\times p}$, and impose $L_{0}=L_{-0}\,\beta^{*}$ for some weights $\beta^{*}\geq 0, \mathbf{1}^T\beta^{*}=1$. In implementation, we use $\hat{\beta}$ from the unpenalized synthetic control fit as a plug-in for $\beta^{*}$.
We estimate the factors and loadings from the  Shijiazhuang panel using the algorithm of \citeA{xu2017generalized}.\footnote{We model the preprocessed data, which was passed through an MA(3) filter and was demeaned.
We fit separate models for sales and market shares. The extended model with time fixed effects was considered but has higher information criteria \cite{bai2002determining}. }

To generate a full panel, we simulate a total of $T=36$ time periods from the factor model, consisting of $T^*=23$ pre-treatment periods and 13 post-treatment periods, as in the application. 

 For our first simulation design, we consider the Gaussian factor model.  Specifically, we simulate the data independently across $t$ from
\[
\left(\begin{array}{c}
Y_{t}\\
\mathbf{X}_{t}
\end{array}\right)\sim N\left(\mathbf{0}_{p+1},\mathbf{L}\mathbf{L}^{T}+\Sigma\right),
\]
where $\mathbf{L}=\left(L_{0},L_1,...,L_{p}\right)^{T}\in\mathbb{R}^{(p+1)\times r}$. Under this design, $Y_t|\mathbf{X}_t$ is Gaussian, and the conditional expectation $E\left[\left.Y_{t}\right|\mathbf{X}_{t}\right]$
is available in closed form (\ref{equ:expectation_expression}), and the true risk is analytically tractable.

\begin{figure}

\begin{center}
\includegraphics[scale=0.33]{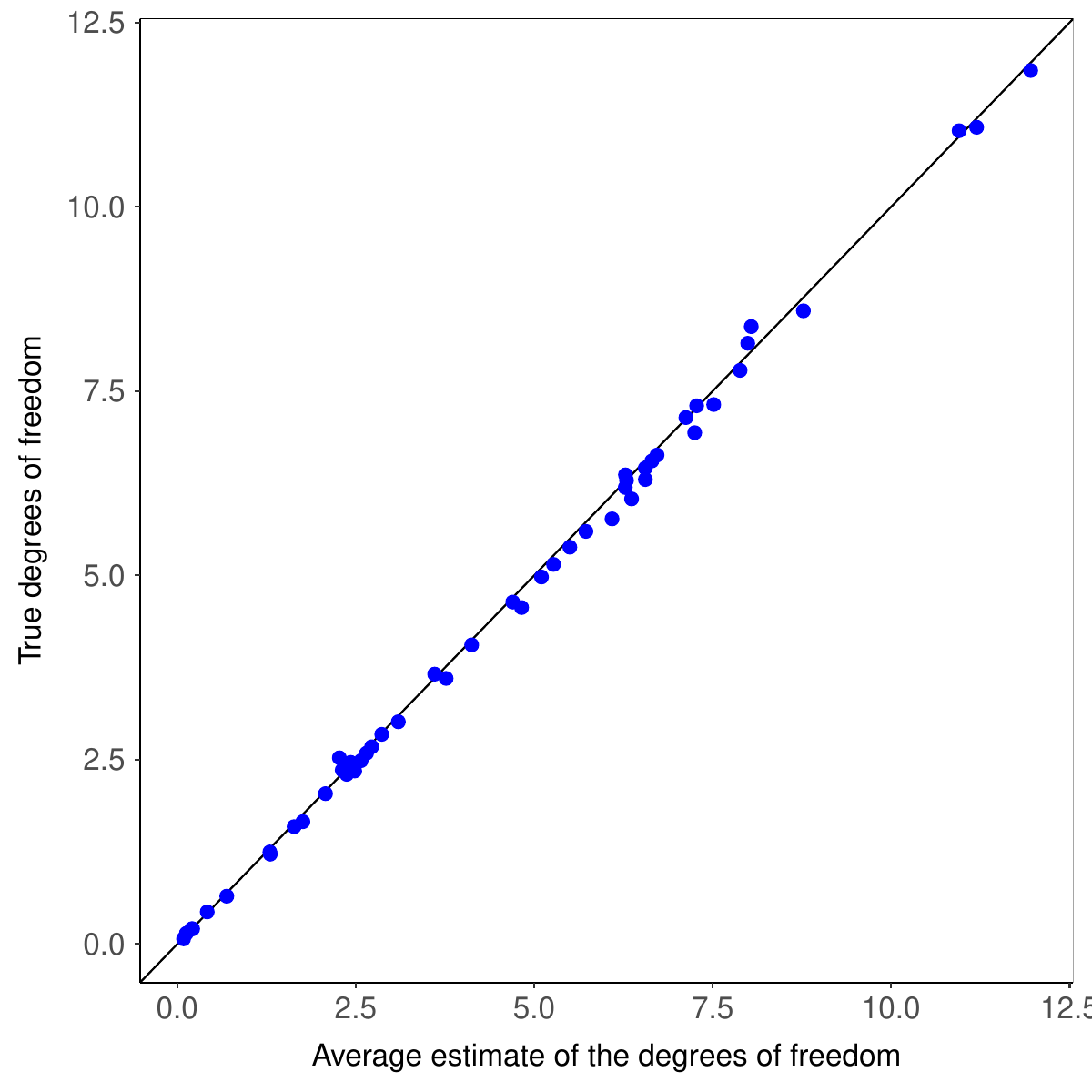}
\includegraphics[scale=0.33]{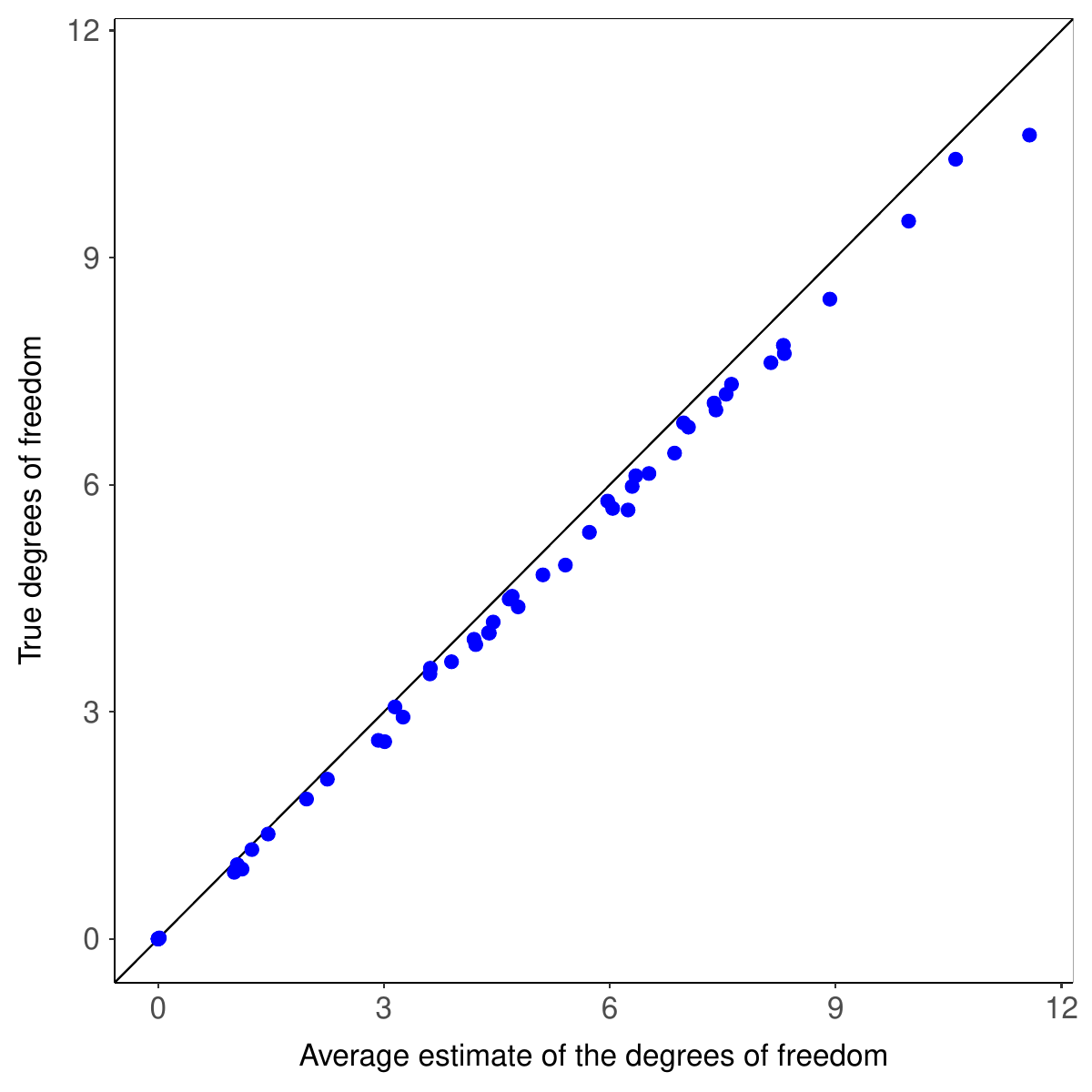}
\end{center}
\begin{center}
\includegraphics[scale=0.33]{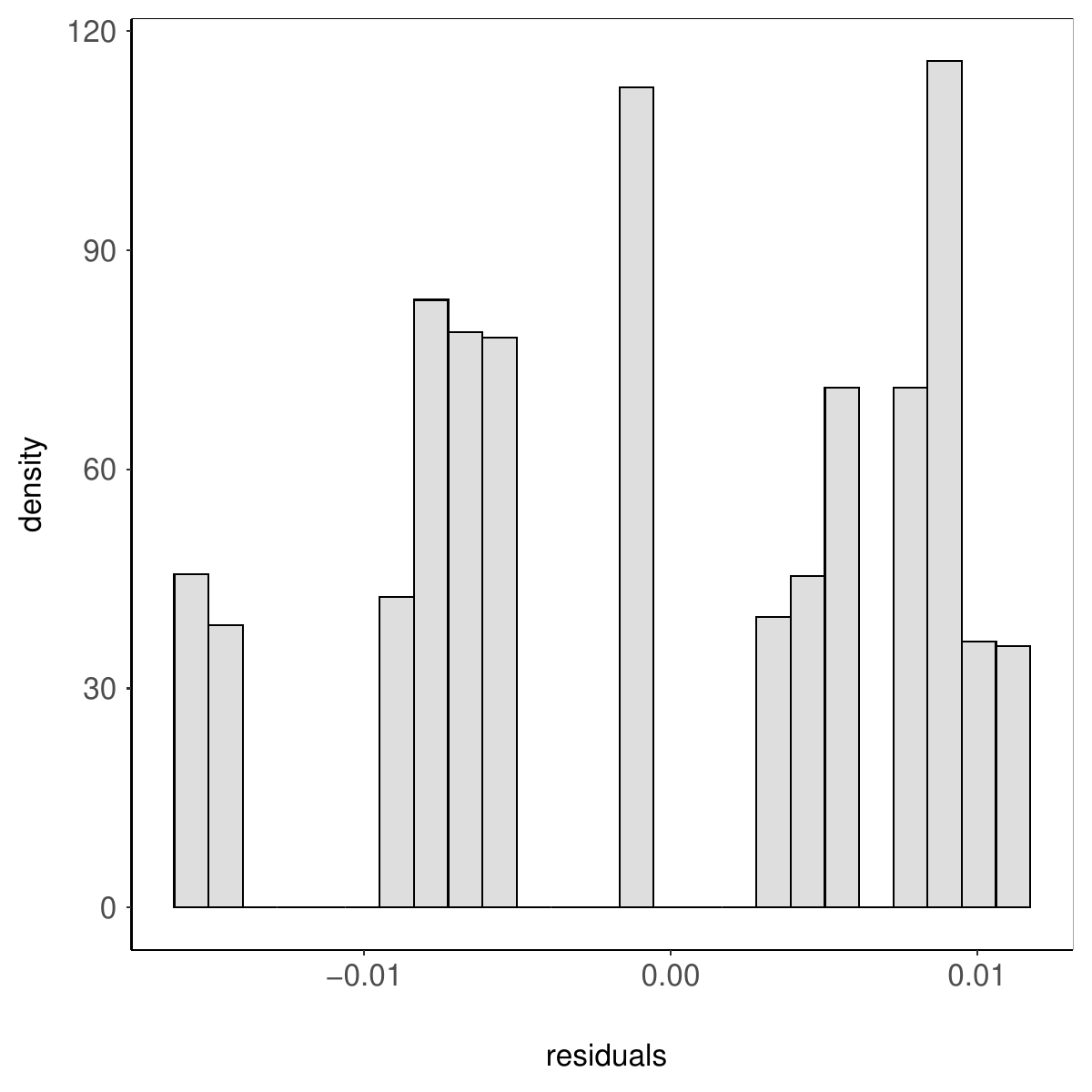}
\includegraphics[scale=0.33]{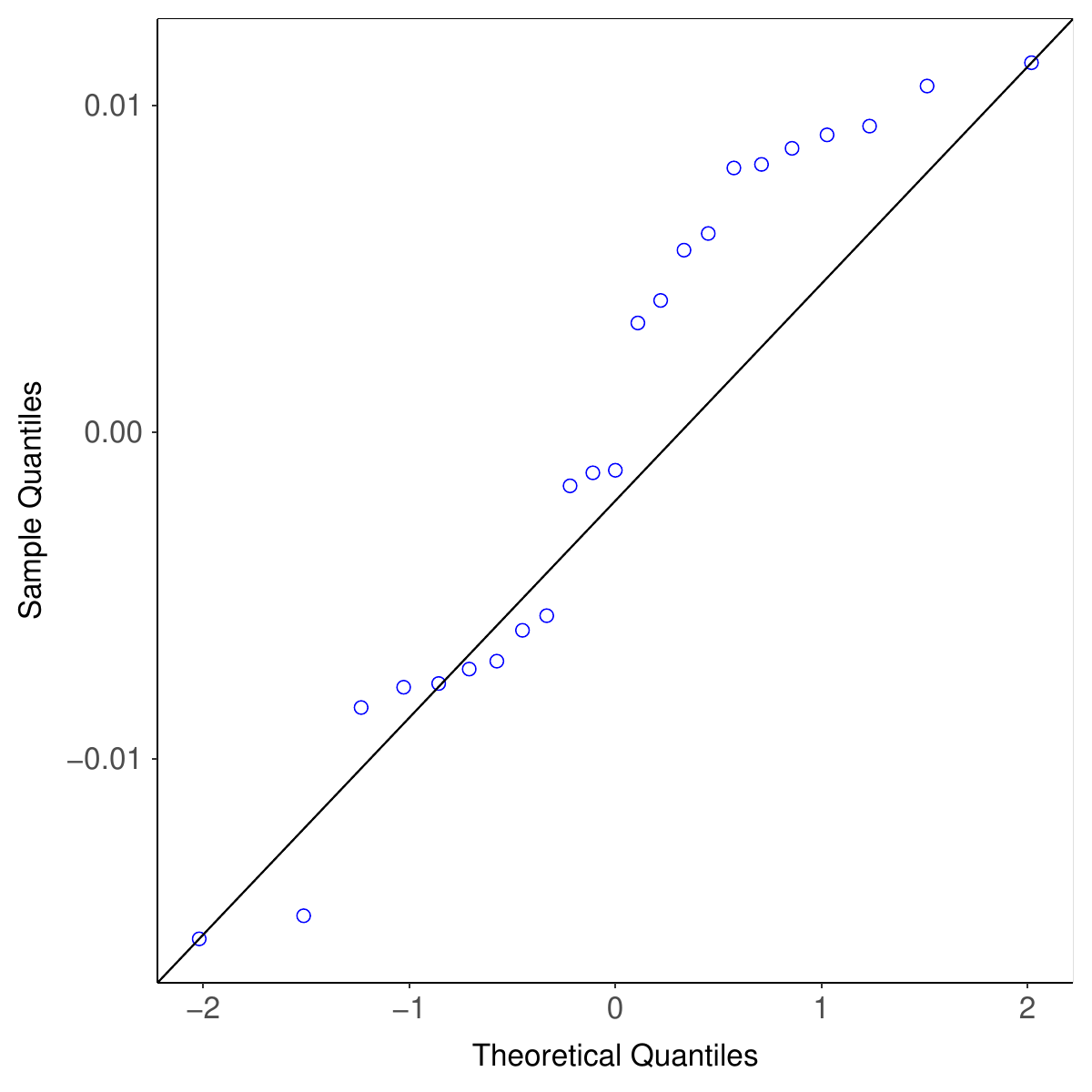}
\end{center}
\footnotesize\textit{} Estimated
degrees of freedom versus true degrees of freedom for the Gaussian factor
model (top left), Estimated degrees of freedom versus true degrees
of freedom for the Gaussian factor model with empirical residuals (top
right), histogram of empirical residuals (bottom left), quantile-quantile
plot of empirical residuals (bottom right). 
\caption{\emph{\label{fig:Study-of-Robustness.}Assessment of robustness to Gaussian assumption}}
\end{figure}

In this simulation design, the assumptions underlying the SURE hold exactly, and the degrees of freedom estimate is unbiased, as illustrated in the top-left panel of Figure \ref{fig:Study-of-Robustness.}.

For the second simulation design, we maintain the Gaussian factor structure for the $p$ donor units, but replace the treated unit's idiosyncratic shocks with draws from the empirical distribution of the fitted residuals. This preserves the common factor structure while inducing non-Gaussianity in $Y_t|\mathbf{X}_t$. Because the donors remain Gaussian and the treated shock remains mean-zero and independent of $\mathbf{X}_t$, the conditional mean $E[Y_t|\mathbf{X}_t]$ remains available in closed-form, and the true risk can still be computed exactly. Appendix \ref{simdetails} provides simulation details.

In Figure \ref{fig:Study-of-Robustness.} we study the robustness
of our estimate of degrees of freedom, and hence of the risk, to the
ubiquitous failure of the Gaussian assumption. We see in the
histogram (bottom-left panel) and quantile-quantile plot (bottom-right panel) that the fitted residuals are
decidedly not Gaussian, but the quantile-quantile plot nevertheless
suggests a moderate enough departure from Gaussianity that our estimate
of degrees of freedom may still be accurate enough to be useful. This
is validated by the top right panel of Figure \ref{fig:Study-of-Robustness.},
which shows that the estimated degrees of freedom still line up well with
the truth, in expectation, when we simulate the errors by drawing them from the empirical distribution of the fitted residuals. 

While the simulation exercise is intrinsically motivated by the need to assess the robustness of SURE to distributional assumptions, it is likewise motivated by the need to select a model selection method for the application whose data generating process the simulation emulates.

\begin{figure}
\begin{center}
\includegraphics[scale=0.30]{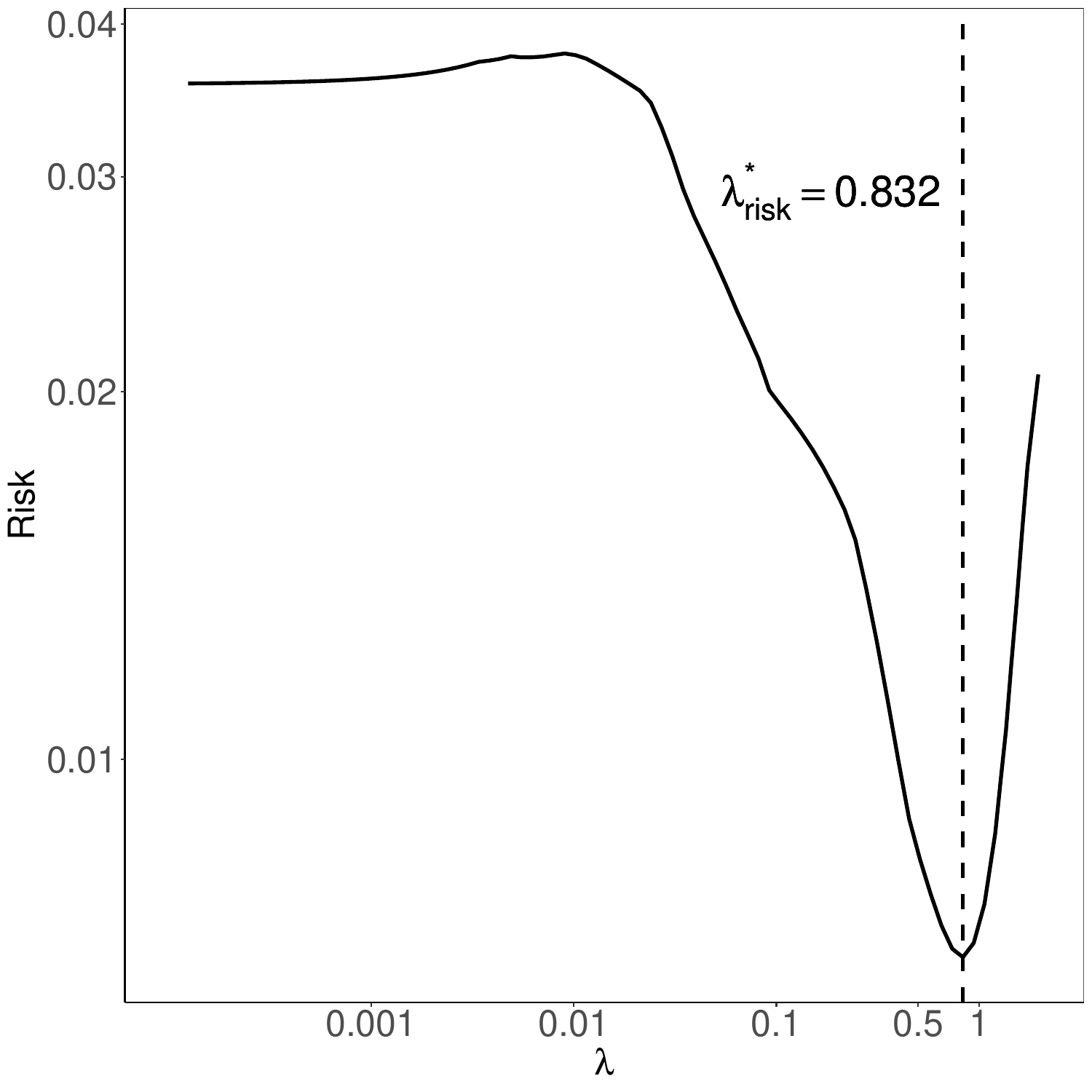}
\includegraphics[scale=0.30]{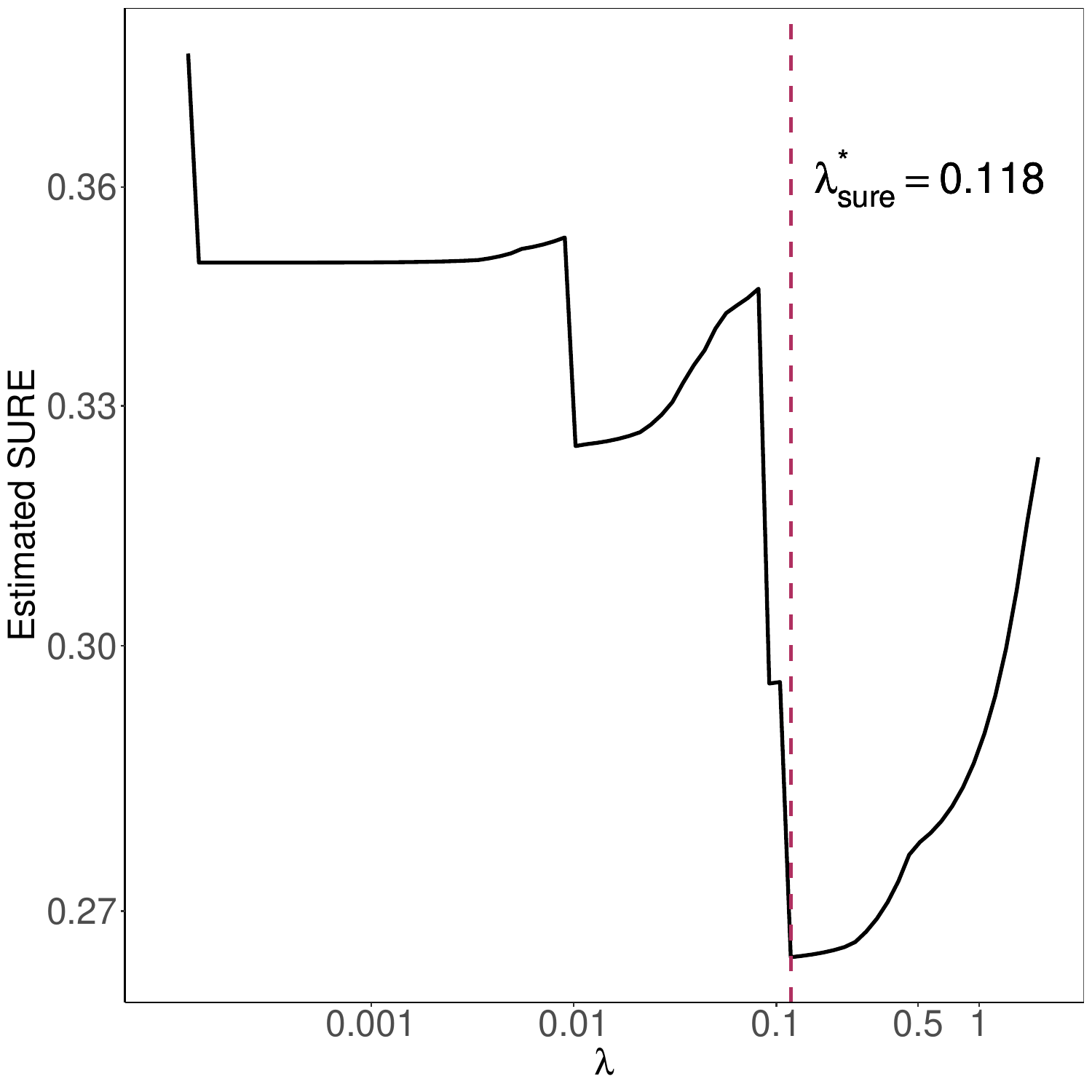}
\end{center}
\begin{center}
\includegraphics[scale=0.30]{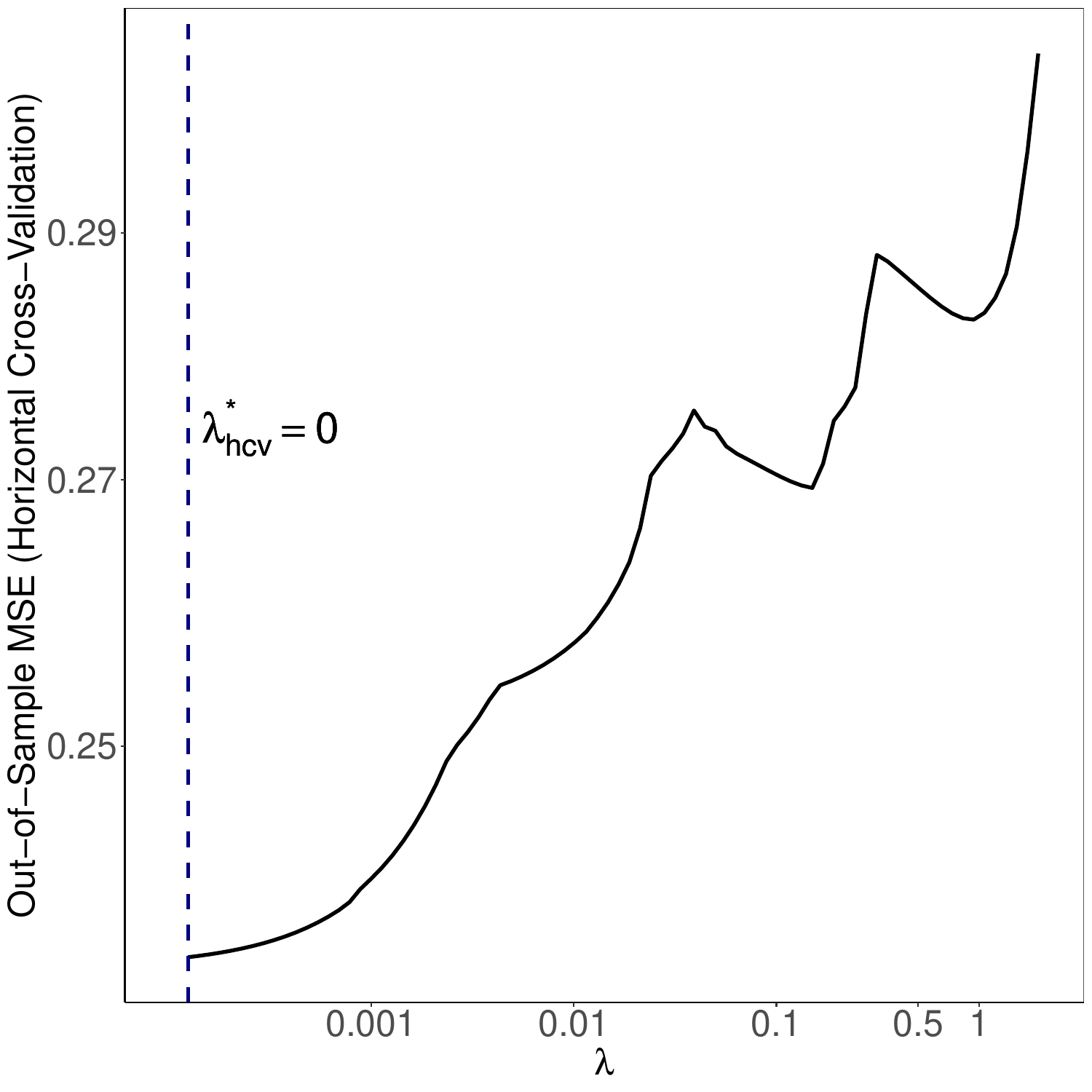}
\includegraphics[scale=0.30]{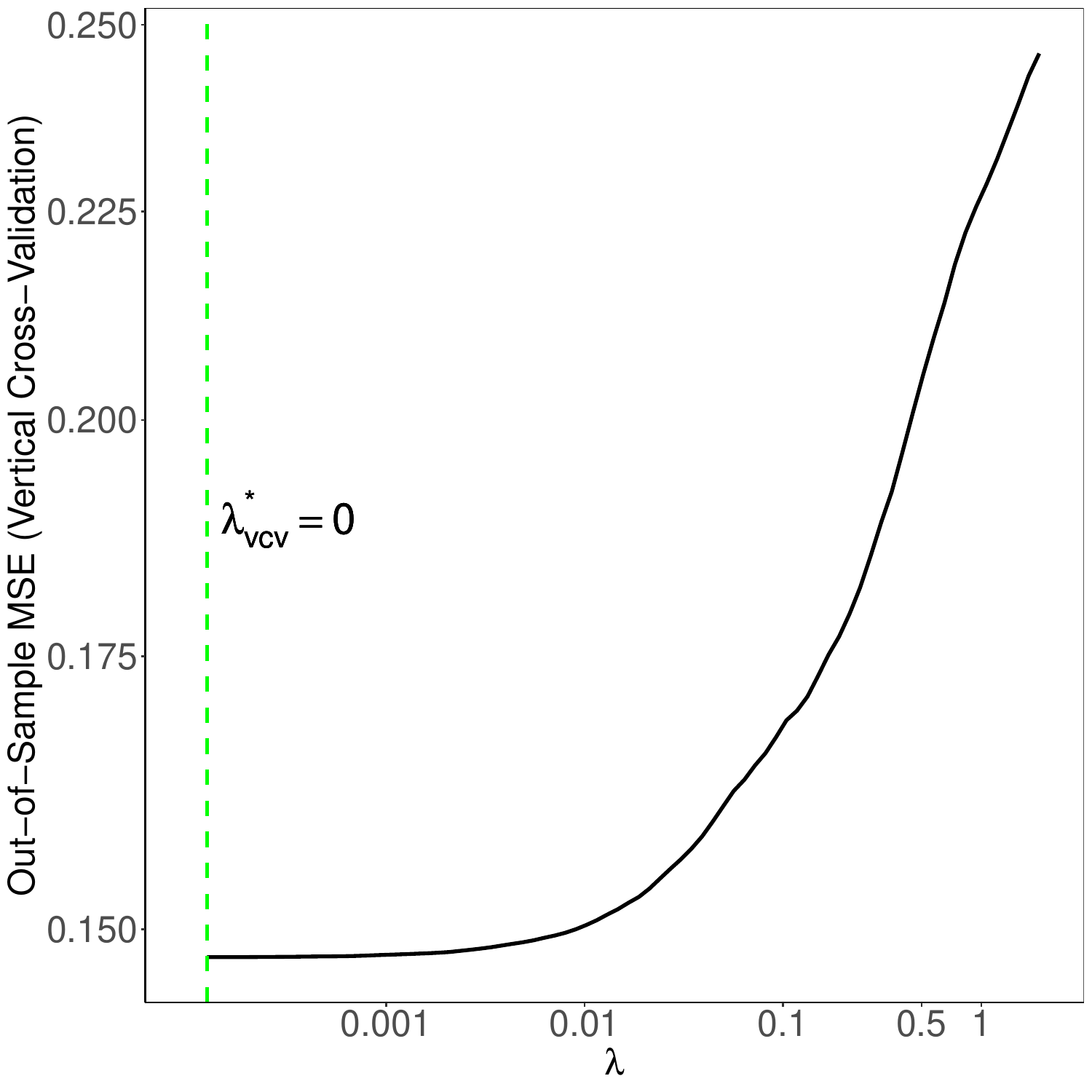}
\end{center}

\footnotesize\textbf{} Risk (top-left), estimated SURE (top-right), out-of-sample mean-squared error in horizontal cross-validation (bottom-left) and out-of-sample mean-squared error in vertical cross-validation (bottom-right). All as a function of $\lambda$ for a single, representative realization of the simulation. The dashed lines indicate the $\lambda$ values that minimize each respective criterion.
\caption{\label{fig:Risk vs lambda Gaussian sim}\emph{Risk, SURE, and Cross-Validation MSE Across \(\lambda\) in a Single Simulation Realization}}
\end{figure}

To that end, Figure \ref{fig:Risk vs lambda Gaussian sim} displays typical output
from this simulation. We see in the top-left panel that the simulation
design produces a U-shape for the true risk. 
This replicates nicely, albeit somewhat more pronouncedly, the plot of estimated risk presented in Figure \ref{Fig:single_model_control_city}.

The information criteria based on SURE capture this true pattern, and select approximately the same model.
Horizontal and vertical cross-validations (see Appendix \ref{subsection:tuning_parameter_selection} for formal descriptions) display substantially different
patterns, and select tuning parameters far from the oracle tuning parameter, i.e., the one selected according to the population risk.


\begin{table}[!h]
\begin{center}
\footnotesize
\renewcommand{\arraystretch}{1.2}
\begin{tabular}{lcccc|cccc}
\toprule
\multicolumn{1}{c}{} & \multicolumn{4}{c|}{Gaussian} & \multicolumn{4}{c}{Empirical}  \\
\midrule
& {RMSE} & {RMSE} & {Median} & {Mean}  & {RMSE} & {RMSE} & {Median} & {Mean} \\
& {$\hat{\tau}_{1}\times 10^{-2}$} & {$\hat{\tau}_{12}\times 10^{-2}$} & {$\hat{\lambda}$} & {$|\hat{\lambda}-\lambda_{risk}|$}  & {$\hat{\tau}_{1}\times 10^{-2}$} & {$\hat{\tau}_{12}\times 10^{-2}$} & {$\hat{\lambda}$} & {$|\hat{\lambda}-\lambda_{risk}|$} \\
\midrule
{Risk}               & 10.347  & 36.113  & 0.192  & 0.000  & 10.446  & 36.123  & 0.192  & 0.000  \\
{IC$_{\text{oracle}}$} & 10.789  & 37.498 & 0.133 & 0.909 & 10.764 & 37.512 & 0.150 & 0.882 \\
{$\widehat{\text{IC}}$}    & 10.768 & 37.462 & 0.277 & 1.007 & 10.802 & 37.538 & 0.313 & 0.979 \\
{CV-Horizontal}      & 10.880 & 37.767 & 0.150 & 1.278 & 10.842 & 37.693 & 0.170 & 1.206 \\
{CV-Vertical}        & 10.883 & 38.065 & 0.000 & 0.796 & 10.931 & 38.030 & 0.000 & 0.768 \\
{Rolling Window}     & 10.869 & 37.691 & 0.192 & 1.133 & 10.828 & 37.665 & 0.217 & 1.083 \\
\bottomrule
\end{tabular}
\end{center}
\footnotesize\textbf{} 
The quantities $\hat{\tau}_{1}$ and $\hat{\tau}_{12}$ are the treatment effect estimates one month and one year after treatment, respectively, using the tuning parameter selected by each procedure. The root mean-squared error (RMSE) is evaluated over 5,000 Monte-Carlo replications.  
“Median $\hat\lambda$’’ is the median selected tuning parameter, and \(|\hat\lambda-\lambda_{\text{risk}}|\) is the mean absolute deviation from the oracle penalty that minimizes the population risk. Formal definitions for the selection procedures are provided in Appendix \ref{subsection:tuning_parameter_selection}. 
\caption{\emph{ Prediction performance and selected tuning parameters from different selection procedures }}
\label{tab:selection_compare}
\end{table}

In Table \ref{tab:selection_compare}, we consider different methods under the
two aforementioned simulation designs. 
The proposed information criteria appear to systematically outperform the vertical and horizontal cross-validation approaches and perform comparably to but marginally better than the rolling-window cross-validation (also detailed in Appendix \ref{subsection:tuning_parameter_selection}), whose better performance amongst cross-validation methods had already been documented
\cite{kellogg2020combining}.

From the simulation exercise, we conclude that the SURE information criterion is more reliable for studying the data at hand and we thus elect it as our model selection method for analyzing the data.

\subsection{Data Analysis}

Having elected the information criteria approach \eqref{eq:information-criteria-estimate} as our model selection method, we may now carry out the regression exercise.

We first investigate in detail the impact of rationing for a single, popular model, the Toyota Highlander. 
This procedure can be automated
and allows for the joint analysis of a selection of models, which we carry out subsequently.

In preprocessing, we pass the data through an MA(3) filter. The start date of the time series is January 2012, and the end date is December 2014.  

\subsubsection*{Analysis for a single model}

We analyze in isolation the impact of rationing on the demand for
the Toyota Highlander. At a high level, the core task is to build a counterfactual;
the time series of demand for the Highlander had there not been rationing.

\begin{figure}
\begin{center}
\includegraphics[scale=0.30]{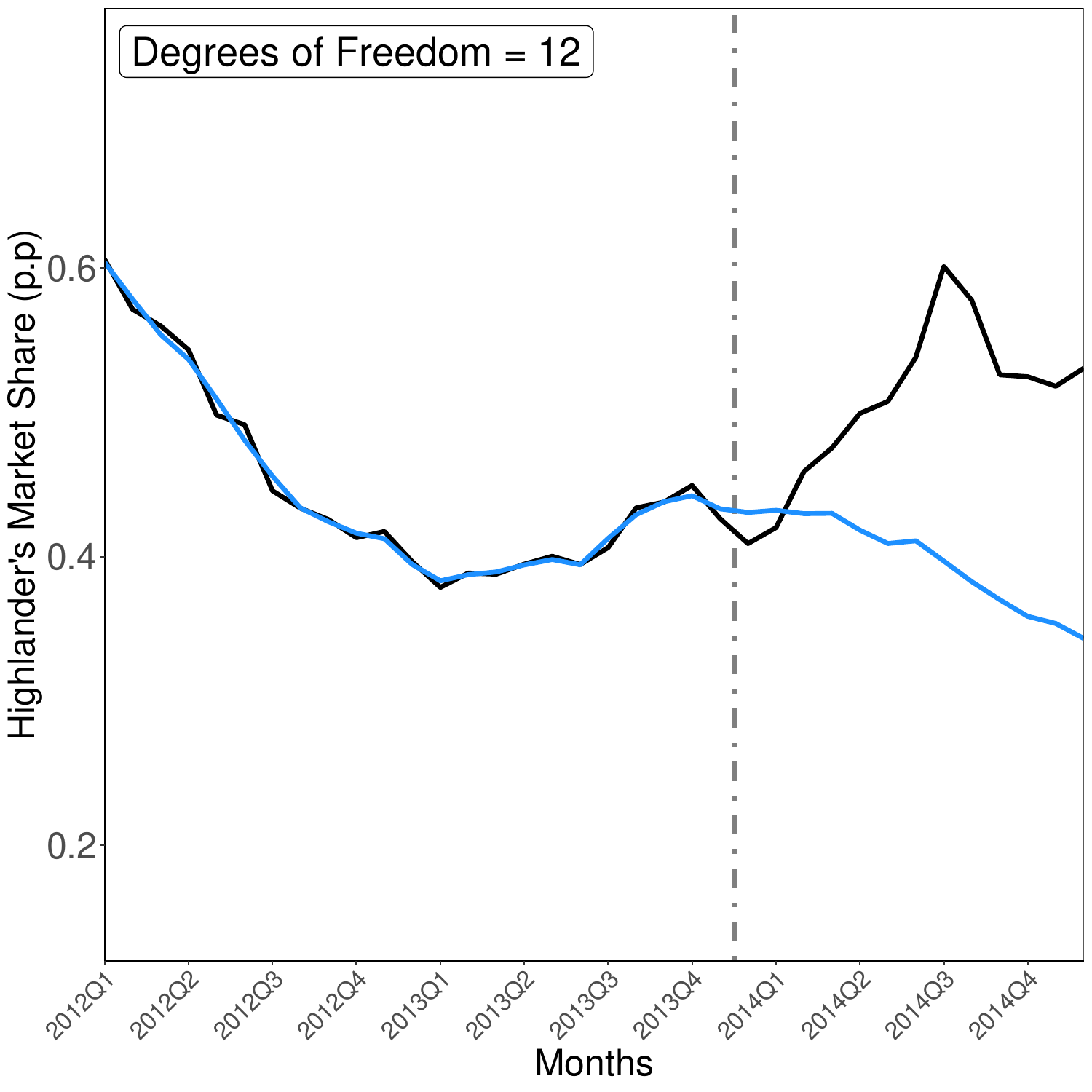}  \ \ \ \ \
\includegraphics[scale=0.30]{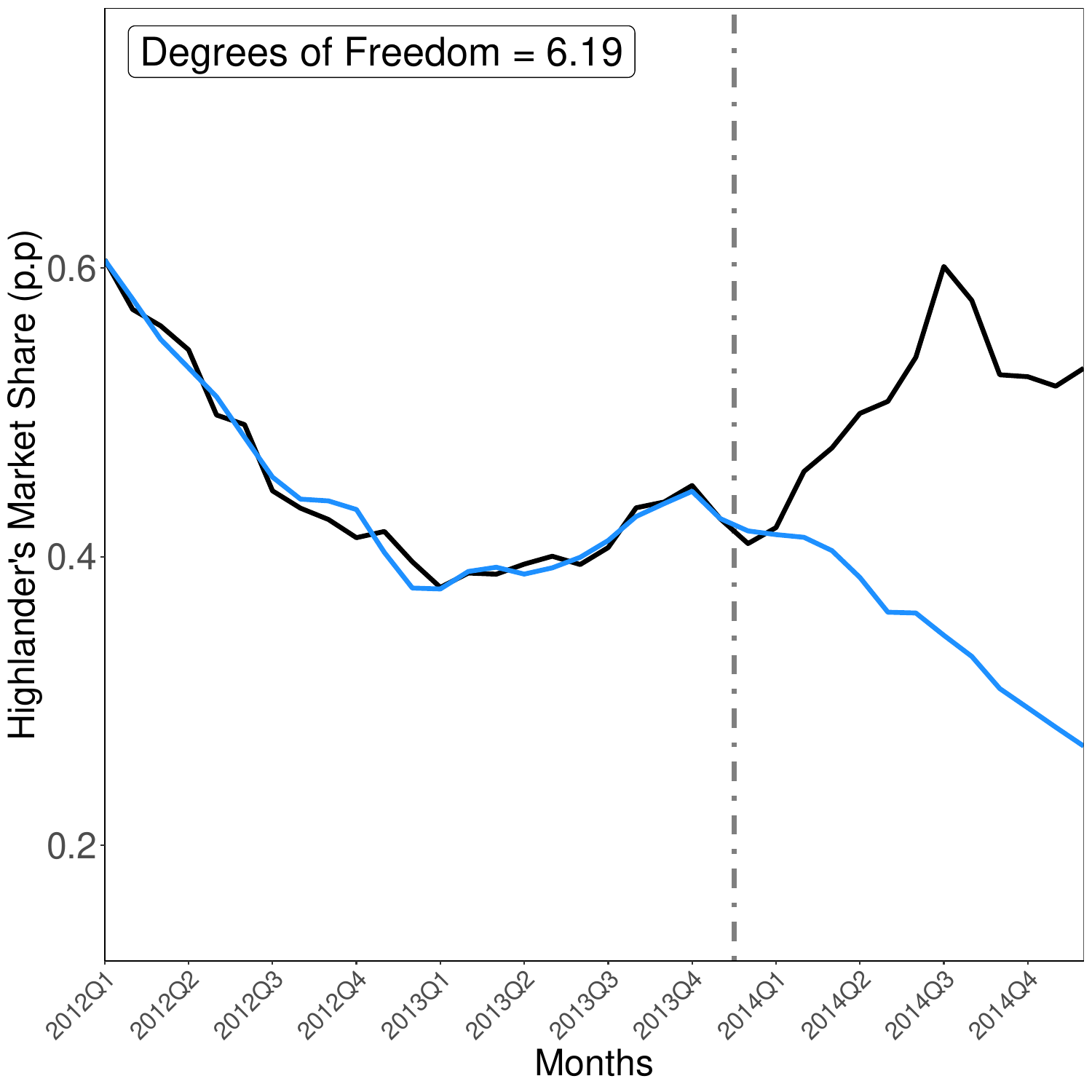}
\end{center}

\begin{center}
\includegraphics[scale=0.30]{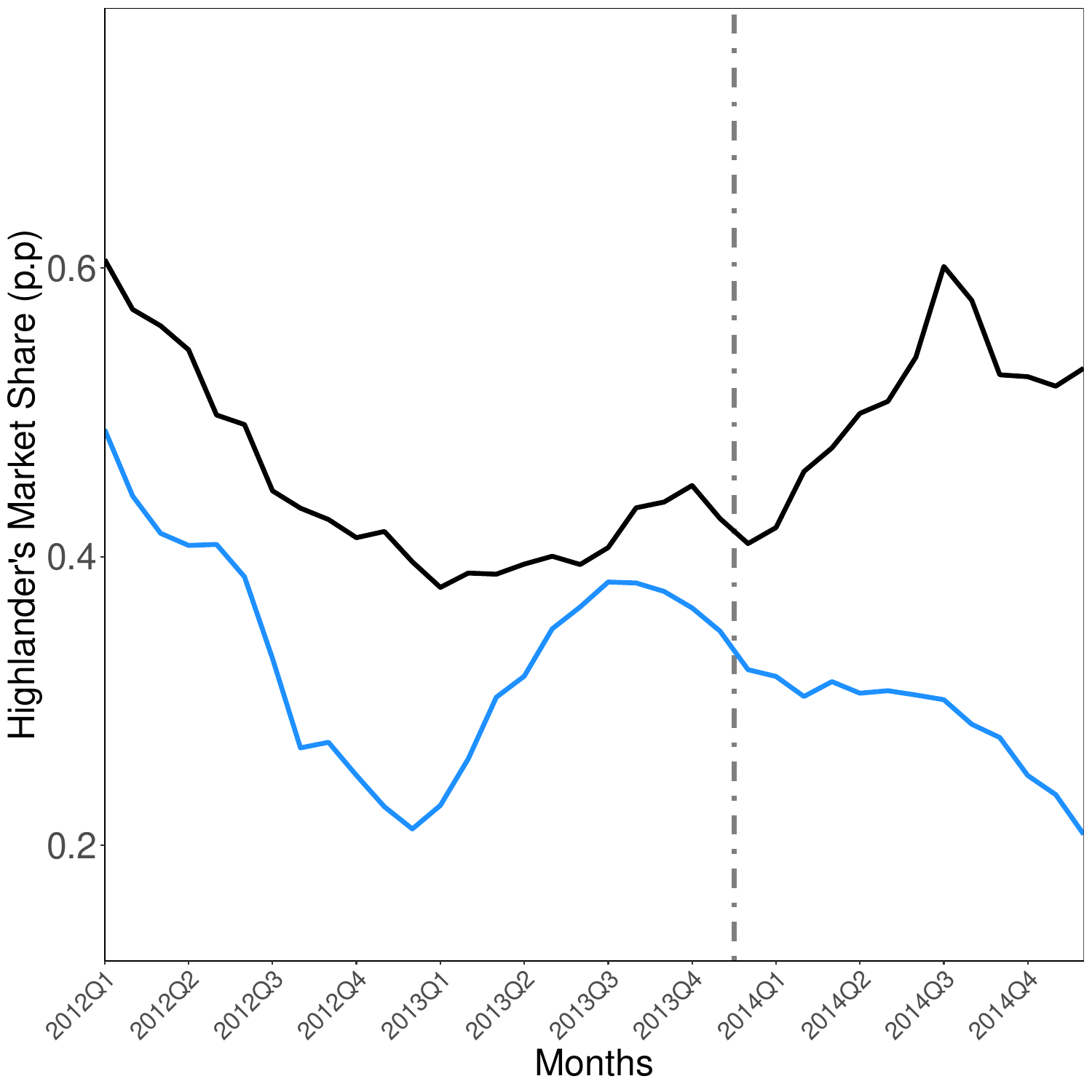}  \ \ \ \ \
\includegraphics[scale=0.30]{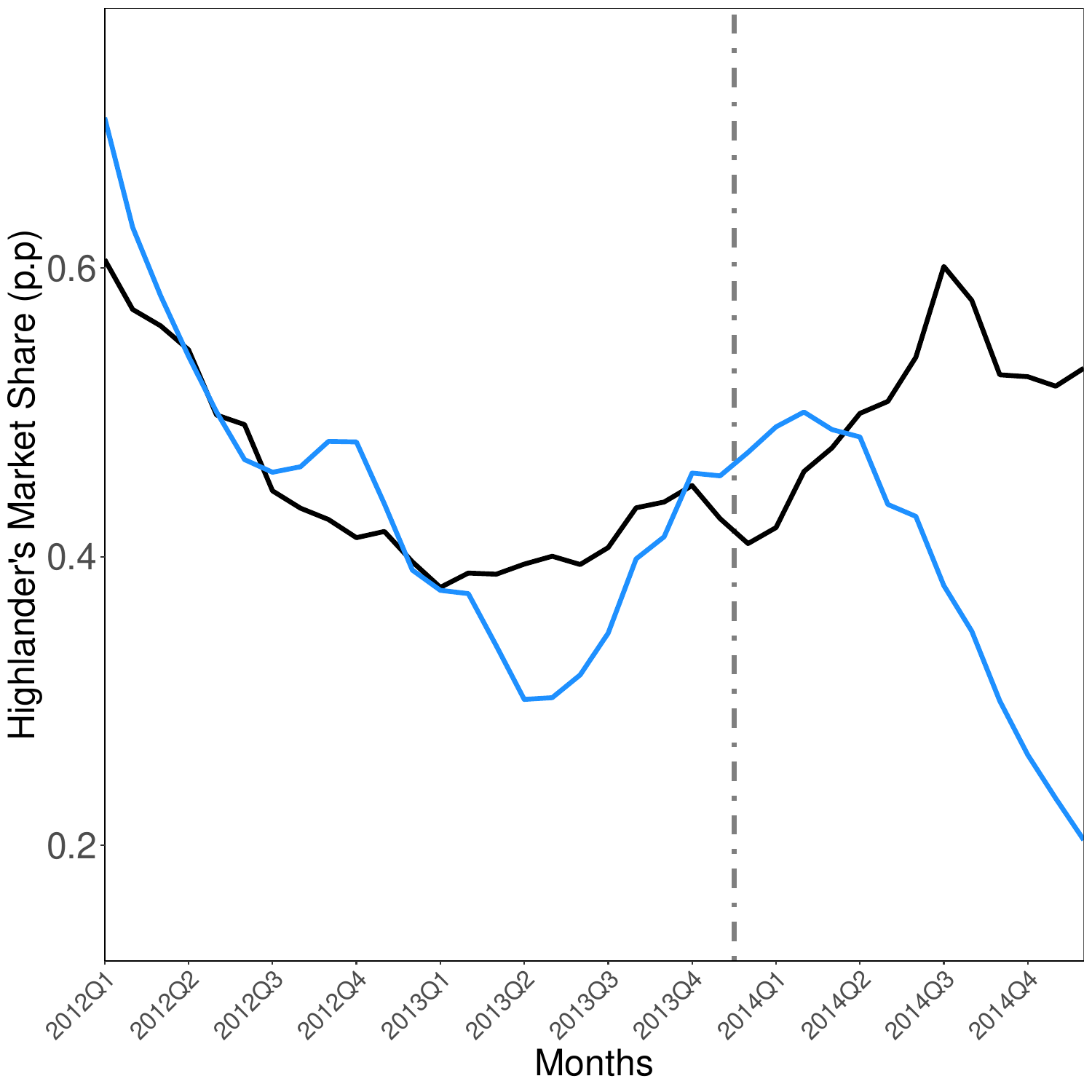}
\end{center}
\footnotesize\textbf{} Unpenalized synthetic controls (top-left), penalized synthetic
controls with the tuning parameter selected by SURE
(top-right), matching by car model (bottom-left), and matching by $\ell_2$ distance
(bottom-right). The black line plots the observed outcome series and the pale blue line plots the estimated counterfactual series.  
\caption{\label{fig:one model fits}\emph{Observed and Estimated Counterfactual Series for the Highlander in Tianjin.}}
\end{figure}

As intuited and anticipated above, pure matching approaches deliver
a poor counterfactual for the to-be-treated unit. Whether we match
the Highlander in Tianjin to the Highlander in Shijiazhuang, or to the nearest
model in Shijiazhuang according to the penalty term, so in the $\ell_{2}$ sense,
we get an unconvincing fit even as assessed by the plot of the time
series. See Figure \ref{fig:one model fits}.

This motivates the use of a synthetic control, a match that is averaged over multiple donors and is thus expected to have lower variance.\footnote{For a more general discussion of methods blending difference-in-differences and synthetic controls, see \citeA{doudchenko2016balancing}.}
Concerns of model flexibility --there are 76 donors after eliminating
models that were either introduced after the beginning of our time
series or discarded before the end-- and noisy to-be-treated series
motivate the use of penalized synthetic controls (\citeNP{abadie2021penalized}). As is generally the case, two general approaches avail to estimate
the tuning parameter of the penalty term: the \emph{cross-validation}
approach and the\emph{ information criteria} approach.
In the simulation exercise of \Cref{sec:simulation}, we found
that the information criteria approach provided a more accurate estimate
of prediction error and, crucially, produced a model that better predicted
treatment effects, especially at short horizons.

It may be that certain time-invariant variables improve the fit when
used as covariates. Two natural covariates are the stock price of
the model's brand and the average --over time-- of its outcome data.
We consider the penalized synthetic control model with these two covariates,
and select both $\lambda$ and $V$ according to our information criteria.
The selected model put full weight on the covariate constructed as
an average of outcome data, and had a worse information criterion than
the model without covariates. We therefore opt for proceeding
without covariates.

The White test for heteroskedasticity (\citeNP{white1980heteroskedasticity}, \citeNP{breusch1979simple}) produced a
 $p$-value of $0.12$. 
 This suggests a moderate
amount of heteroskedasticity and, in light of our simulation study, leaves us sufficiently confident that the theory will remain by and large reliable.

\begin{figure}
\begin{center}
\includegraphics[scale=0.30]{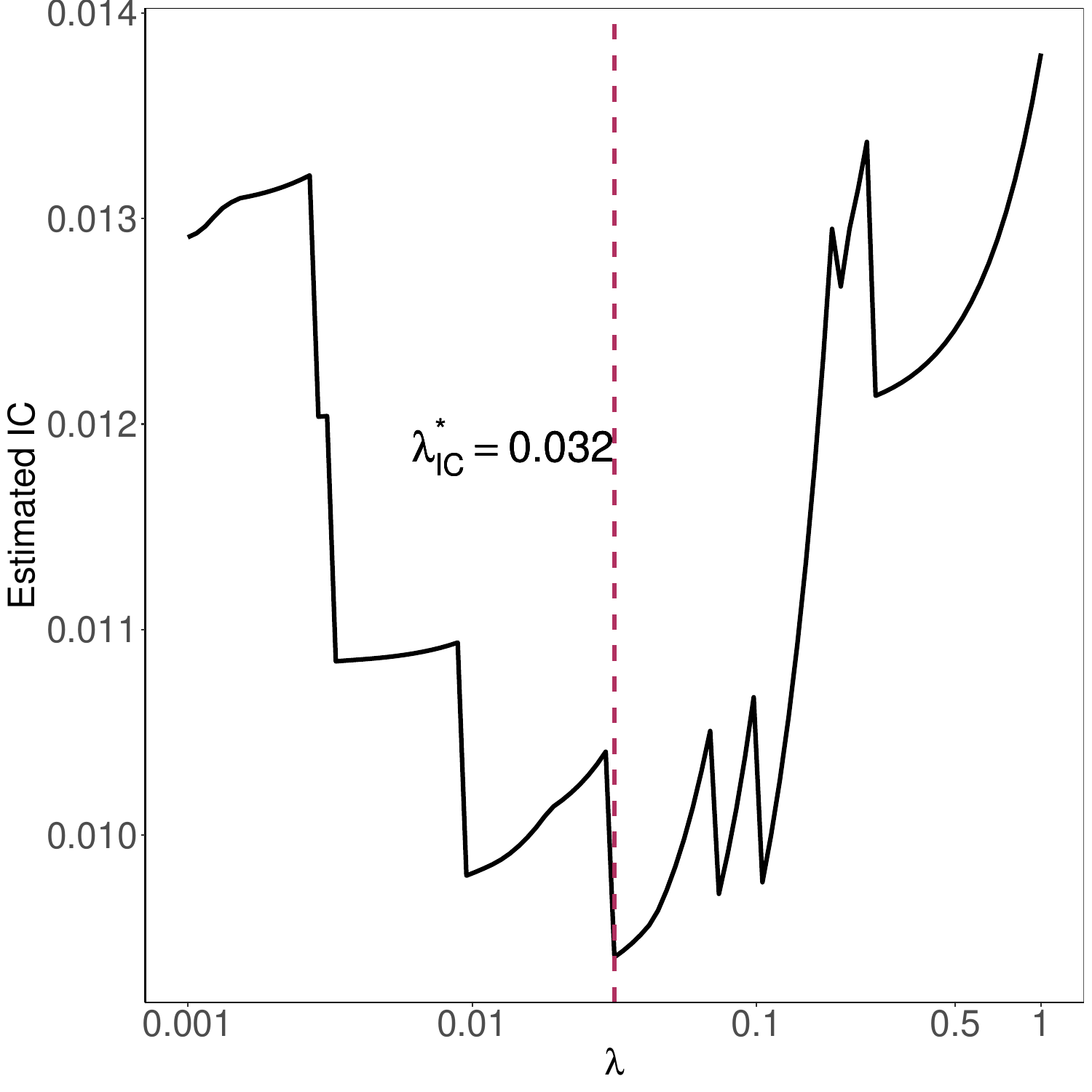}  \ \ \ \ \
\includegraphics[scale=0.30]{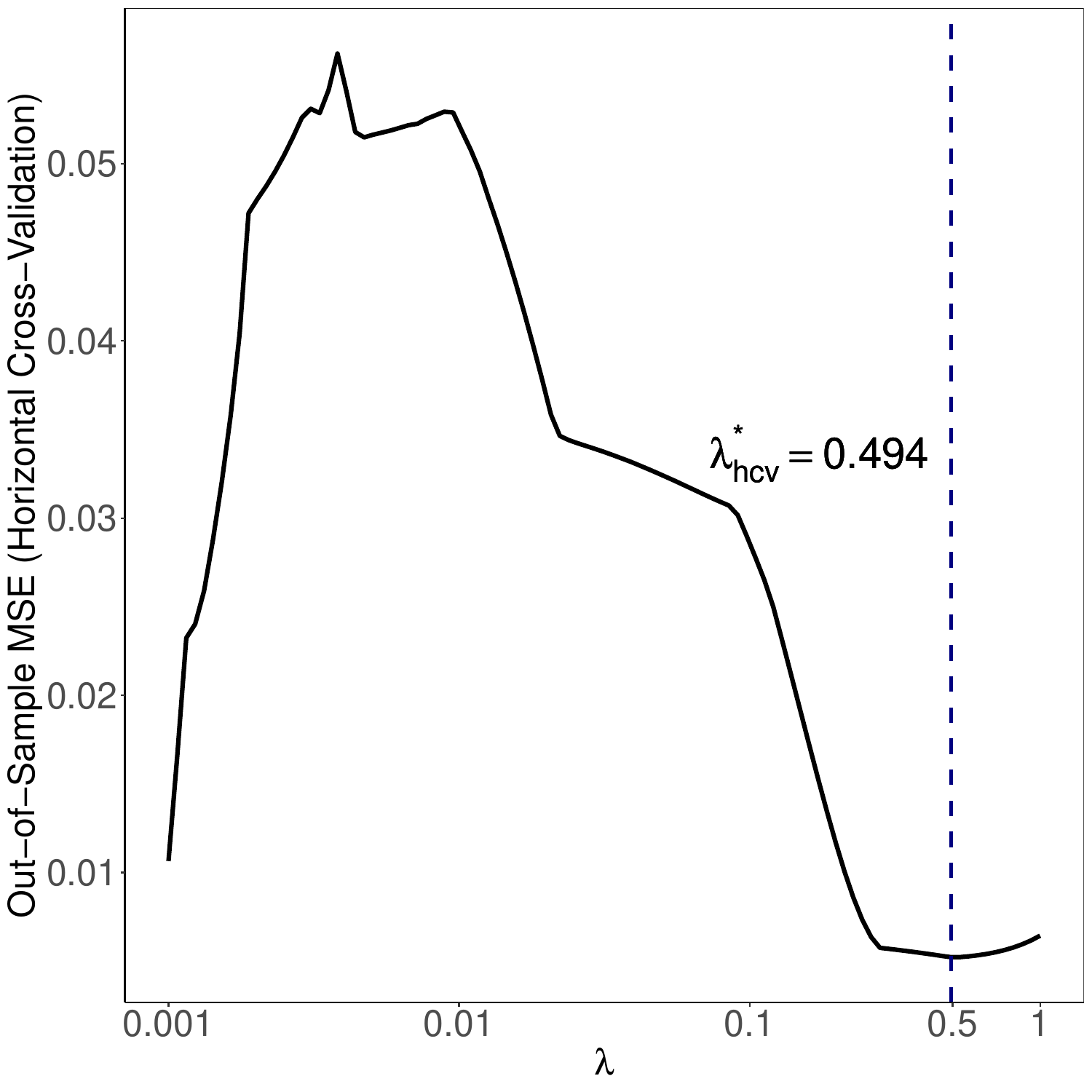}
\end{center}

\begin{center}
\includegraphics[scale=0.30]{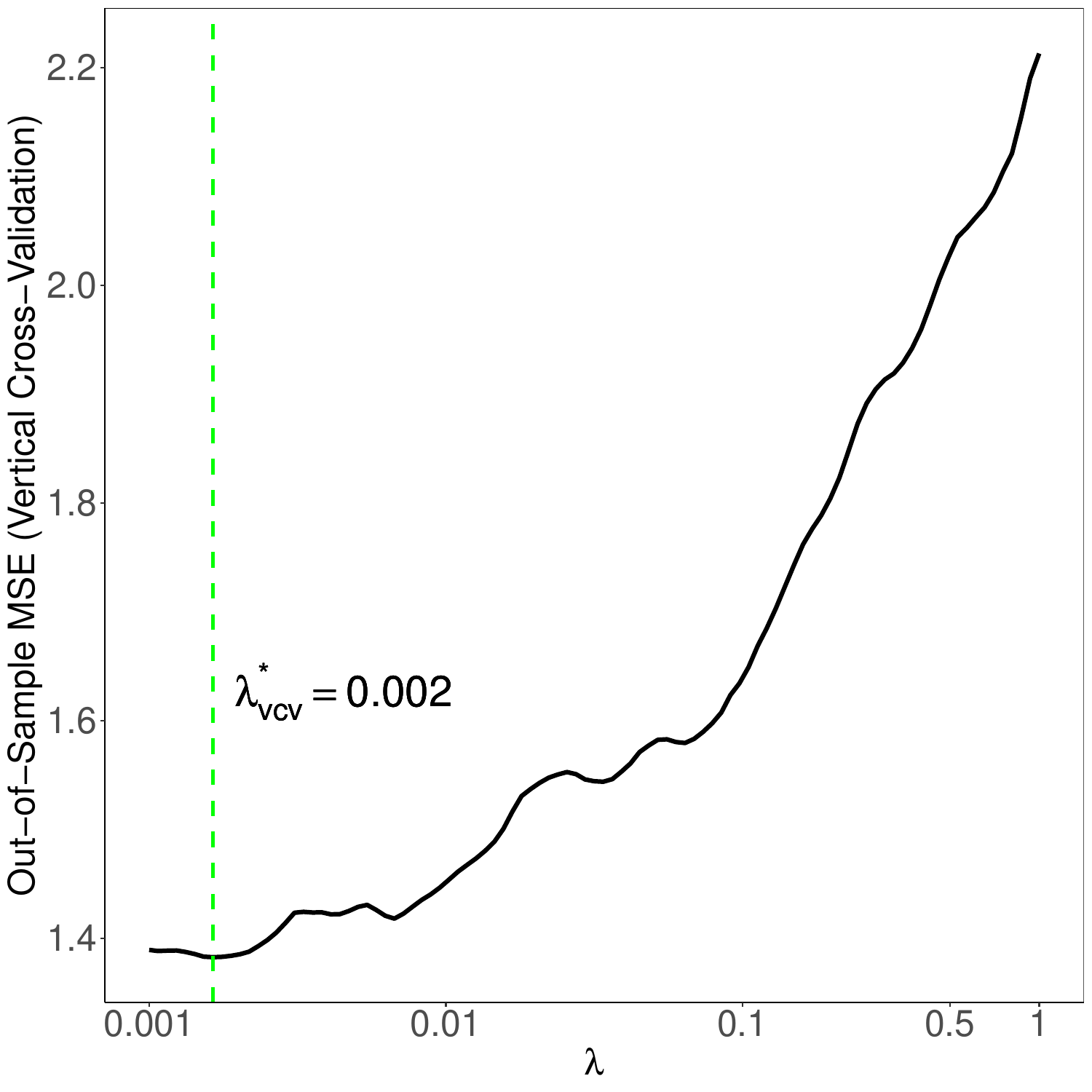}  \ \ \ \ \
\includegraphics[scale=0.30]{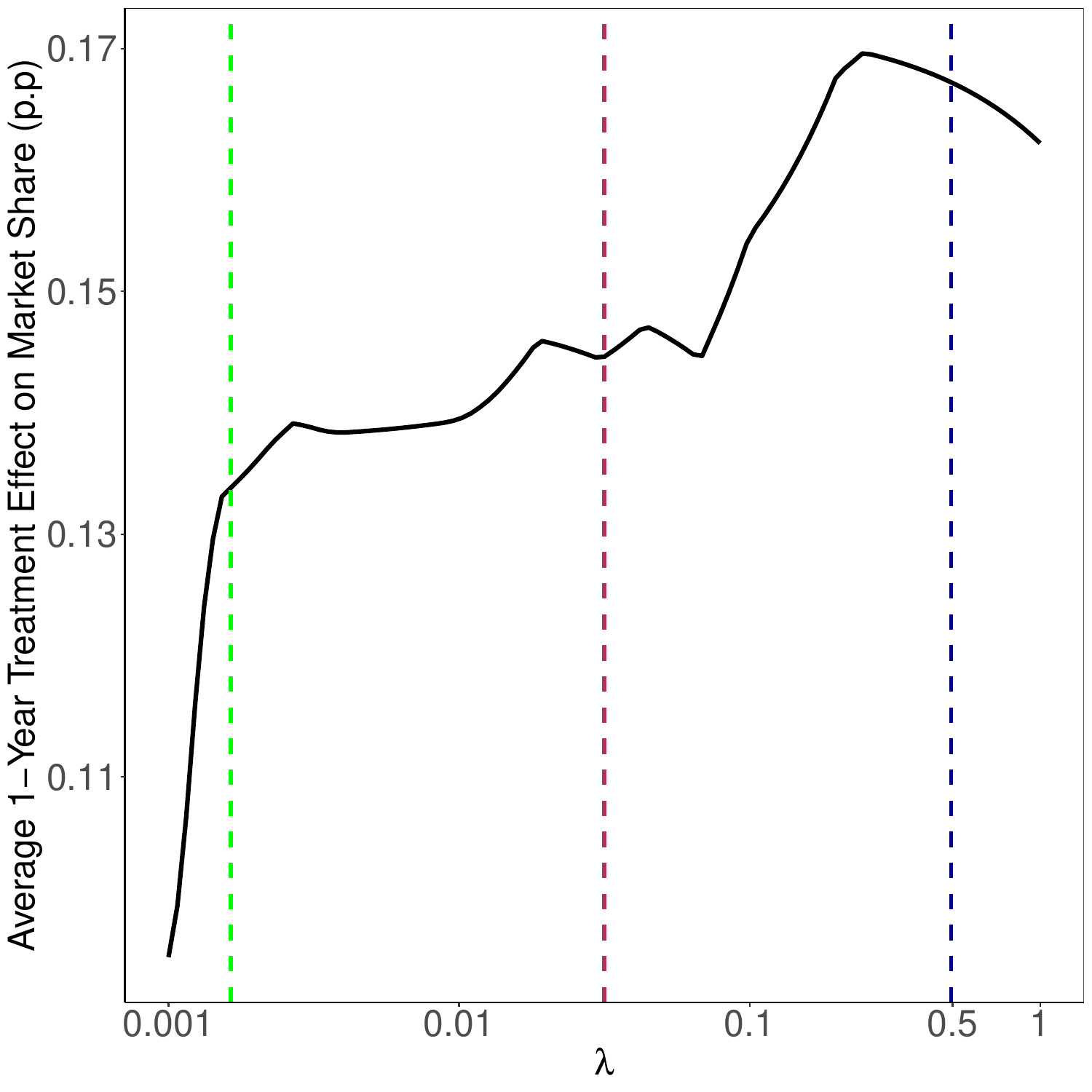}
\end{center}
\footnotesize\textbf{} Estimated information criterion (top-left), horizontal cross-validation loss (top-right), vertical cross-validation loss (bottom-left), estimated one-year treatment effect (bottom-right), each plotted as a function of \(\lambda\).  
Vertical lines indicate the tuning parameter selected by different methods: \(\lambda^{\ast}_{\mathrm{VCV}}\) (green), \(\lambda^{\ast}_{\mathrm{IC}}\) (red), and \(\lambda^{\ast}_{\mathrm{HCV}}\) (blue).
\caption{\label{fig:real data plots 1}\emph{Tuning Parameter Selection for PSCM and Estimated Treatment Effects for the Highlander in Tianjin}}
\end{figure}

Remark that the choice of tuning parameter selection method is consequential.
As is well exemplified in the case of the Highlander, two different selection methods can yield substantially different tuning parameters, leading to different
treatment effect estimates. As is immediate from Figure \ref{fig:real data plots 1},
the estimate of prediction mean-squared error is essentially monotone in $\lambda$
according to cross-validation, detecting none of the expected overfitting.
The plot of the estimated risk versus $\lambda$, according
to the information criterion estimate, however presents the U-shape
typical of overfitting scenarios; penalization for smaller values
of $\lambda$ reduces overfitting and decreases the prediction error,
but for too large values of $\lambda$ the regression underfits and
the prediction error increases.

The implied difference in the treatment effect estimate is economically
important. For instance, while the unpenalized synthetic control method (which corresponds to $\lambda=0$) estimates an increase in relative demand of 20\% for the Highlander, the information criteria-based estimate predicts an increase of 36\%.


This difference in output from different model selection methodologies ought to be qualified. The conceptual motivation for penalizing is that we are confident, or have an \emph{a priori} belief, that ``nearby'' donors are ``better'' donors; indeed, that is the motivation for matching in the first place.
A more heavily penalized synthetic control estimate is less prone to use ``far away'' donors whose fluctuations may coincidentally cancel and produce a good in-sample fit without capturing signal.
For instance, the linear combination of a luxury minivan, say the Buick GL8, and an ultra economical hatchback, say the Zotye Z100, may give a better in-sample fit for the Highlander than does the linear combination of the CR-V and Highlander itself but, especially since there are many more donors than training periods, we would suspect that this is an instance of overfitting via model selection.

A larger tuning parameter forces the estimator towards the matching
estimator, and away from perhaps coincidental linear combinations
of donors producing a tighter in-sample fit. In a way that echoes the model selection interpretation with the lasso --which forces the estimated regression
function towards zero or another value considered \emph{a priori}
plausible-- the more penalized estimate is more ``conservative''
in a desirable sense.

 When penalizing, to what extent are we relying more on the ``nearby'' donors? 
The penalty  $\sqrt{\sum_{j=1}^{p}\hat{\beta}_{j}\left\Vert \mathbf{Y}-\mathbf{X}_{j}\right\Vert _{2}^{2}}$,
where $\hat{\beta}_{j}$ is a function of $\lambda$, equals 0.92 when no penalty is applied,
and equals 0.37 for $\lambda_{\mathrm{IC}}^{*}=0.3$. Hence, in the $\ell_{2}$
sense, the relevant donors are ``on average'' two and a half times
farther from the to-be-treated unit in the unpenalized synthetic control estimate.

Our main conclusion from the single model analysis is thus that proportional sales for the Toyota Highlander increased substantially
due to the introduction of rationing.
This is quite interesting. The Highlander is a mid-range car
and it was not \emph{ex ante} obvious that it could be a common choice for auction or lottery winners. 
We do not attempt to differentiate
between the two types of winners and leave this for further research.
Note however that such a question may be tackled using the optimal
transport reduced form methodology developed in \citeA{daljord2021black}.

\begin{figure}[!h]
\centering
\includegraphics[scale=0.375]{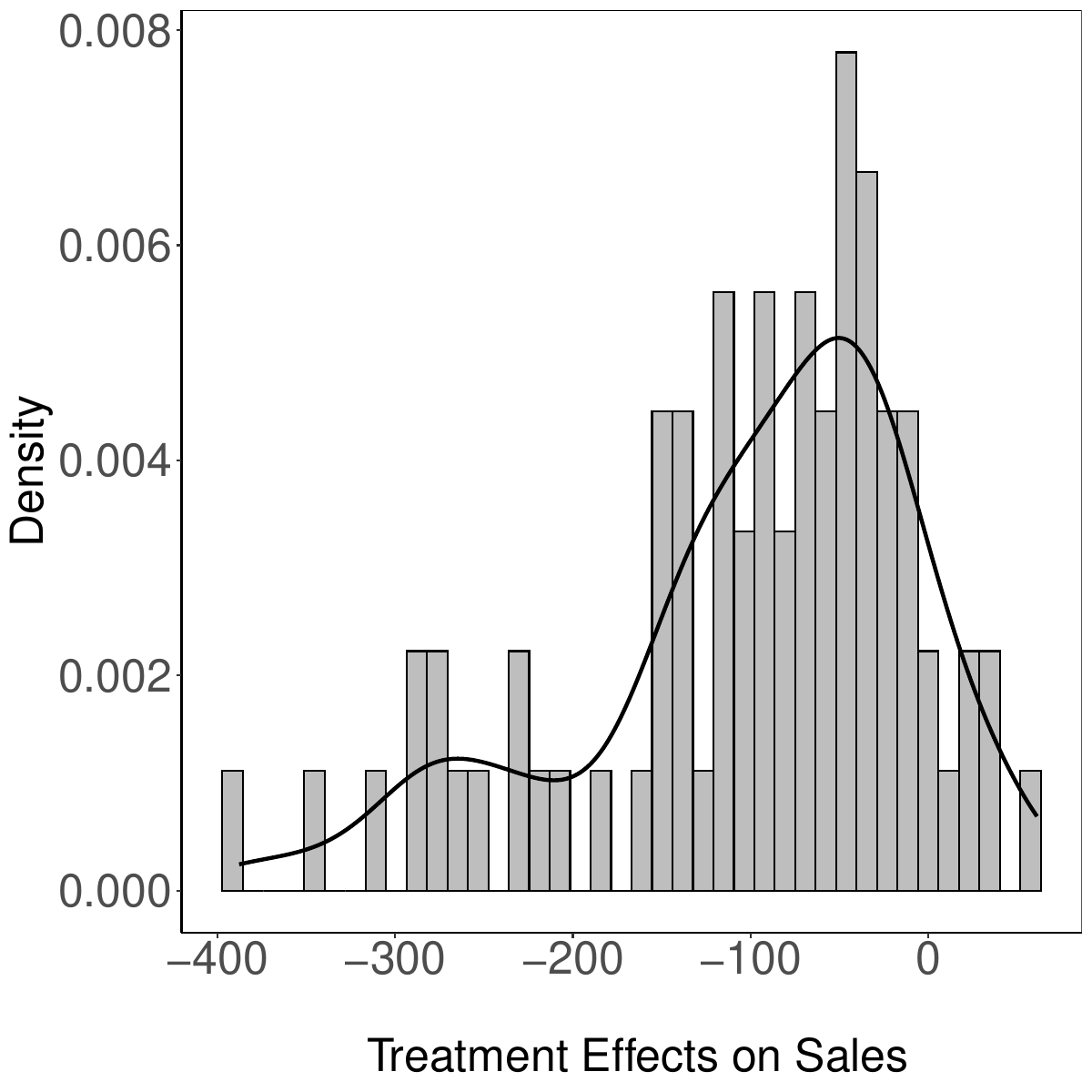}
\includegraphics[scale=0.375]{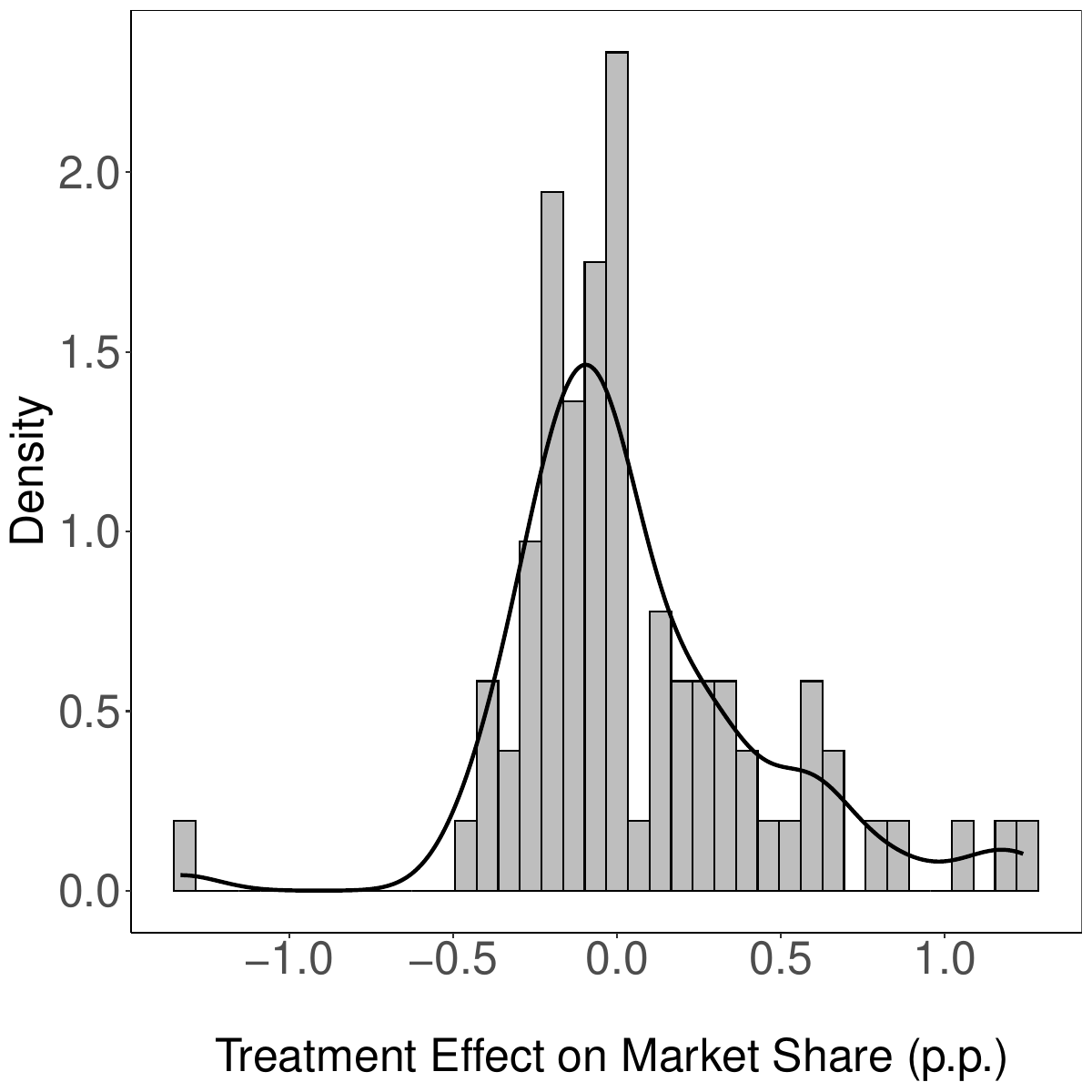}
\begin{flushleft}
\footnotesize Distribution of one-year treatment effects on the number of units sold (left panel) and   
 distribution of one-year treatment effects on market shares (right panel).  
Kernel densities are superimposed on each histogram.
\end{flushleft}
\vspace{-1em}
\caption{\emph{Distribution of car-model-specific treatment effects one year after rationing}}
\label{fig:distribution_effects}
\end{figure}

\subsubsection*{Joint analysis for multiple models}

We wish to investigate the heterogeneity in treatment effects across
different car models. Did luxury cars indeed fare well under the new quota? What about low-end cars? To accommodate such considerations,
we consider simultaneously the treatment effect estimates of multiple
car models.

\begin{figure}
\begin{center}
\includegraphics[scale=0.37]{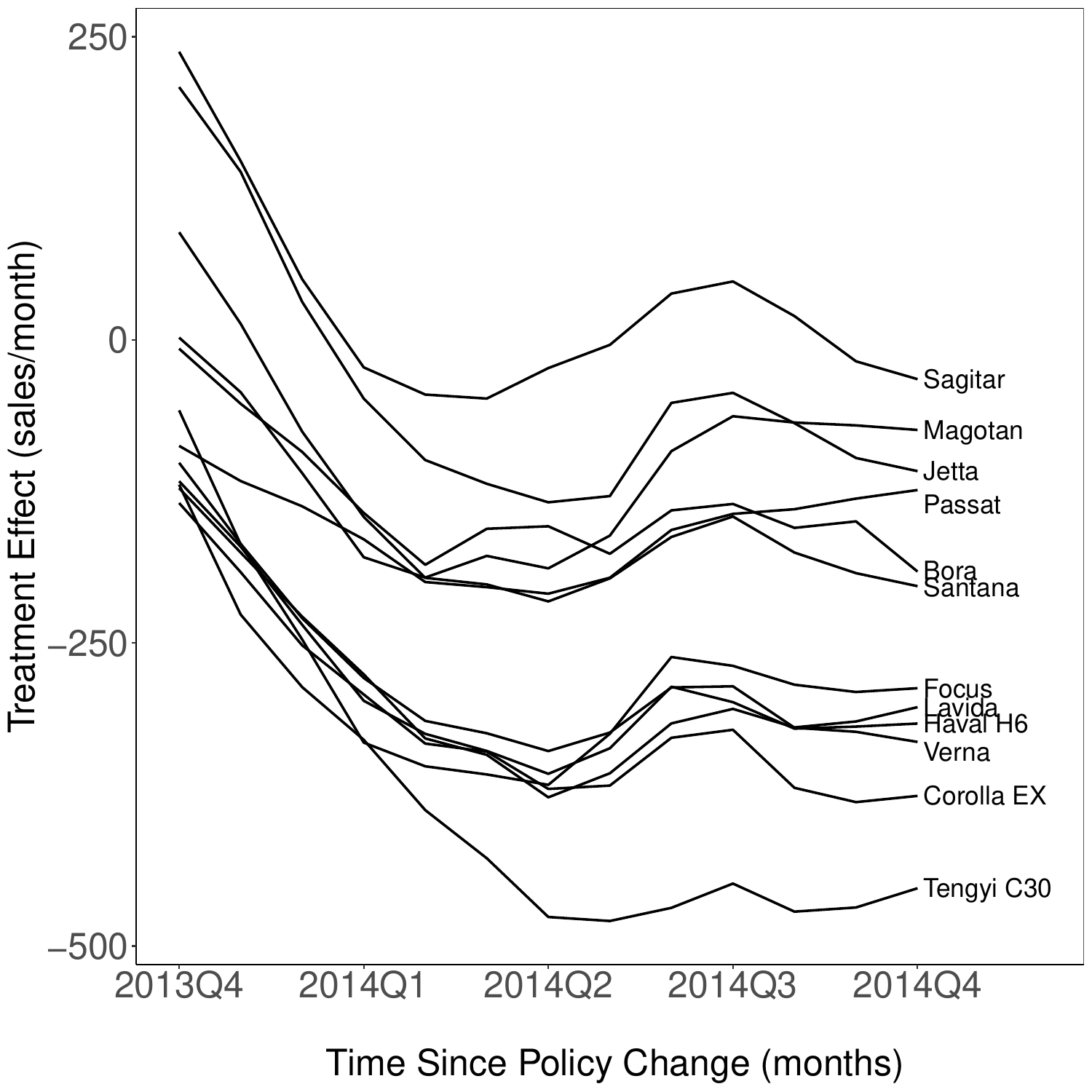}
\end{center}
\caption{\emph{\label{fig:Heterogeneous-treatment-effects}Time Series of Car-Model-Specific Treatment Effect Estimates.}}
\end{figure}

We restrict the analysis to the 78 models that sold at least 2,000 units over the sample window.
We automate the procedure carried out for the Toyota Highlander. 
For each model, we estimate the penalized synthetic control model using the tuning parameter selected by the estimated SURE information criterion and recover the full counterfactual path of the model.

We first assess the plausibility of the homoskedasticity assumption underlying our suggested risk approximation (\ref{eq:information-criteria-estimate}).  
Applying the White test described above, we fail to reject homoskedasticity at the 5\% level for 47 of the 78 models when the outcome is market share (65 at the 1\% level) and for 41 models when the outcome is the number of units sold (59 at the 1\% level).  Although heteroskedasticity cannot be ruled out entirely, it seems too modest to overturn the qualitative conclusions.

With all individual treatment effects in hand, we can produce their
histogram and assess visually the distribution of treatment effects.
Figure~\ref{fig:distribution_effects} plots the empirical distribution of one-year treatment effects across models.  The left panel shows effects on sales levels; almost all mass lies below zero.  
This is expected since the rationing decreased total sales.  The right panel plots effects on market shares.  
Here the distribution is skewed; most models experience a slight loss in relative share, but the right tail is longer and thinner, capturing the fact that a group of models gained a larger market share.

Of course this does not indicate which specific cars are being sold
in greater or lesser proportions.
To accomplish that, Figure \ref{fig:Heterogeneous-treatment-effects} displays the treatment effect paths for the 13 models whose  cumulative sales exceeded 10,000.  All series trend downward, as expected given the sharp contraction in license supply, yet the magnitude of the decline varies markedly: upper-mainstream sedans such as the \emph{Magotan} and \emph{Sagitar} lost far fewer units than budget models such as the \emph{Tengyi C30} or the \emph{Corolla EX}.  

\begin{figure}
\begin{center}
\includegraphics[scale=0.37]{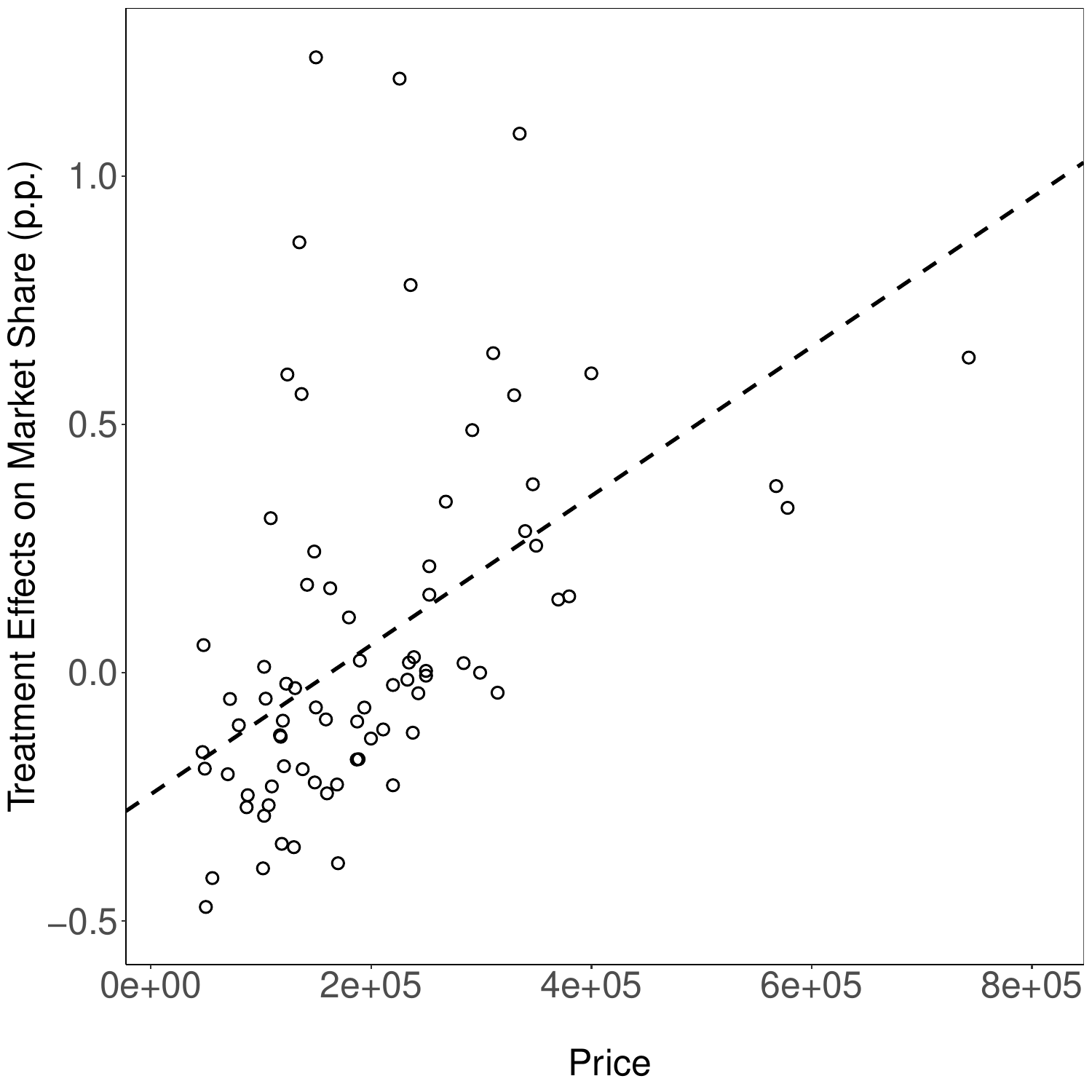}
\end{center}

\caption{\emph{\label{fig:effect-versus-price}Treatment effect versus price,
given in proportional sales. The dashed line gives the OLS fit.}}
\end{figure}

Figure \ref{fig:effect-versus-price}  relates each model’s one-year market-share treatment effect to its average pre-policy price (MSRP).  Higher-priced vehicles experienced larger gains (or smaller losses) in relative share.  An OLS fit yields a positive slope of \(1.5\times10^{-6}\),
implying that a ¥100,000 increase in price is associated with a 0.15-percentage-point rise in post-rationing market share.  
The empirical analysis thus suggests that mid- to high-priced models fared better under rationing, which is consistent with the fact that rationed plates were allocated —through auction or secondary markets— to higher-income households.

This confirms the expected
pattern. More expensive cars tend to see a smaller decrease
in sales. The cars with the most extreme reduction in relative demand
tend to be the cheapest.

\section{Conclusion}
\label{section5}

When used in high-dimensional settings, regression methods are often
appended with a penalty term to avoid overfitting, with the lasso being a popular example.  
The synthetic control method is no exception, and penalized extensions have been developed to deal with high-dimensional settings. 
Such extensions require model selection, either according to cross-validation or to an information criterion.
While information criteria are commonplace for the lasso, and the default in some packages, such methodology was missing for the synthetic control method. 
We have developed novel theory and methodology in order to carry out model selection according to an information criterion. 
We have argued that the information criteria approach is more reliable than cross-validation in our application of interest.

The herein developed theory delivers degrees of freedom estimates for the synthetic control method. 
These in turn produce reassuring theoretical guarantees that the good in-sample fit of the method in early, seminal applications (e.g., \citeNP{abadie2003economic}, \citeNP{abadie2010synthetic}, \citeNP{abadie2015comparative})  was due to the information content and not the flexibility of the synthetic control model.


In analyzing Chinese car sales data, we find that the synthetic control method can be profitably used for filtering when good but noisy matches are available.  In our application, this called for the use of penalized variants of SCM in order to avoid overfitting, and of an information criterion to select the correct tuning parameter.  Thus equipped, we were able to produce model specific treatment effect paths for a selection of cars.

\newpage

\begin{appendices}

\section{Lemmas and Proofs of Main Results}

\paragraph{Notation.} We collect additional notation used throughout the Appendix. We write $\mathbf{1}_k$ for the $k$-vector of ones and $\mathbf{I}_k$ for the $k \times k$ identity matrix. We define $\mathbf{M}_n = \mathbf{I}_n - \frac{1}{n}\mathbf{1}_n\mathbf{1}_n^T$ and $\tilde{\mathbf{X}} = \mathbf{M}_n \mathbf{X}$, where $\tilde{\mathbf{X}}_{\mathcal{A}} = \mathbf{M}_n \mathbf{X}_{\mathcal{A}}$ . For a solution $\hat{\beta}(\mathbf{Y})$ of a synthetic control problem, $\mathcal{A} = \mathcal{A}(\mathbf{Y}) := \{j : \hat{\beta}_j > 0\}$ denotes its active set, and $\mathcal{A}^{\perp} = \{1,\ldots,p\}\setminus\mathcal{A}$. We write $\mathrm{col}(\mathbf{X})$ and $\mathrm{null}(\mathbf{X})$ for the column space and null space of $\mathbf{X}$ respectively, and $\mathbf{X}^{+}$ for the Moore-Penrose pseudoinverse. Finally, $\Pi_{\mathcal{A}}=\mathbf{X}_{\mathcal{A}}(\mathbf{X}_{\mathcal{A}}^T\mathbf{X}_{\mathcal{A}})^{+}\mathbf{X}_{\mathcal{A}}^T$ denotes the orthogonal projection onto $\mathrm{col}(\mathbf{X}_{\mathcal{A}})$.

\paragraph{Proof roadmap.} The proofs proceed in three steps. In \Cref{subsectionA1}, we derive closed-form expressions for the divergence of unpenalized and penalized least-squares problems under linear equality constraints. In \Cref{subsectionA3}, we establish the regularity of synthetic control solutions needed to apply these formulas: the active set is locally stable and the divergence is invariant across optimal solutions.  
We also give a mild condition under which the optimizer is unique. In \Cref{subsectionA4}, we combine these ingredients to prove the main results.

\subsection{Divergence Expressions of Least Squares Problem with Linear Equality Constraints}
\label{subsectionA1}
The divergence obtains by differentiating a tractable expression for
the fitted values. Our proof strategy will be to reformulate all instances
as a problem of least-squares minimization subject to linear equality
constraints. This is convenient because the divergence for such a
problem obtains from elementary manipulations.

\begin{proposition}
Let $\hat{\beta}$ be a solution of
the unpenalized linearly constrained least-squares problem
\begin{equation}
\min_{\beta}\frac{1}{2}\left\Vert \mathbf{Y}-\mathbf{X}\beta\right\Vert _{2}^{2}
\label{equ:constrained_LS_1}
\end{equation}
subject to
\begin{equation}
\mathbf{D}\beta=\mathbf{Z},
\label{equ:constrained_LS_2}
\end{equation}
where $\mathbf{X}\in\mathbb{R}^{n\times p}$ is arbitrary and $\mathbf{D}\in\mathbb{R}^{h\times p}$,
$h\le p$ has full row rank. Then, the divergence matrix of $\hat{\mathbf{Y}}=\mathbf{X}\hat{\beta}$ has the closed-form solution
\[
\nabla\hat{\mathbf{Y}}=\left( \mathbf{X}\mathbf{M}_D \right)\left( \mathbf{X}\mathbf{M}_D \right)^{+} \quad  \text{with} \quad \mathbf{M}_D:=\mathbf{I}_p-\mathbf{D}^T(\mathbf{D}\mathbf{D}^T)^{-1}\mathbf{D}
\]
and its trace is given by
\[
\mathrm{Tr}(\nabla\hat{\mathbf{Y}})=\mathrm{rank}(\tilde{\mathbf{X}})-h \quad  \text{with} \quad \tilde{\mathbf{X}}=(\mathbf{X}^T,\mathbf{D}^T)^T\in\mathbb{R}^{(n+h)\times p}.
\]
\label{proposition:div_unpenalized_constrained_LS}
\end{proposition}
\vspace{-3em}
\begin{proof}[Proof of \Cref{proposition:div_unpenalized_constrained_LS}]
First, we note that, for a given optimal solution $\beta_0$,  any feasible solution $\beta$ satisfying $\mathbf{D}\beta=\mathbf{Z}$ can be written as $\beta=\beta_0+v$, where $v\in\mathrm{null}(\mathbf{D})$. 
Then, the problem (\ref{equ:constrained_LS_1})-(\ref{equ:constrained_LS_2}) becomes
\begin{equation}
\min_{v\in\mathrm{null}(\mathbf{D})} \ \frac{1}{2}\left\Vert \mathbf{Y}-\mathbf{X}\beta_0-\mathbf{X}v\right\Vert _{2}^{2}.
\label{equ:constrained_LS_reduced}
\end{equation}

Since $\mathbf{D}$ has full row rank, $(\mathbf{D}\mathbf{D}^T)^{-1}$ exists. We can therefore express the orthogonal projector onto $\mathrm{null}(\mathbf{D})$ as $\mathbf{M}_D:=\mathbf{I}_p-\mathbf{D}^T(\mathbf{D}\mathbf{D}^T)^{-1}\mathbf{D}$. Let $A=\mathbf{X}\mathbf{M}_D$, we have $\mathbf{X}\mathrm{null}(\mathbf{D})=\mathbf{X}\mathrm{range}(\mathbf{M}_D)=\mathrm{range}(\mathbf{X}\mathbf{M}_D)=\mathrm{col}(A)$. The minimizer in $s:=\mathbf{X}v$ is therefore the orthogonal projection of $\mathbf{Y}-\mathbf{X}\beta_0$ onto $\mathrm{col}(A)$, so the fitted shift is $\mathbf{X}\hat{v}=\left( \mathbf{X}\mathbf{M}_D \right)\left(\mathbf{X}\mathbf{M}_D\right)^{+}(\mathbf{Y}-\mathbf{X}\beta_0)$ where we use the fact that $AA^{+}$ is the orthogonal projection onto $\mathrm{col}(A)$.

Thus
\begin{equation*}
\hat{\mathbf{Y}}=\mathbf{X}\hat{\beta}=\mathbf{X}\beta_0+\mathbf{X}\hat{v}=\mathbf{X}\beta_0+\left( \mathbf{X}\mathbf{M}_D \right)\left( \mathbf{X}\mathbf{M}_D \right)^{+}\left( \mathbf{Y}-\mathbf{X}\beta_0 \right),
\end{equation*}
so the divergence matrix is
\begin{equation*}
\nabla \hat{\mathbf{Y}}=\left( \mathbf{X}\mathbf{M}_D \right)\left( \mathbf{X}\mathbf{M}_D \right)^{+}.
\end{equation*}

Because $AA^+$ is an orthogonal projector, $\mathrm{Tr}(AA^{+})=\mathrm{rank}(A)$, so $\mathrm{Tr}(\nabla\hat{\mathbf{Y}})=\mathrm{rank}(\mathbf{X}\mathbf{M}_D)$. For the rank identity, note that $\mathrm{null}(\tilde{\mathbf{X}})=\mathrm{null}(\mathbf{X})\cap \mathrm{null}(\mathbf{D})$ for $\tilde{\mathbf{X}}=(\mathbf{X}^T,\mathbf{D}^T)^T$, hence $\mathrm{rank}(\tilde{\mathbf{X}})=p-\mathrm{dim}(\mathrm{null}(\mathbf{X})\cap \mathrm{null}(\mathbf{D}))$. Also, by Equation 4.5.1 in \citeA{meyer2023matrix}, we have that
\begin{align*}
\mathrm{rank}(\mathbf{X}\mathbf{M}_D)&=\mathrm{rank}(\mathbf{M}_D)-\mathrm{dim}(\mathrm{null}(\mathbf{X})\cap \mathrm{null}(\mathbf{D}))
\end{align*}
Because $\mathbf{D}$ has full row rank $h$, $\mathrm{rank}(\mathbf{M}_D)=p-h$. Subtracting gives that $\mathrm{rank}(\mathbf{X}\mathbf{M}_D)=\mathrm{rank}(\tilde{\mathbf{X}})-h$ and therefore $\mathrm{Tr}(\nabla\hat{\mathbf{Y}})=\mathrm{rank}(\tilde{\mathbf{X}})-h$, as desired.
\end{proof}

We collect the divergence of the penalized constrained least-squares problem separately.  In this case, we do not obtain a simplified closed form for the trace. 

\begin{proposition}
Let $\hat{\beta}$ be the solution of
the penalized linearly constrained least-squares problem
\begin{equation}
\min_{\beta}\frac{1}{2}\left\Vert \mathbf{Y}-\mathbf{X}\beta\right\Vert _{2}^{2}+\frac{1}{2}
\lambda\sum_{j=1}^p \beta_j \left\Vert \mathbf{Y}-\mathbf{X}_j \right\Vert_{2}^{2}
\label{psc1}
\end{equation}
subject to
\begin{equation}
\mathbf{D}\beta=\mathbf{Z}, 
\label{psc2}
\end{equation}
where $\mathbf{X}\in\mathbb{R}^{n\times p}$ has full column rank, $\mathbf{D}\in\mathbb{R}^{h\times p}$ has full row rank and $\mathbf{X}_j$ is the $j$-th column of $\mathbf{X}$. The divergence of $\mathbf{X}\hat{\beta}$ has the closed-form expression
\begin{align*}
\nabla\mathbf{X}\hat{\beta}=&\mathbf{X}(\mathbf{X}^T\mathbf{X})^{-1}\mathbf{X}^T-\mathbf{X}(\mathbf{X}^T\mathbf{X})^{-1}\mathbf{D}^T\Bigg{(}\mathbf{D}(\mathbf{X}^T\mathbf{X})^{-1}\mathbf{D}^T\Bigg{)}^{-1}\mathbf{D}(\mathbf{X}^T\mathbf{X})^{-1}\mathbf{X}^T\\
&-\lambda\mathbf{X}(\mathbf{X}^T\mathbf{X})^{-1}(\mathbf{1}_p\mathbf{Y}^T-\mathbf{X}^T)+\lambda\mathbf{X}(\mathbf{X}^T\mathbf{X})^{-1}\mathbf{D}^T\Bigg{(}\mathbf{D}(\mathbf{X}^T\mathbf{X})^{-1}\mathbf{D}^T\Bigg{)}^{-1}\mathbf{D}(\mathbf{X}^T\mathbf{X})^{-1}\left(\mathbf{1}_p\mathbf{Y}^T-\mathbf{X}^T\right).
\end{align*}
\label{proposition:div_penalized_constrained_LS}
\end{proposition}
\vspace{-3em}
\begin{proof}[Proof of \Cref{proposition:div_penalized_constrained_LS}]
The problem (\ref{psc1})-(\ref{psc2}) has the Lagrangian function
\begin{equation*}
\mathcal{L}=\frac{1}{2}\|\mathbf{Y}-\mathbf{X}\beta \|_2^2+\frac{1}{2}\lambda\sum\limits_{j=1}^p \beta_j\|\mathbf{Y}-\mathbf{X}_j\|_2^2+\xi^T(\mathbf{D}\beta-\mathbf{Z})
\end{equation*}
where $\xi\in\mathbb{R}^h$, and KKT conditions are
\begin{align*}
\cfrac{\partial\mathcal{L}}{\partial\beta}=-\mathbf{X}^T(\mathbf{Y}-\mathbf{X}\beta)+\frac{1}{2}\lambda \mathcal{Q}+\mathbf{D}^T\xi=\mathbf{0}, \ \text{where} \ \mathcal{Q} =
\left(\begin{matrix}
\|\mathbf{Y}-\mathbf{X}_1\|_2^2\\
\vdots \\
\|\mathbf{Y}-\mathbf{X}_p\|_2^2
\end{matrix}\right).
\end{align*}

The optimality conditions can be rewritten in matrix form
\begin{align*}
\left(\begin{matrix}
\mathbf{X}^T\mathbf{X} & \mathbf{D}^T \\
\mathbf{D}         &  \mathbf{0}
\end{matrix}\right)
\left(\begin{matrix}
\hat{\beta}\\
\hat{\xi}
\end{matrix}\right)=
\left(\begin{matrix}
\mathbf{X}^T\mathbf{Y}-\frac{1}{2}\lambda\mathcal{Q}\\
\mathbf{Z}
\end{matrix}\right).
\end{align*}

Using the same approach as in the unpenalized problem, we obtain the closed-form expression
\begin{align*}
\hat{\beta}=&(\mathbf{X}^T\mathbf{X})^{-1}\mathbf{X}^T\mathbf{Y}-(\mathbf{X}^T\mathbf{X})^{-1}\mathbf{D}^T\Bigg{(}\mathbf{D}(\mathbf{X}^T\mathbf{X})^{-1}\mathbf{D}^T\Bigg{)}^{-1}\mathbf{D}(\mathbf{X}^T\mathbf{X})^{-1}\mathbf{X}^T\mathbf{Y}\\
&+(\mathbf{X}^T\mathbf{X})^{-1}\mathbf{D}^T\Bigg{(} \mathbf{D}(\mathbf{X}^T\mathbf{X})^{-1}\mathbf{D}^T \Bigg{)}^{-1}\mathbf{Z}
-\frac{1}{2}\lambda(\mathbf{X}^T\mathbf{X})^{-1}\mathcal{Q}\\
&+ \frac{1}{2}\lambda(\mathbf{X}^T\mathbf{X})^{-1}\mathbf{D}^T\Bigg{(}\mathbf{D}(\mathbf{X}^T\mathbf{X})^{-1}\mathbf{D}^T\Bigg{)}^{-1}\mathbf{D}(\mathbf{X}^T\mathbf{X})^{-1}\mathcal{Q}.
\end{align*}

Note that $\nabla\mathcal{Q}(\mathbf{Y})=2\mathbf{1}_p\mathbf{Y}^T-2\mathbf{X}^T$. The closed-form expression for the divergence of the penalized linearly constrained least-squares problem obtains as
\begin{align*}
\cfrac{\partial\mathbf{X}\hat{\beta}}{\partial\mathbf{Y}}=&\mathbf{X}(\mathbf{X}^T\mathbf{X})^{-1}\mathbf{X}^T-\mathbf{X}(\mathbf{X}^T\mathbf{X})^{-1}\mathbf{D}^T\Bigg{(}\mathbf{D}(\mathbf{X}^T\mathbf{X})^{-1}\mathbf{D}^T\Bigg{)}^{-1}\mathbf{D}(\mathbf{X}^T\mathbf{X})^{-1}\mathbf{X}^T\\
&-\lambda\mathbf{X}(\mathbf{X}^T\mathbf{X})^{-1}(\mathbf{1}_p\mathbf{Y}^T-\mathbf{X}^T)+\lambda\mathbf{X}(\mathbf{X}^T\mathbf{X})^{-1}\mathbf{D}^T\Bigg{(}\mathbf{D}(\mathbf{X}^T\mathbf{X})^{-1}\mathbf{D}^T\Bigg{)}^{-1}\mathbf{D}(\mathbf{X}^T\mathbf{X})^{-1}\left(\mathbf{1}_p\mathbf{Y}^T-\mathbf{X}^T\right).
\end{align*}
as claimed.
\end{proof}

The 
case of the penalized SCM without covariates 
is a special case with
simpler expression for the divergence.
In that case, $\mathbf{D}=\mathbf{1}_p^T$ and $\mathbf{Z}=1$, and we  have the more concise expression
\allowdisplaybreaks
\begin{align*}
\cfrac{\partial\mathbf{X}\hat{\beta}}{\partial\mathbf{Y}}=&\mathbf{X}(\mathbf{X}^T\mathbf{X})^{-1}\mathbf{X}^T-\mathbf{X}(\mathbf{X}^T\mathbf{X})^{-1}\mathbf{1}_p\Bigg{(}\mathbf{1}_p^T(\mathbf{X}^T\mathbf{X})^{-1}\mathbf{1}_p\Bigg{)}^{-1}\mathbf{1}_p^T(\mathbf{X}^T\mathbf{X})^{-1}\mathbf{X}^T\\
&-\lambda\mathbf{X}(\mathbf{X}^T\mathbf{X})^{-1}\mathbf{1}_p\mathbf{Y}^T+\lambda\mathbf{X}(\mathbf{X}^T\mathbf{X})^{-1}\mathbf{1}_p\Bigg{(}\mathbf{1}_p^T(\mathbf{X}^T\mathbf{X})^{-1}\mathbf{1}_p\Bigg{)}^{-1}\mathbf{1}_p^T(\mathbf{X}^T\mathbf{X})^{-1}\mathbf{1}_p\mathbf{Y}^T
\\
&+\lambda\mathbf{X}(\mathbf{X}^T\mathbf{X})^{-1}\mathbf{X}^T-\lambda\mathbf{X}(\mathbf{X}^T\mathbf{X})^{-1}\mathbf{1}_p\Bigg{(}\mathbf{1}_p^T(\mathbf{X}^T\mathbf{X})^{-1}\mathbf{1}_p\Bigg{)}^{-1}\mathbf{1}_p^T(\mathbf{X}^T\mathbf{X})^{-1}\mathbf{X}^T
\\
=&(1+\lambda)\left[\mathbf{X}(\mathbf{X}^T\mathbf{X})^{-1}\mathbf{X}^T-\cfrac{1}{\mathbf{1}_p^T(\mathbf{X}^T\mathbf{X})^{-1}\mathbf{1}_p }\mathbf{X}(\mathbf{X}^T\mathbf{X})^{-1}\mathbf{1}_p\mathbf{1}_p^T(\mathbf{X}^T\mathbf{X})^{-1}\mathbf{X}^T\right].
\end{align*}


Finally, we treat the constrained ridge regression problem, which includes an intercept term.
\begin{proposition}
\label{prop:div_const_ridge}
Let $(\hat{\beta},\hat{\beta}_0)$ denote the solution to the constrained ridge regression problem
\begin{equation*}
\min_{\beta_{0},\beta} \   \left\Vert \mathbf{Y}-\beta_{0}\mathbf{1}_{n}-\mathbf{X}\beta\right\Vert _{2}^{2}+\lambda\left\Vert \beta\right\Vert _{2}^{2}
\end{equation*}
subject to
\begin{equation*}
\mathbf{D}\beta=\mathbf{Z},
\end{equation*}
where $\mathbf{X}\in\mathbb{R}^{n\times p}$ is arbitrary  and $\mathbf{D}\in\mathbb{R}^{h\times p}$,
$h\le p$, has full row rank. Then, for $\lambda> 0$, the constrained ridge regression fit $\hat{ \mathbf{Y}}=\mathbf{X}\hat{\beta}+\hat{\beta}_{0}\mathbf{1}_n$ has divergence
\begin{align*}
\nabla\hat{\mathbf{Y}}=\tilde{\mathbf{X}}\mathbf{Q}^{-1}\tilde{\mathbf{X}}^T
+\frac{1}{n}\mathbf{1}_n\mathbf{1}_n^T-\tilde{\mathbf{X}}\mathbf{Q}^{-1}\mathbf{D}^T\left( \mathbf{D}\mathbf{Q}^{-1}\mathbf{D}^T \right)^{-1}\mathbf{D}\mathbf{Q}^{-1}\tilde{\mathbf{X}}^T.
\end{align*}
where  $\tilde{\mathbf{X}}=(\mathbf{I}_n-\frac{1}{n}\mathbf{1}_n\mathbf{1}_n^T)\mathbf{X}$ and $\mathbf{Q}=\tilde{\mathbf{X}}^T\tilde{\mathbf{X}}+\lambda\mathbf{I}_{p}$.
\label{propdivcrsc}
\end{proposition}

\begin{proof}[Proof of \Cref{propdivcrsc}]
The corresponding Lagrangian of the constrained ridge regression problem is
\begin{equation*}
\mathcal{L}=\left\Vert \mathbf{Y}-\beta_{0}\mathbf{1}_{n}-\mathbf{X}\beta\right\Vert _{2}^{2}+\lambda\left\Vert \beta\right\Vert _{2}^{2}+\xi^T(\mathbf{D}\beta-\mathbf{Z}),
\end{equation*}
where $\xi\in\mathbb{R}^h$ and the KKT conditions are
\begin{align*}
\cfrac{\partial\mathcal{L}}{\partial\beta}&=-2\mathbf{X}^T(\mathbf{Y}-\beta_0\mathbf{1}_n-\mathbf{X}\beta)+2\lambda\beta+\mathbf{D}^T\xi=\mathbf{0}_p,\\[7pt]
\cfrac{\partial\mathcal{L}}{\partial\beta_0}&=\mathbf{1}_n^T(\mathbf{Y}-\beta_0\mathbf{1}_n-\mathbf{X}\beta)=0,\\[7pt]
\cfrac{\partial\mathcal{L}}{\partial\xi} &= \mathbf{D}\beta-\mathbf{Z}=\mathbf{0}_h.
\end{align*}

The optimality conditions can be rewritten in the matrix form
\begin{align*}
\left(\begin{matrix}
\mathbf{X}^T\mathbf{X}+\lambda\mathbf{I}_p & \mathbf{X}^T\mathbf{1}_n & \mathbf{D}^T \\[7pt]
\mathbf{1}_n^T\mathbf{X} & n & \mathbf{0}\\[7pt]
\mathbf{D}         &  \mathbf{0} & \mathbf{0}
\end{matrix}\right)
\left(\begin{matrix}
\beta\\[7pt]
\beta_0\\[7pt]
\xi/2
\end{matrix}\right)=
\left(\begin{matrix}
\mathbf{X}^T\mathbf{Y}\\[7pt]
\mathbf{1}_n^T\mathbf{Y}\\[7pt]
\mathbf{Z}
\end{matrix}\right).
\end{align*}

Recall that $\tilde{\mathbf{X}}=(\mathbf{I}_n-\frac{1}{n}\mathbf{1}_n\mathbf{1}_n^T)\mathbf{X}$, $\mathbf{Q}=\tilde{\mathbf{X}}^T\tilde{\mathbf{X}}+\lambda\mathbf{I}_{p}$. We solve this system for $(\hat{\beta},\hat{\beta}_0)$:
\begin{align*}
\left(
\begin{matrix}
\hat{\beta}\\[7pt]
\hat{\beta}_0
\end{matrix}
\right)=&\left(
\begin{matrix}
\mathbf{Q}^{-1} & -\mathbf{Q}^{-1}\frac{1}{n}\mathbf{X}^T\mathbf{1}_n\\[7pt]
-\frac{1}{n}\mathbf{1}_n^T\mathbf{X}\mathbf{Q}^{-1} & \frac{1}{n}+\frac{1}{n^2}\mathbf{1}_n^T\mathbf{X}\mathbf{Q}^{-1}\mathbf{X}^T\mathbf{1}_n
\end{matrix}
\right)\left(
\begin{matrix}
\mathbf{X}^T\mathbf{Y}\\[7pt]
\mathbf{1}_n^T\mathbf{Y}
\end{matrix}
\right)\\[7pt]
&-\left(
\begin{matrix}
\mathbf{Q}^{-1}\mathbf{D}^T\\[7pt]
-\frac{1}{n}\mathbf{1}_n^T\mathbf{X}\mathbf{Q}^{-1}\mathbf{D}^T
\end{matrix}
\right)\left(
\mathbf{D}\mathbf{Q}^{-1}\mathbf{D}^T
\right)^{-1}\left(
\begin{matrix}
\mathbf{D}\mathbf{Q}^{-1}   \\[7pt]
-\frac{1}{n}\mathbf{D}\mathbf{Q}^{-1}\mathbf{X}^T\mathbf{1}_n
\end{matrix}
\right)^T\left(
\begin{matrix}
\mathbf{X}^T\mathbf{Y}\\[7pt]
\mathbf{1}_n^T\mathbf{Y}
\end{matrix}
\right)\\[7pt]
&+\left(
\begin{matrix}
\mathbf{Q}^{-1}\mathbf{D}^T\\[7pt]
-\frac{1}{n}\mathbf{1}_n^T\mathbf{X}\mathbf{Q}^{-1}\mathbf{D}^T
\end{matrix}
\right)\left(
\mathbf{D}\mathbf{Q}^{-1}\mathbf{D}^T
\right)^{-1}\mathbf{Z}.
\end{align*}
Now we can obtain the closed-form expression for the divergence of the constrained ridge regression problem:
\begin{align*}
\frac{\partial \hat{\mathbf{Y}}}{\partial \mathbf{Y}}=&\left(\mathbf{I}_n-\frac{1}{n}\mathbf{1}_n\mathbf{1}_n^T\right)\mathbf{X}\mathbf{Q}^{-1}\mathbf{X}^T\left(\mathbf{I}_n-\frac{1}{n}\mathbf{1}_n\mathbf{1}_n^T\right)
+\frac{1}{n}\mathbf{1}_n\mathbf{1}_n^T\\[7pt]
&-\left( \mathbf{X}\mathbf{Q}^{-1}\mathbf{D}^T-\frac{1}{n}\mathbf{1}_n\mathbf{1}_n^T\mathbf{X}\mathbf{Q}^{-1}\mathbf{D}^T \right)
\left( \mathbf{D}\mathbf{Q}^{-1}\mathbf{D}^T \right)^{-1}\left( \mathbf{D}\mathbf{Q}^{-1}\mathbf{X}^T-\frac{1}{n}\mathbf{D}\mathbf{Q}^{-1}\mathbf{X}^T\mathbf{1}_n\mathbf{1}_n^T \right)\\[7pt]
=&\tilde{\mathbf{X}}\mathbf{Q}^{-1}\tilde{\mathbf{X}}^T
+\frac{1}{n}\mathbf{1}_n\mathbf{1}_n^T-\tilde{\mathbf{X}}\mathbf{Q}^{-1}\mathbf{D}^T\left( \mathbf{D}\mathbf{Q}^{-1}\mathbf{D}^T \right)^{-1}\mathbf{D}\mathbf{Q}^{-1}\tilde{\mathbf{X}}^T,
\end{align*}
as claimed.
\end{proof}

\subsection{Regularity Conditions of Synthetic Control Solutions}

\label{subsectionA3}


The divergence formulas in \Cref{subsectionA1} are derived for a fixed active set $\mathcal{A}$. To apply them, 
we show that, outside a Lebesgue-null set, the active set is locally stable. As a result, the regression fit admits a local representation with fixed $\mathcal{A}$ and hence falls under \Cref{subsectionA1}. 
Furthermore, we establish the rank invariance of the divergence across optimal solutions, and we give a mild condition under which the synthetic control optimizer is unique.


Throughout this subsection, we focus on the following optimization problem:
\begin{equation}
\min_{\beta} \ \left\Vert \mathbf{Y}-\mathbf{X}\beta\right\Vert _{2}^{2}
\label{a2_problem_1}
\end{equation}
subject to
\begin{equation}
\mathbf{D}\beta=\mathbf{Z}, \quad \beta\geq 0.
\label{a2_problem_2}
\end{equation}
 This formulation covers the unpenalized synthetic control method, with or without covariates. Although this is a special case of our general result, 
 it presents the argument in full mathematical depth, albeit with more digestible notation. 
 The general case is conceptually identical.

Let $\hat{\beta}(\mathbf{Y})$ be a solution of problem (\ref{a2_problem_1})-(\ref{a2_problem_2}), $\mathcal{A}(\mathbf{Y})$ its active set, and let $\mu(\mathbf{Y})$ be the Lagrangian multiplier vector corresponding to the primal non-negativity constraint, where the notation is modified to emphasize the dependence on $\mathbf{Y}$.

\subsubsection{Local Stability of the Active Set}

First, we show that the active set $\mathcal{A}(\mathbf{Y})$ is locally constant in $\mathbf{Y}$ outside a Lebesgue-null set. In such a neighborhood, the regression fit admits a representation with fixed $\mathcal{A}$, so the divergence formulas in \Cref{subsectionA1} apply.

\begin{lemma}
For the problem (\ref{a2_problem_1})-(\ref{a2_problem_2}), there is a measure zero set $\mathcal{N}\subset\mathbb{R}^n$ such that for all $\mathbf{Y}\notin\mathcal{N}$ and any optimal solution $\hat{\beta}(\mathbf{Y})$ with active set $\mathcal{A}(\mathbf{Y})$, there is a neighborhood $U$ such that for every $\mathbf{Y}'\in U$, the problem admits an optimal solution $\hat{\beta}(\mathbf{Y}')$ with $\mathcal{A}(\mathbf{Y}')=\mathcal{A}(\mathbf{Y})$.
\label{lemma2}
\end{lemma}
\begin{remark}
In our problem (\ref{a2_problem_1})-(\ref{a2_problem_2}), the optimizer $\hat{\beta}(\mathbf{Y})$ may not be unique. Moreover, even after fixing an active set $\mathcal{A}$, the KKT system may be singular, so one cannot in general obtain a locally unique affine representation $\mathbf{Y}\mapsto\beta_{\mathcal{A}}(\mathbf{Y})$ without additional non-degeneracy assumptions. For this reason, we use an argument different from that used in Lemma 5 of \citeA{tibshirani2012degrees}.
\end{remark}

\begin{proof}[Proof of \Cref{lemma2}]

The proof proceeds as follows.  First, we construct polyhedral regions $R_{\mathcal{A},m}$ whose points admit an optimal solution with active set~$\mathcal{A}$.  Second, we define $\mathcal{N}$ as the countable union of their boundaries, which has Lebesgue measure zero.  Third, we show that every $\mathbf{Y}\notin\mathcal{N}$ lies in the interior of some $R_{\mathcal{A},m}$, so the solution corresponding to
any outcome vector in a sufficiently small neighborhood of $\mathbf{Y}$ has the same active set.

For an index set $\mathcal{A}\subseteq\{1,...,p\}$ and an integer $m\geq 1$, let $R_{\mathcal{A},m}\subseteq\mathbb{R}^n$ be the set of all $\mathbf{Y}\in\mathbb{R}^n$ for which there exist $\beta_{\mathcal{A}}\in\mathbb{R}^{|\mathcal{A}|}$ and $\xi\in\mathbb{R}^h$ satisfying
\begin{align}
\mathbf{D}_{\mathcal{A}}\beta_{\mathcal{A}}=\mathbf{Z}, \quad \quad \beta_{\mathcal{A}}\geq \frac{1}{m}\mathbf{1}_{|\mathcal{A}|}, \label{eq:RAm_feas}\\[7pt]
2\mathbf{X}_{\mathcal{A}}^T\left(\mathbf{Y}-\mathbf{X}_{\mathcal{A}}\beta_{\mathcal{A}}\right)-\mathbf{D}_{\mathcal{A}}^T\xi=\mathbf{0} ,\label{eq:RAm_statA} \\[7pt]
2\mathbf{X}_{\mathcal{A}^\perp}^T\left(\mathbf{Y}-\mathbf{X}_{\mathcal{A}}\beta_{\mathcal{A}}\right)-\mathbf{D}_{\mathcal{A}^\perp}^T\xi\leq \mathbf{0}. \label{eq:RAm_statAc}
\end{align}
Together, \eqref{eq:RAm_feas}--\eqref{eq:RAm_statAc} are the KKT conditions of problem \eqref{a2_problem_1}--\eqref{a2_problem_2} restricted to active set~$\mathcal{A}$, with the strict positivity of the active coordinates tightened to $\beta_{\mathcal{A}}\geq m^{-1}\mathbf{1}$.

We now verify that every $\mathbf{Y}\in R_{\mathcal{A},m}$ admits an optimal solution whose active set is exactly~$\mathcal{A}$. Choose any \((\beta_{\mathcal A},\xi)\) satisfying
\Cref{eq:RAm_feas,eq:RAm_statA,eq:RAm_statAc}. Extend \(\beta\in\mathbb R^p\) by setting
\(\beta_{\mathcal A^{\perp}}=\mathbf 0\), and define
\(\mu\in\mathbb R^p\) by \(\mu_{\mathcal A}=\mathbf 0\) and
\begin{equation*}
\mu_{\mathcal A^{\perp}}
:=2\mathbf X_{\mathcal A^{\perp}}^\top(\mathbf Y-\mathbf X_{\mathcal A}\beta_{\mathcal A})
 -\mathbf D_{\mathcal A^{\perp}}^\top\xi \le \mathbf 0.
\end{equation*}
Then, the triple $(\beta,\xi,\mu)$ satisfies the KKT conditions for (\ref{a2_problem_1})-(\ref{a2_problem_2}). Since the objective is convex and the constraints are convex, these KKT conditions are sufficient for optimality. Then $\beta_{\mathcal{A}}(\mathbf{Y})\geq m^{-1}\mathbf{1}_{|\mathcal{A}|}$ implies that this optimal solution has active set \(\mathcal A\).

Next, observe that $R_{\mathcal{A},m}$ is the projection onto the $\mathbf{Y}$-coordinates of a polyhedron in $(\mathbf{Y},\beta_{\mathcal{A}},\xi)$-space defined by the linear constraints
\eqref{eq:RAm_feas}--\eqref{eq:RAm_statAc}.  Hence, $R_{\mathcal{A},m}$ is a polyhedron in $\mathbb{R}^n$ (\citeNP[~Chapter 12.2]{schrijver1998theory}). Therefore its boundary $\partial R_{\mathcal{A},m}$ is contained in a finite union of faces of dimension at most $n-1$, and thus has Lebesgue measure zero.

Define
\begin{equation*}
\mathcal{N}:=\bigcup_{\mathcal{A}}\bigcup_{m\geq 1} \partial R_{\mathcal{A},m}.
\end{equation*}
This is a countable union of Lebesgue-null sets, hence $\mathcal{N}$ has Lebesgue measure zero.

Take $\mathbf{Y}\notin\mathcal{N}$ and an optimal solution $\hat{\beta}(\mathbf{Y})$ of the problem (\ref{a2_problem_1})-(\ref{a2_problem_2}).
Let $\mathcal{A}=\mathcal{A}(\mathbf{Y})$ so that $\hat{\beta}_{\mathcal{A}}(\mathbf{Y})>0$ and $\hat{\beta}_{\mathcal{A}^{\perp}}(\mathbf{Y})=0$. 
Define  $\delta(\mathbf{Y}):=\min_{i\in\mathcal{A}}\hat{\beta}_i(\mathbf{Y})>0$ and choose an integer $m\geq 1$ such that $m^{-1}<\delta(\mathbf{Y})$. Let $(\hat{\xi}(\mathbf{Y}),\hat{\mu}(\mathbf{Y}))$ be any KKT multipliers associated with $\hat{\beta}(\mathbf{Y})$.
Then, $(\hat{\beta}_{\mathcal{A}}(\mathbf{Y}),\hat{\xi}(\mathbf{Y}))$ satisfies the defining constraints (\ref{eq:RAm_feas})-(\ref{eq:RAm_statAc}), so $\mathbf{Y}\in R_{\mathcal{A},m}$. But $\mathbf{Y}\notin \partial R_{\mathcal{A},m}$, so $\mathbf{Y}$ lies in the interior of $R_{\mathcal{A},m}$. 
Hence, there exists a neighborhood $U$ of $\mathbf{Y}$ such that $U\subseteq R_{\mathcal{A},m}$.

For every $\mathbf{Y}'\in U$, because $\mathbf{Y}'\in R_{\mathcal{A},m}$, the problem admits an optimal solution $\hat{\beta}(\mathbf{Y}')$ with $\hat{\beta}_{\mathcal{A}}(\mathbf{Y}')\geq m^{-1}\mathbf{1}_{|\mathcal{A}|}$ and $\hat{\beta}_{\mathcal{A}^{\perp}}(\mathbf{Y}')=0$. Therefore, for each $\mathbf Y'\in U$, there exists an optimal solution $\hat{\beta}(\mathbf{Y}')$ whose active set equals $\mathcal{A}$.
\end{proof}

In other words, \Cref{lemma2} shows that, outside a Lebesgue-null set, $\mathbf{Y}$ lies in the interior of a region where one can choose an optimal solution with a locally constant set of binding nonnegativity constraints. This is closely related to strict-complementarity for polyhedral constraints.
\begin{remark}
The proof mechanics above extend with essentially no changes to the other synthetic control variants in \Cref{section2} provided that (i) the feasible set is polyhedral (e.g.\ the simplex constraints); and (ii) after fixing an active set $\mathcal{A}$ (and, when an $\ell_1$ term is present, also fixing the sign pattern on $\mathcal{A}$), the KKT conditions can be written as finitely many linear equalities or inequalities.
\end{remark}

\subsubsection{Rank Invariance of the Divergence}

The synthetic control problem may admit multiple optimal solutions $\hat{\beta}$, and the associated active sets can differ. In this subsection, we show that such non-uniqueness does not affect the divergence representation. We show that, outside a Lebesgue-null set, the divergence is invariant across all optimal solutions, and is therefore well-defined. 

\begin{lemma}[nested idempotents, \citeNP{meyer2023matrix}, Exercise 5.9.18]
Let $P,Q\in\mathbb{R}^{n\times n}$ be idempotent matrices, so $P^2=P, Q^2=Q$, and suppose that $PQ=QP=Q$, which is equivalent to $\mathrm{col}(Q)\subseteq\mathrm{col}(P)$ and $\mathrm{null}(P)\subseteq \mathrm{null}(Q)$. Then, $\mathrm{col}(P-Q)=\mathrm{col}(P)\cap \mathrm{null}(Q)$ and therefore $\mathrm{rank}(P-Q)=\dim(\mathrm{col}(P)\cap\mathrm{null}(Q))=\mathrm{rank}(P)-\mathrm{rank}(Q)$.
\label{nested_idempotents}
\end{lemma}


\begin{lemma}[rank invariance]
Consider the problem (\ref{a2_problem_1})-(\ref{a2_problem_2}). For almost every $\mathbf{Y}\in\mathbb{R}^{n}$,  $\mathrm{rank}(\nabla \mathbf{X}\hat{\beta})$ is invariant with respect to the optimal solution $\hat{\beta}(\mathbf{Y})$, which need not be unique. 

\label{lemma_rank_invariance}
\end{lemma}

\begin{proof}[Proof of \Cref{lemma_rank_invariance}]

By \Cref{lemma2}, the active set $\mathcal{A}$ is locally stable. That is to say, for almost every $\mathbf{Y}$, there is a neighborhood $U$
of $\mathbf{Y}$ such that
\[
\mathbf{X}\hat{\beta}\left(\mathbf{Y}'\right)=\Pi_{\mathcal{A}}\mathbf{Y}'-\mathcal{C}_{\mathcal{A}}\mathbf{Y}'+\mathbf{X}_{\mathcal{A}}\left(\mathbf{X}_{\mathcal{A}}^{T}\mathbf{X}_{\mathcal{A}}\right)^{+}\mathbf{D}_{\mathcal{A}}^{T}\left(\mathbf{D}_{\mathcal{A}}\left(\mathbf{X}_{\mathcal{A}}^{T}\mathbf{X}_{\mathcal{A}}\right)^{+}\mathbf{D}_{\mathcal{A}}^{T}\right)^{+}\mathbf{Z}
\]
for all $\mathbf{Y}' \in U$, where we define the orthogonal projections $\Pi_{\mathcal{A}}=\mathbf{X}_{\mathcal{A}}(\mathbf{X}_{\mathcal{A}}^{T}\mathbf{X}_{\mathcal{A}})^{+}\mathbf{X}_{\mathcal{A}}^{T}$ and
\begin{equation*}
\mathcal{C}_{\mathcal{A}}=\mathbf{X}_{\mathcal{A}}\left(\mathbf{X}_{\mathcal{A}}^{T}\mathbf{X}_{\mathcal{A}}\right)^{+}\mathbf{D}_{\mathcal{A}}^{T}\left(\mathbf{D}_{\mathcal{A}}\left(\mathbf{X}_{\mathcal{A}}^{T}\mathbf{X}_{\mathcal{A}}\right)^{+}\mathbf{D}_{\mathcal{A}}^{T}\right)^{+}\mathbf{D}_{\mathcal{A}}\left(\mathbf{X}_{\mathcal{A}}^{T}\mathbf{X}_{\mathcal{A}}\right)^{+}\mathbf{X}_{\mathcal{A}}^{T}
\end{equation*}
for all $\mathbf{Y}'\in U$, for fixed active set $\mathcal{A}$.

Now suppose that $\mathcal{A}^{*}$ is the active set corresponding to another optimal solution $\hat{\beta}^{*}\left(\mathbf{Y}\right)$ at $\mathbf{Y}$. Then, likewise, $\mathcal{A}^{*}$ is locally stable, and there is a neighborhood
$U^{*}$ of $\mathbf{Y}$ where
\[
\mathbf{X}\hat{\beta}^{*}\left(\mathbf{Y}'\right)=\Pi_{\mathcal{A^{*}}}\mathbf{Y}'-\mathcal{C}_{\mathcal{A}^*}\mathbf{Y}'+\mathbf{X}_{\mathcal{A}^{*}}\left(\mathbf{X}_{\mathcal{A}^{*}}^{T}\mathbf{X}_{\mathcal{A^{*}}}\right)^{+}\mathbf{D}_{\mathcal{A^{*}}}^{T}\left(\mathbf{D}_{\mathcal{A}^{*}}\left(\mathbf{X}_{\mathcal{A}^{*}}^{T}\mathbf{X}_{\mathcal{A^{*}}}\right)^{+}\mathbf{D}_{\mathcal{A}^{*}}^{T}\right)^{+}\mathbf{Z}
\]
for all $\mathbf{Y}'\in U^{*}$, for fixed active set $\mathcal{A}^{*}$.

By the uniqueness of the fitted values, the right-hand side of both previous displays
are equal for any $\mathbf{Y}'\in U\cap U^{*}$. Now notice that,
since $U\cap U^{*}$ is open, for any $\mathbf{W}\in \mathcal{S}_{\mathcal{A}}:=
\mathrm{col}(\mathbf{X}_{\mathcal{A}})
\;\cap\;
{\mathrm{null}}\!\left(
\mathbf{D}_{\mathcal{A}}
(\mathbf{X}_{\mathcal{A}}^{T}\mathbf{X}_{\mathcal{A}})^{+}
\mathbf{X}_{\mathcal{A}}^{T}
\right)$,
there exists $\epsilon>0$ such that $\mathbf{Y}+\epsilon\mathbf{W}\in U\cap U^{*}$.
Therefore, by equating the right-hand side of both previous displays
and plugging in $\mathbf{Y}'=\mathbf{Y}+\epsilon\mathbf{W}$ and $\mathbf{Y}$ and differencing, we get
\[
\Pi_{\mathcal{A}}\mathbf{W}-\mathcal{C}_{\mathcal{A}}\mathbf{W}=\Pi_{\mathcal{A^{*}}}\mathbf{W}-\mathcal{C}_{\mathcal{A}^*}\mathbf{W}.
\]
Since $\mathbf{W}\in\mathcal{S}_{\mathcal{A}}$, we have $\Pi_{\mathcal{A}}\mathbf{W}=\mathbf{W}$ and $\mathcal{C}_{\mathcal{A}}\mathbf{W}=\mathbf{0}$, hence
\[
\mathbf{W}=\Pi_{\mathcal{A^{*}}}\mathbf{W}-\mathcal{C}_{\mathcal{A}^*}\mathbf{W}.
\]
Left-multiplying $\Pi_{\mathcal{A^{*}}}$ on both sides gives that $\mathcal{C}_{\mathcal{A}^*}\mathbf{W}=\mathbf{0}$, and thus $\mathbf{W}-\Pi_{\mathcal{A^{*}}}\mathbf{W}=\mathbf{0}$. Therefore $\mathbf{W}\in \mathcal{S}_{\mathcal{A}^{*}}:=\mathrm{col}(\mathbf{X}_{\mathcal{A}^{*}})
\;\cap\;
{\mathrm{null}}\!\left(
\mathbf{D}_{\mathcal{A}^{*}}
(\mathbf{X}_{\mathcal{A}^{*}}^{T}\mathbf{X}_{\mathcal{A}^{*}})^{+}
\mathbf{X}_{\mathcal{A}^{*}}^{T}
\right)$, meaning that $\mathcal{S}_{\mathcal{A}}\subseteq\mathcal{S}_{\mathcal{A}^{*}}$. By symmetry, $\mathcal{S}_{\mathcal{A}^{*}}\subseteq\mathcal{S}_{\mathcal{A}}$. We conclude that $\mathcal{S}_{\mathcal{A}}=\mathcal{S}_{\mathcal{A}^{*}}$. By \Cref{nested_idempotents}, this implies that
\begin{align*}
\operatorname{rank}\!\big(\Pi_{\mathcal A}-\mathcal{C}_{\mathcal{A}}\big)
=\dim(\mathcal{S}_{\mathcal{A}})
=\dim( \mathcal{S}_{\mathcal{A}^{*}} )
=\operatorname{rank}\!\big(\Pi_{\mathcal A^{*}}-\mathcal{C}_{\mathcal{A}^*}\big),
\end{align*}
that is $\mathrm{rank}(\nabla \mathbf{X}\hat{\beta})=\mathrm{rank}(\nabla \mathbf{X}\hat{\beta}^{*})$.
\end{proof}
\begin{remark}
As shown in the proof, $\mathrm{rank}(\nabla \mathbf{X}\hat{\beta})$ depends on $\hat{\beta}(\mathbf{Y})$ only through $\mathcal{A}(\mathbf{Y})$, hence invariance with respect to $\hat{\beta}(\mathbf{Y})$ follows directly from invariance with respect to $\mathcal{A}(\mathbf{Y})$.
\end{remark}

\subsubsection{Uniqueness of the Solution}

Rank invariance guarantees that the degrees of freedom are well defined even when the optimization problem admits multiple minimizers. We now strengthen this conclusion by giving a mild condition under which the synthetic control solution is unique.

\begin{assumption}[General position]
Let $\tilde{\mathbf{X}}=(\mathbf{X}^T,\mathbf{D}^T)^T\in\mathbb{R}^{(n+h)\times p}$ with columns $\{\tilde{\mathbf{X}}_1,...,\tilde{\mathbf{X}}_p\}$. We say that $\tilde{\mathbf{X}}$ is in \textit{general position} if for any integer $k<\min\{n+h,p\}$ and every collection of $k+1$ distinct indices $j_1,...,j_{k+1}\in\{1,...,p\}$, we have $\tilde{\mathbf{X}}_{j_{k+1}}\notin\mathrm{span}\{\tilde{\mathbf{X}}_{j_1},...,\tilde{\mathbf{X}}_{j_{k}} \}$.
\label{asm:general_position}
\end{assumption}

\begin{remark}
\Cref{asm:general_position} holds with probability one when $\tilde{\mathbf{X}}$ has a joint distribution that is absolutely continuous with respect to Lebesgue measure on $\mathbb{R}^{(n+h)\times p}$, see \citeA{tibshirani2013uniqueness} for an analogous argument in the lasso setting.
\label{remark1}
\end{remark}

\begin{lemma}
Consider the problem (\ref{a2_problem_1})-(\ref{a2_problem_2}). Let \Cref{asm:general_position} hold. There exists an optimal solution with active set $\mathcal{A}=\{j: \hat{\beta}_j(\mathbf{Y})>0\}$ such that $\mathrm{rank}(\tilde{\mathbf{X}}_{\mathcal{A}})=|\mathcal{A}|\leq n+h$. If we additionally assume that $\mathbf{Y}\notin \mathcal{H}:=\{ \mathbf{X}\beta: \mathbf{D}\beta=\mathbf{Z}, \beta\geq 0 \}$, then the synthetic control solution $\hat{\beta}(\mathbf{Y})$ is unique and $\mathrm{rank}(\tilde{\mathbf{X}}_{\mathcal{A}})=|\mathcal{A}|\leq n+h-1$.
\label{uniqueness_scm_without}
\end{lemma}
\begin{proof}[Proof of \Cref{uniqueness_scm_without}]

First, we prove the existence of an optimal solution with $|\mathcal{A}|\leq n+h$. For any feasible solution $\beta$, we have $\tilde{\mathbf{X}}\beta$ belongs to the convex cone generated by $\{\tilde{\mathbf{X}}_1,...,\tilde{\mathbf{X}}_p\}\subset \mathbb{R}^{n+h}$. 
By the Carath\'eodory Theorem (\citeNP{rockafellar1997convex}, Section 17), there exists an optimal solution $\hat{\beta}$ with the corresponding active set $\mathcal{A}$ such that $|\mathcal{A}|\leq n+h$. Under \Cref{asm:general_position}, any collection of at most $n+h$ columns of $\tilde{\mathbf{X}}$ is linearly independent: if a set of size $m\leq \min\{n+h,p\}$ were dependent, then some column would lie in the span of the other $m-1$ columns, contradicting general position with $k=m-1$. Hence $\mathrm{rank}(\tilde{\mathbf{X}}_{\mathcal{A}})=|\mathcal{A}|\leq n+h$, proving the first claim.

Second, we show that if $\mathbf{Y}\notin \mathcal{H}$, then $|\mathcal{A}|\leq n+h-1$. Let $\bm{\varepsilon}=\mathbf{Y}-\mathbf{X}\hat{\beta}\neq \mathbf{0}$ at the solution $\hat{\beta}$, and let $\xi$ be the Lagrangian multiplier for $\mathbf{D}\beta=\mathbf{Z}$. The KKT condition of the problem (\ref{a2_problem_1})-(\ref{a2_problem_2}) gives that
\begin{equation}
2\mathbf{X}_i^T{\bm\varepsilon}-\mathbf{D}_i^T\xi=0, \ \ \ \text{for all } i\in\mathcal{A},
\label{equ:foc_scm}
\end{equation}
where $\mathbf{X}_i$ denotes the $i$-th column of $\mathbf{X}$. Stack $\bm{\omega}:=(2{\bm\varepsilon},-\xi)\in\mathbb{R}^{n+h}$, then the KKT condition (\ref{equ:foc_scm}) is equivalent to ${\bm \omega}^T\tilde{\mathbf{X}}_i=0$ for all $i\in\mathcal{A}$. Since ${\bm\omega}\neq 0$, the set $\{\tilde{\mathbf{X}}_i: i\in\mathcal{A}\}$ lies in the linear subspace $\mathcal{W}=\{u\in\mathbb{R}^{n+h}: {\bm\omega}^{T}u=0\}$ which has dimension $n+h-1$. Hence $\mathrm{rank}(\tilde{\mathbf{X}}_{\mathcal{A}})\leq n+h-1$. From Step 1 and invoking the general position, $\mathrm{rank}(\tilde{\mathbf{X}}_{\mathcal{A}})=|
\mathcal{A}|\leq n+h-1$.

Finally, we show that $\hat{\beta}$ is unique. Suppose for contradiction there exist two distinct solutions $\hat{\beta}^{(1)}\neq \hat{\beta}^{(2)}$. For any $t\in(0,1)$, the convex combination $\hat{\beta}^{(t)}=t\hat{\beta}^{(1)}+(1-t)\hat{\beta}^{(2)}$ is feasible and, by the convexity of the objective function, also optimal. Let $\mathcal{S}:=\mathrm{supp}(\hat{\beta}^{(1)})\cup \mathrm{supp}(\hat{\beta}^{(2)})$. If $j\in\mathcal{S}$, then at least one of $\hat{\beta}_j^{(1)}$ and $\hat{\beta}_j^{(2)}$ is strictly positive, so $\hat{\beta}_j^{(t)}>0$. Conversely, if $j\notin \mathcal{S}$, then $\hat{\beta}_j^{(1)}=\hat{\beta}_j^{(2)}=0$, so $\hat{\beta}_j^{(t)}=0$. Hence, $\mathrm{supp}(\hat{\beta}^{(t)})=\mathcal{S}$. Since $\mathbf{Y}\notin\mathcal{H}$, Step 2 applies to $\hat{\beta}^{(t)}$, implying that $\mathrm{rank}(\tilde{\mathbf{X}}_{\mathcal{S}})=|\mathcal{S}|$.

Now we consider $\delta:=\hat{\beta}^{(1)}-\hat{\beta}^{(2)}\neq 0$. By feasibility of $\hat{\beta}^{(1)}$ and $\hat{\beta}^{(2)}$, we have that $\mathbf{D}\delta=\mathbf{0}$. Next, we claim that $\mathbf{X}\delta=0$. Define $g(s):=\|\mathbf{Y}-\mathbf{X}(\hat{\beta}^{(2)}+s\delta)\|_2^2$ for $s\in\mathbb{R}$. Since both $\hat{\beta}^{(1)}$ and $\hat{\beta}^{(2)}$ are optimal, $g(0)=g(1)$ equals the minimum value. 
Remark that, by expanding, $g(s)=\|\mathbf{Y}-\mathbf{X}\hat{\beta}^{(2)}\|_2^2-2s(\mathbf{Y}-\mathbf{X}\hat{\beta}^{(2)})^T\mathbf{X}\delta+s^2\|\mathbf{X}\delta\|_2^2$. If $\|\mathbf{X}\delta\|_2^2>0$, then $g$ is strictly convex and cannot have two distinct points be minimizers. Then it must be that $\|\mathbf{X}\delta\|_2^2=0$, i.e. $\mathbf{X}\delta=0$. Combining with $\mathbf{D}\delta=0$ yields that $\tilde{\mathbf{X}}\delta=0$. Since $\delta_j=0$ for all $j\notin\mathcal{S}$, we have that $\delta_{\mathcal{S}^c}=0$ and therefore $0=\tilde{\mathbf{X}}\delta=\tilde{\mathbf{X}}_{\mathcal{S}}\delta_{\mathcal{S}}$. Because  $\mathrm{rank}(\tilde{\mathbf{X}}_{\mathcal{S}})=|\mathcal{S}|$, it follows that $\delta_{\mathcal{S}}=0$. Hence $\delta=0$, contradicting $\hat{\beta}^{(1)}\neq \hat{\beta}^{(2)}$. We conclude that the solution $\hat{\beta}(\mathbf{Y})$ is unique.
\end{proof}

\subsection{Proofs of Main Results}
\label{subsectionA4}

We now combine the divergence formulas in \Cref{subsectionA1} with the regularity results of \Cref{subsectionA3} to derive our main results. The common argument is as follows. By \Cref{lemma2}, the active set $\mathcal{A}(\mathbf{Y})$ is locally stable for almost every $\mathbf{Y}$. On such a neighborhood, the synthetic control problem reduces to a least-squares problem with linear equality constraints. The divergence on this neighborhood therefore follows from the corresponding formula in \Cref{subsectionA1}. \Cref{lemma_rank_invariance} ensures that the divergence representation is well-defined.

As previously investigated (\citeNP{zou2007degrees}, \citeNP{tibshirani2012degrees}, \citeNP{meyer2000degrees}), by applying Stein's lemma, we may compute degrees of freedom. All that is left to check is continuity and almost differentiability. Lipschitz continuity suffices for these and is immediate from our locally stable closed-form representation of the fitted values. The degrees of freedom expression then obtains as the expectation of the trace of the divergence. For each synthetic control estimator, we provide the divergence for the regression fit, compute its trace, and take expectations to obtain the degrees of freedom.

\subsubsection{Unpenalized SCM without Covariates}

\begin{proposition}[Divergence of SCM without covariates]
Let $\hat{\beta}_{\mathrm{sc}}(\mathbf{Y})$ be a solution of the synthetic control problem (\ref{eq:SCwithoutCOVbeginning})-(\ref{SCwithoutCOVend}) and define $\mathcal{A}=\mathcal{A}\left(\mathbf{Y}\right):=\{ j : \hat{\beta}_j(\mathbf{Y}) > 0 \}$ as the active support corresponding to that solution. For almost every $\mathbf{Y}$, the fitted values vector $\hat{\mathbf{Y}}_{\mathrm{sc}}=\mathbf{X}\hat{\beta}_{\mathrm{sc}}$ has the divergence
\begin{align}
\label{eq:traceSCMnoCov}
\nabla\hat{\mathbf{Y}}_{\mathrm{sc}}=\left( \mathbf{X}_{\mathcal{A}}\mathbf{M}_{\mathcal{A}} \right)\left(\mathbf{X}_{\mathcal{A}}\mathbf{M}_{\mathcal{A}}\right)^{+}, \quad \text{where} \
\mathbf{M}_{\mathcal{A}}=\mathbf{I}_{|\mathcal{A}|}-\frac{1}{|\mathcal{A}|}\mathbf{1}_{|\mathcal{A}|}\mathbf{1}_{|\mathcal{A}|}^T.
\end{align}
\label{prop:div_scm_without}
\end{proposition}
\vspace{-3em}

\begin{proof}[Proof of \Cref{prop:div_scm_without}]
By \Cref{lemma2}, the active set $\mathcal{A}(\mathbf{Y})$ is locally stable for almost every $\mathbf{Y}$. On a neighborhood of such a $\mathbf{Y}$, the problem (\ref{eq:SCwithoutCOVbeginning})-(\ref{SCwithoutCOVend}) reduces to a constrained least-squares problem on the fixed active set $\mathcal{A}$. Applying \Cref{proposition:div_unpenalized_constrained_LS} with design matrix $\mathbf{X}_{\mathcal{A}}$ and constraint matrix $\mathbf{D}=\mathbf{1}_{|\mathcal{A}|}^T$ yields the desired divergence matrix.
\end{proof}

\paragraph{Proof of \Cref{proposition4}.} Under (\ref{GaussianAssumption}) and (\ref{eq:Ahomoskedastic}), Stein's Lemma gives
$\mathrm{df}(\hat{\mathbf{Y}}_{\mathrm{sc}})
  =\operatorname{E}[\mathrm{Tr}(\nabla\hat{\mathbf{Y}}_{\mathrm{sc}}) \ | \ \mathbf{X}]$.
The trace of~(\ref{eq:traceSCMnoCov}) equals
$\mathrm{rank}(\tilde{\mathbf{X}}_{\mathcal{A}})-1$ by
\Cref{proposition:div_unpenalized_constrained_LS}, delivering the desired degrees of freedom formula. \Cref{lemma_rank_invariance} ensures that this expression is well-defined for almost every $\mathbf{Y}$. \qed

\begin{corollary}[Uniqueness of SCM without covariates]
Consider the synthetic control problem (\ref{eq:SCwithoutCOVbeginning})-(\ref{SCwithoutCOVend}). Let $\tilde{\mathbf{X}}=(\mathbf{X}^\top,\mathbf{1}_p)^\top\in\mathbb{R}^{(n+1)\times p}$ satisfy \Cref{asm:general_position}.
Then, for any $\mathbf{Y}\in\mathbb{R}^n$, there exists an optimal solution with active set
$\mathcal{A}=\{j:\hat\beta_j(\mathbf{Y})>0\}$ such that $\mathrm{rank}(\tilde{\mathbf{X}}_{\mathcal{A}})=|\mathcal{A}|\le n+1.$ Moreover, if $\mathbf{Y}\notin \mathrm{conv}\{\mathbf{X}_1,\ldots,\mathbf{X}_p\}$, then the synthetic control solution $\hat\beta(\mathbf{Y})$ is unique and $|\mathcal{A}|\leq n$.
\label{corollary:uniquness_SCM}
\end{corollary}
\begin{remark}
\Cref{corollary:uniquness_SCM} formalizes the geometric discussion of sparsity and uniqueness in \citeA{abadie2021using}. There, the two cases are treated separately and informally: when $\mathbf{Y}$ lies outside the convex hull and the donor columns satisfy a general position condition, the synthetic control solution is declared unique and sparse with at most $n$ nonzero weights; when $\mathbf{Y}$ lies inside the convex hull, uniqueness may fail, though sparse solutions with at most $n+1$ nonzero weights are noted to exist. \Cref{corollary:uniquness_SCM} shows that, under \Cref{asm:general_position}, every outcome vector $\mathbf{Y}$ admits an optimal solution supported on at most $n+1$ donors whose augmented columns are affinely independent --regardless of whether $\mathbf{Y}$ falls inside or outside the convex hull. When $\mathbf{Y}\notin \mathrm{conv}\{\mathbf{X}_1,\ldots,\mathbf{X}_p\}$ --the typical setting in applications, as \citeA{abadie2021using} notes-- \Cref{corollary:uniquness_SCM} also shows that the general position condition further guarantees that the synthetic control solution $\hat{\beta}(\mathbf{Y})$ is unique.
\end{remark}

\paragraph{Proof of \Cref{corollary:dof}.} It follows immediately from \Cref{proposition4} and \Cref{corollary:uniquness_SCM}. \qed

\begin{remark}[The smallest active set]
Assume the conditions of \Cref{proposition4} hold and $\mathbf{X}$ has a distribution that is absolutely continuous with respect to Lebesgue measure on $\mathbb{R}^{n\times p}$. Then with probability one over $\mathbf{X}$, we have $\mathrm{df}(\mathbf{X}\hat{\beta}_{\mathrm{sc}})=E_{Y|X}\left|\mathcal{A}^{*}(\mathbf{Y})\right|-1$, where $|\mathcal{A}^{*}(\mathbf{Y})|$ is the \textit{smallest} cardinality among all active sets of synthetic control solutions at $\mathbf{Y}$. This can be obtained from \Cref{corollary:uniquness_SCM} and \Cref{lemma_rank_invariance}. Analogous results for the lasso, stated in terms of the smallest active set, can be found in \citeA{dossal2013degrees} and \citeA[Page 1214]{tibshirani2012degrees}.
\end{remark}

\subsubsection{SCM with Covariates}

Define the augmented matrix $\tilde{\mathbf{D}}=(\mathbf{D}^T,\mathbf{1}_p)^T$ and
let $\mathbb{S}=\{ \beta: \mathbf{1}_p^T\beta=1, \beta\geq 0 \}$ and $\mathcal{H}(\mathbf{D})=\{\mathbf{D}\beta: \beta\in\mathbb{S}\}$.

When $\mathbf{Z}\in\mathcal{H}(\mathbf{D})$, there exists $\bar{\beta}\in\mathbb{S}$ with $\mathbf{D}\bar{\beta}=\mathbf{Z}$, so the minimum value of the inner problem is 0. Since $V\succ 0$ , $\|\mathbf{Z}-\mathbf{D}\beta\|_V^2=0$ holds if and only if $\mathbf{D}\beta=\mathbf{Z}$. Therefore $\arg\min_{\beta'\in\mathbb{S} } \ \left\Vert \mathbf{Z}-\mathbf{D}\beta'\right\Vert_{V}^2=\{\beta'\in\mathbb{S}: \mathbf{D}\beta'=\mathbf{Z} \}.$ In this case, the two-level synthetic control problem reduces to the single constrained least-squares problem
\begin{equation*}
\min_{\beta} \ \left\Vert \mathbf{Y}-\mathbf{X}\beta\right\Vert _{2}^{2}
\end{equation*}
subject to
\begin{equation*}
\mathbf{D}\beta=\mathbf{Z}, \quad \mathbf{1}^T\beta=1, \quad \beta\geq 0.
\end{equation*}
We give the divergence expression of this optimization problem in \Cref{prop:div_scm_with}.

\begin{proposition}[Divergence of SCM with covariates]
Fix $\mathbf{Z}\in\mathbb{R}^{n_{\mathrm{cov}}}$. Let $(\mathbf{X},\mathbf{D})$ be a random design matrix, where
$\mathbf{D}\in\mathbb{R}^{n_{\mathrm{cov}}\times p}$ has an absolutely continuous distribution.
Work conditional on $(\mathbf{X},\mathbf{D})$, and assume that the realized $\mathbf{D}$ has the full row rank
and $\mathbf{Z}\in \mathrm{relint}\,\mathcal{H}(\mathbf{D})$. Let $\hat{\beta}(\mathbf{Y})$ be a solution of the synthetic control problem (\ref{eq:SCwithCOVbeginning})-(\ref{eq:SCwithCOVend}), and define $\mathcal{A}=\mathcal{A}\left(\mathbf{Y}\right):=\{ j : \hat{\beta}_j(\mathbf{Y}) > 0 \}$ as the active support corresponding to that solution. Then for almost every $\mathbf{Y}$, the fitted values vector $\hat{\mathbf{Y}}=\mathbf{X}\hat{\beta}$ has the divergence
\[
\nabla\hat{\mathbf{Y}}=\left( \mathbf{X}_{\mathcal{A}}\mathbf{M}_{\tilde{D}_{\mathcal{A}}} \right)\left(\mathbf{X}_{\mathcal{A}}\mathbf{M}_{\tilde{D}_{\mathcal{A}}}\right)^{+}, \quad
\mathbf{M}_{\tilde{D}_{\mathcal{A}}}=\mathbf{I}_{|\mathcal{A}|}-\tilde{\mathbf{D}}_{\mathcal{A}}^T\left(\tilde{\mathbf{D}}_{\mathcal{A}}\tilde{\mathbf{D}}_{\mathcal{A}}^T\right)^{+}\tilde{\mathbf{D}}_{\mathcal{A}}
\]
where $\tilde{\mathbf{D}}_{\mathcal{A}}=(\mathbf{D}_{\mathcal{A}}^T, \mathbf{1}_{|\mathcal{A}|})^T\in \mathbb{R}^{(n_{\mathrm{cov}}+1)\times |\mathcal{A}|}$.
\label{prop:div_scm_with}
\end{proposition}

\paragraph{Proofs of \Cref{proposition7} and \Cref{cor:card_with_cov}.}
Taking the conditional expectation of the trace of the divergence expression in Proposition \ref{prop:div_scm_with} yields \Cref{proposition7}. Combining this result with \Cref{uniqueness_scm_without} immediately establishes \Cref{cor:card_with_cov}. \qed

\paragraph{Proof of \Cref{prop:df_SCM_cov_notin}.} When $\mathbf{Z}\notin\mathcal{H}(\mathbf{D})$, since $V\succ 0$, the inner problem is strictly convex, its regression fit $\mathbf{D}\hat{\beta}_{\mathrm{in}}$ is uniquely determined as the $V$-matrix projection of $\mathbf{Z}$ onto the convex set $\mathcal{H}(\mathbf{D})$. Under the general position assumption, by analogous arguments in \Cref{uniqueness_scm_without}, the inner minimizer $\hat{\beta}_{\mathrm{in}}$ is unique. Consequently, the regression fit in the outer problem $\hat{\mathbf{Y}}(\mathbf{Y})$ is constant in $\mathbf{Y}$, and therefore the degrees of freedom is zero. \qed

\subsubsection{Penalized SCM}

Claims in the main text assume uniform continuity with respect to Lebesgue measure, which is a lower-level sufficient condition for a general position assumption that is needed in the proofs.  We give the latter assumption below and produce the arguments in terms of that assumption for greater clarity and generality.

\begin{assumption}[\citeNP{abadie2021penalized}]
Define the augmented matrix $\tilde{\mathbf{X}}\in\mathbb{R}^{p\times (n+2)}$ whose $j$th row is $(\mathbf{X}_j^T, 1, \|\mathbf{Y}-\mathbf{X}_j\|_2^2)$. Assume that every submatrix obtained by selecting any subset of rows of $\tilde{\mathbf{X}}$ with at most $n+2$ rows has full row rank.
\label{assumption_abadie_uniqueness}
\end{assumption}
Under \Cref{assumption_abadie_uniqueness}, Theorem 1 in \citeA{abadie2021penalized} shows that for $\lambda>0$, the penalized SC problem (\ref{eq:PenalizedCSbeginning})-(\ref{eq:PenalizedSCend}) has a unique solution, and the corresponding active set satisfies $|\mathcal{A}|\leq n+1$.

\begin{proposition}[Divergence of penalized SCM]
Suppose \Cref{assumption_abadie_uniqueness} holds. 
Let $\hat{\beta}_{\mathrm{psc}}(\mathbf{Y})$ be the solution of (\ref{eq:PenalizedCSbeginning})-(\ref{eq:PenalizedSCend}) and let $\mathcal{A}(\mathbf{Y})=\{j:\hat{\beta}_{\mathrm{psc},j}>0\}$. 
Then, with probability one over $\mathbf{X}$, for $\mathbf{Y}|\mathbf{X}$-almost every $\mathbf{Y}$, the fitted value of the penalized synthetic control method (\ref{eq:PenalizedCSbeginning})-(\ref{eq:PenalizedSCend}) has divergence
\begin{equation*}
\nabla\mathbf{X}\hat{\beta}_{\mathrm{psc}}=(1+\lambda)\left[\Pi_{\mathcal{A}}-\cfrac{1}{\mathbf{1}_{|\mathcal{A}|}^T(\mathbf{X}_{\mathcal{A}}^T\mathbf{X}_{\mathcal{A}})^{+}\mathbf{1}_{|\mathcal{A}|} }\mathbf{X}_{\mathcal{A}}(\mathbf{X}_{\mathcal{A}}^T\mathbf{X}_{\mathcal{A}})^{+}\mathbf{1}_{|\mathcal{A}|}\mathbf{1}_{|\mathcal{A}|}^T(\mathbf{X}_{\mathcal{A}}^T\mathbf{X}_{\mathcal{A}})^{+}\mathbf{X}_{\mathcal{A}}^T\right].
\end{equation*}
\label{prop:div_pscm}
\end{proposition}
\vspace{-2.5em}

\paragraph{Proof of \Cref{proposition5}.} It follows by taking the conditional expectation of the trace of the divergence expression in \Cref{prop:div_pscm}. \qed

\subsubsection{Constrained Ridge SCM}

\begin{proposition}[Divergence of Constrained Ridge SCM]
Fix $\lambda> 0$, for almost every $\mathbf{Y}\in\mathbb{R}^n$, the fitted values vector $\hat{\mathbf{Y}}_{\mathrm{crsc}}=\mathbf{X}\hat{\beta}_{\mathrm{crsc}}+\hat{\beta}_{0,\mathrm{crsc}}\mathbf{1}_n$ of the constrained ridge synthetic control problem (\ref{eq:CR objective})-(\ref{eq:CR constraint})  has the divergence
\begin{align*}
\nabla\hat{\mathbf{Y}}_{\mathrm{crsc}}=&\tilde{\mathbf{X}}_{\mathcal{A}}\mathbf{Q}^{-1}\tilde{\mathbf{X}}_{\mathcal{A}}^T
+\frac{1}{n}\mathbf{1}_n\mathbf{1}_n^T-\tilde{\mathbf{X}}_{\mathcal{A}}\mathbf{Q}^{-1}\mathbf{1}_{|\mathcal{A}|}\left( \mathbf{1}_{|\mathcal{A}|}^T\mathbf{Q}^{-1}\mathbf{1}_{|\mathcal{A}|} \right)^{-1}\mathbf{1}_{|\mathcal{A}|}^T\mathbf{Q}^{-1}\tilde{\mathbf{X}}_{\mathcal{A}}^T,
\end{align*}
where $\mathcal{A}$ corresponds to a solution to (\ref{eq:CR objective})-(\ref{eq:CR constraint}), $\tilde{\mathbf{X}}_{\mathcal{A}}=(\mathbf{I}_n-\frac{1}{n}\mathbf{1}_n\mathbf{1}_n^T)\mathbf{X}_{\mathcal{A}}$, $\mathbf{Q}=\tilde{\mathbf{X}}_{\mathcal{A}}^T\tilde{\mathbf{X}}_{\mathcal{A}}+\lambda\mathbf{I}_{|\mathcal{A}|}$.
\label{prop:div_ridge_scm}
\end{proposition}

\paragraph{Proof of \Cref{prodfcrsc}.}
Because $\lambda>0$, the objective (\ref{eq:CR objective}) is strongly convex in $(\beta_0,\beta)$. With linear constraint (\ref{eq:CR constraint}), this implies the optimizer $(\hat{\beta}_{\mathrm{crsc}}(\mathbf{Y}), \hat{\beta}_{0,\mathrm{crsc}}(\mathbf{Y}))$ is unique for each $\mathbf{Y}$.

Recall that the singular value decomposition of $\tilde{\mathbf{X}}_{\mathcal{A}}$ is $\tilde{\mathbf{X}}_{\mathcal{A}} = \mathbf{U}\mathbf{S}\mathbf{V}^T$, where $\mathbf{U}\in\mathbb{R}^{n\times|\mathcal{A}|}$, $\mathbf{V}\in\mathbb{R}^{|\mathcal{A}|\times|\mathcal{A}|}$ are orthogonal matrices and $\mathbf{S}=\mathrm{diag}(s_1,\ldots,s_{|\mathcal{A}|})$ with $s_1\geq\cdots\geq s_{|\mathcal{A}|}\geq 0$. Then $\mathbf{Q}=\mathbf{V}(\mathbf{S}^2+\lambda\mathbf{I}_{|\mathcal{A}|})\mathbf{V}^T$. By Stein's Lemma, the degrees of freedom of the constrained ridge SCM are given by taking the conditional expectation of the trace  of the divergence expression in \Cref{prop:div_ridge_scm}:
\begin{equation}
\begin{aligned}
&\mathrm{df}\left(\hat{\mathbf{Y}}_{\mathrm{crsc}}\right)
=\operatorname{E}_{Y|X}\left[\mathrm{Tr}\left(\nabla\hat{\mathbf{Y}}_{\mathrm{crsc}}\right)\right] \\[7pt]
=&\operatorname{E}_{Y|X}\left[\mathrm{Tr}\left( \tilde{\mathbf{X}}_{\mathcal{A}}\mathbf{Q}^{-1}\tilde{\mathbf{X}}_{\mathcal{A}}^T\right)\right]
+1-\operatorname{E}_{Y|X}\left[\mathrm{Tr}\left(\tilde{\mathbf{X}}_{\mathcal{A}}\mathbf{Q}^{-1}\mathbf{1}_{|\mathcal{A}|}\left( \mathbf{1}_{|\mathcal{A}|}^T\mathbf{Q}^{-1}\mathbf{1}_{|\mathcal{A}|} \right)^{-1}\mathbf{1}_{|\mathcal{A}|}^T\mathbf{Q}^{-1}\tilde{\mathbf{X}}_{\mathcal{A}}^T \right)\right].
\end{aligned}
\label{equ:df_cr_scm}
\end{equation}
For the first term in \Cref{equ:df_cr_scm}, we plug in the singular value decomposition of $\tilde{\mathbf{X}}_{\mathcal{A}} = \mathbf{U}\mathbf{S}\mathbf{V}^T$ and obtain that
\allowdisplaybreaks
\begin{align*}
\mathrm{Tr}\left( \tilde{\mathbf{X}}_{\mathcal{A}}\mathbf{Q}^{-1}\tilde{\mathbf{X}}_{\mathcal{A}}^T\right)&=
\mathrm{Tr}\left[\mathbf{U}\mathbf{S}\mathbf{V}^T\Big{(} \mathbf{V}(\mathbf{S}^2+\lambda\mathbf{I}_{|\mathcal{A}|})\mathbf{V}^T \Big{)}^{-1}\mathbf{V}\mathbf{S}\mathbf{U}^T \right]\\
&=\mathrm{Tr}\left[\mathbf{S}\Big{(} \mathbf{S}^2+\lambda\mathbf{I}_{|\mathcal{A}|} \Big{)}^{-1}\mathbf{S} \right]=\sum_{i=1}^{|\mathcal{A}|}
\frac{s_i^2}{s_i^2+\lambda}.
\end{align*}
To compute the last term in \Cref{equ:df_cr_scm}, we proceed similarly:
\allowdisplaybreaks
\begin{align*}
&\ \mathrm{Tr}\left(\tilde{\mathbf{X}}_{\mathcal{A}}\mathbf{Q}^{-1}\mathbf{1}_{|\mathcal{A}|}\left( \mathbf{1}_{|\mathcal{A}|}^T\mathbf{Q}^{-1}\mathbf{1}_{|\mathcal{A}|} \right)^{-1}\mathbf{1}_{|\mathcal{A}|}^T\mathbf{Q}^{-1}\tilde{\mathbf{X}}_{\mathcal{A}}^T \right)\\[7pt]
=&\ \cfrac{\mathrm{Tr}\left( (\tilde{\mathbf{X}}_{\mathcal{A}}^T\tilde{\mathbf{X}}_{\mathcal{A}}+\lambda\mathbf{I}_{|\mathcal{A}|})^{-1}\mathbf{1}_{|\mathcal{A}|}\mathbf{1}_{|\mathcal{A}|}^T(\tilde{\mathbf{X}}_{\mathcal{A}}^T\tilde{\mathbf{X}}_{\mathcal{A}}+\lambda\mathbf{I}_{|\mathcal{A}|})^{-1}(\tilde{\mathbf{X}}_{\mathcal{A}}^T \tilde{\mathbf{X}}_{\mathcal{A}}+\lambda\mathbf{I}_{|\mathcal{A}|}-\lambda\mathbf{I}_{|\mathcal{A}|})\right)}{\mathbf{1}_{|\mathcal{A}|}^T(\tilde{\mathbf{X}}_{\mathcal{A}}^T\tilde{\mathbf{X}}_{\mathcal{A}}+\lambda\mathbf{I}_{|\mathcal{A}|})^{-1}\mathbf{1}_{|\mathcal{A}|}}\\[7pt]
=&\ 1-\lambda\cfrac{ \mathbf{1}_{|\mathcal{A}|}^T(\tilde{\mathbf{X}}_{\mathcal{A}}^T\tilde{\mathbf{X}}_{\mathcal{A}}+\lambda\mathbf{I}_{|\mathcal{A}|})^{-2}\mathbf{1}_{|\mathcal{A}|}}{\mathbf{1}_{|\mathcal{A}|}^T(\tilde{\mathbf{X}}_{\mathcal{A}}^T\tilde{\mathbf{X}}_{\mathcal{A}}+\lambda\mathbf{I}_{|\mathcal{A}|})^{-1}\mathbf{1}_{|\mathcal{A}|}}
=1-\lambda
\cfrac{\mathbf{1}_{|\mathcal{A}|}^T\left(  \mathbf{V}\left( \mathbf{S}^2+\lambda\mathbf{I}_{|\mathcal{A}|} \right)^2\mathbf{V}^T \right)^{-1}\mathbf{1}_{|\mathcal{A}|}}{\mathbf{1}_{|\mathcal{A}|}^T\left(  \mathbf{V}\left( \mathbf{S}^2+\lambda\mathbf{I}_{|\mathcal{A}|} \right)\mathbf{V}^T \right)^{-1}\mathbf{1}_{|\mathcal{A}|}}.
\end{align*}
Then, the degrees of freedom formula (\ref{dfcrsc}) is obtained, as desired. \qed

\subsubsection{Elastic Net SCM}

\begin{proposition}[Divergence of Elastic Net SCM]
For almost every $\mathbf{Y}$, the fitted values vector $\hat{\mathbf{Y}}_{\mathrm{elast}}=\mathbf{X}\hat{\beta}_{\mathrm{elast}}+\hat{\beta}_{0,\mathrm{elast}}\mathbf{1}_n$ of the elastic net synthetic control (\ref{eq:EN}) has the divergence
\begin{align*}
\nabla\hat{\mathbf{Y}}_{\mathrm{elast}}=&\frac{1}{n} \mathbf{1}_n \mathbf{1}_n^T+\tilde{\mathbf{X}}_{\mathcal{A}}\left(\tilde{\mathbf{X}}_{\mathcal{A}}^{T}\tilde{\mathbf{X}}_{\mathcal{A}}+\lambda_{2}\mathbf{I}_{|\mathcal{A}|}\right)^{-1}\tilde{\mathbf{X}}_{\mathcal{A}}^{T},
\end{align*}
where $\mathcal{A}$ corresponds to a solution to (\ref{eq:EN}), $\tilde{\mathbf{X}}_{\mathcal{A}}=(\mathbf{I}_n-\frac{1}{n}\mathbf{1}_n\mathbf{1}_n^T)\mathbf{X}_{\mathcal{A}}$.
\label{proposition_divergence_elast}
\end{proposition}

\begin{proof}[Proof of \Cref{proposition_divergence_elast}]
By Lemma 1 of \citeA{zou2005regularization}, the elastic net synthetic control problem (\ref{eq:EN}) is equivalent to the following lasso problem:
\begin{equation*}
\min_{\beta_0, \beta} \ \
\left\Vert \left(
\begin{matrix}
\mathbf{Y} \\
\mathbf{0}_p
\end{matrix}
\right) - \left(
\begin{matrix}
\mathbf{1}_n & \mathbf{X} \\
\mathbf{0}_p & \sqrt{\lambda_2}\mathbf{I}_p
\end{matrix}
\right)
\begin{pmatrix}
\beta_0 \\
\beta
\end{pmatrix} \right\Vert_2^2
+ \lambda_1 \|\beta\|_1.
\end{equation*}
By the divergence expression of the lasso problem in terms of the active set (\citeNP{tibshirani2012degrees}), we have that
\begin{align*}
\nabla \hat{\mathbf{Y}}_{\mathrm{elast}}
&=
\begin{pmatrix}
\mathbf{1}_n & \mathbf{X}_{\mathcal{A}}
\end{pmatrix}
\begin{pmatrix}
n & \mathbf{1}_n^T \mathbf{X}_{\mathcal{A}} \\
\mathbf{X}_{\mathcal{A}}^T \mathbf{1}_n & \mathbf{X}_{\mathcal{A}}^T \mathbf{X}_{\mathcal{A}} + \lambda_2 \mathbf{I}_{|\mathcal{A}|}
\end{pmatrix}^{-1}
\begin{pmatrix}
\mathbf{1}_n^T  \\
\mathbf{X}_{\mathcal{A}}^T
\end{pmatrix}\\[7pt]
&=
\frac{1}{n} \mathbf{1}_n \mathbf{1}_n^T
+
\tilde{\mathbf{X}}_{\mathcal{A}}
\left( \tilde{\mathbf{X}}_{\mathcal{A}}^T \tilde{\mathbf{X}}_{\mathcal{A}} + \lambda_2 \mathbf{I}_{|\mathcal{A}|} \right)^{-1}
\tilde{\mathbf{X}}_{\mathcal{A}}^T
\end{align*}
as claimed.
\end{proof}

\paragraph{Proof of \Cref{propescm}.} It follows by taking the conditional expectation of the trace of the divergence expression in \Cref{proposition_divergence_elast} and applying the same SVD simplification used in the proof of \Cref{prodfcrsc}. \qed

\subsubsection{Connection to \protect\citeA{chen2020degrees}}



Our proof strategy works with the active donor set, establishes its local stability, and differentiates the closed-form fitted values on the fixed active set. Our approach produces a closed-form divergence matrix, which is the key input for the heteroskedasticity-robust information criteria. A different method, based on the general result of \citeA{chen2020degrees}, treats the regression fit as the projection of $\mathbf{Y}$ onto a polyhedron and expresses the divergence through the \textit{active constraints} at the solution. Their Theorem~3.2 encompasses the case of least squares problems with linear equality constraints. We now give an alternative derivation of Proposition~\ref{proposition4} by verifying the conditions of Theorem~3.2 of \citeA{chen2020degrees}, confirming that the two approaches yield the same degrees of freedom under conditional homoskedasticity.


Introduce $\theta = \mathbf{X}\beta \in \mathbb{R}^n$ and rewrite
(\ref{eq:SCwithoutCOVbeginning})--(\ref{SCwithoutCOVend}) as
\[
\underset{\theta,\beta}{\min} \ \left\Vert \theta-\mathbf{Y}\right\Vert _{2}^{2} \quad \text{subject to} \quad A\beta+B\theta\le\mathbf{c},
\]
where
\[
A=\left(\begin{array}{c}
\mathbf{X}\\[3pt]
-\mathbf{X}\\[3pt]
\mathbf{1}^T\\[3pt]
-\mathbf{1}^T \\[3pt]
-I_{p\times p}
\end{array}\right),\quad B=\left(\begin{array}{c}
I_{n\times n}\\[3pt]
-I_{n\times n}\\[3pt]
\mathbf{0}_{n}^T\\[3pt]
\mathbf{0}_{n}^T \\[3pt]
\mathbf{0}_{p\times n}
\end{array}\right),\quad \mathbf{c}=\left(\begin{array}{c}
\mathbf{0}_{n}\\[3pt]
\mathbf{0}_{n}\\[3pt]
1\\[3pt]
-1 \\[3pt]
\mathbf{0}_p
\end{array}\right).\\[7pt]
\]
The first $2n$ rows encode $\mathbf{X}\beta = \theta$, the next two encode $\mathbf{1}_p^T\beta = 1$, and the last $p$ encode $\beta \ge 0$.

Let $(\hat{\theta}(\mathbf{Y}),\hat{\beta}(\mathbf{Y}))$ be any optimal pair. Define $\mathcal{J}(\mathbf{Y})=\{i: \langle A_i, \hat{\beta}(\mathbf{Y}) \rangle+\langle B_i,\hat{\theta}(\mathbf{Y})\rangle=\mathbf{c}_i \}$, and let $\mathcal{I}(\mathbf{Y})\subseteq \mathcal{J}(\mathbf{Y})$ be the index set of maximal independent rows of the matrix $[A_{\mathcal{J}}, B_{\mathcal{J}}]$. Their Theorem 3.2 gives that for a.e. $\mathbf{Y}$, $\mathrm{df}(\mathbf{X}\hat{\beta}_{\mathrm{sc}})=n-|\mathcal{I}(\mathbf{Y})|+\mathrm{rank}(A_{\mathcal{I}(\mathbf{Y})})$.

We now connect their result with our degrees of freedom with active set representation. Define that $\mathcal{A}(\mathbf{Y})=\{j: \hat{\beta}_j(\mathbf{Y})>0\}$ and $\mathcal{A}^{\perp}(\mathbf{Y})=\{j: \hat{\beta}_j(\mathbf{Y})=0\}$. Write $e_i$ for the $i$-th standard basis vector. Choose $\mathcal{I}(\mathbf{Y})$ to contain: (i) the $n$ rows from $\mathbf{X}\beta - \theta \le 0$,
(ii) the single row $\mathbf{1}_p^T\beta \le 1$,
and (iii) all $|\mathcal{A}^\perp|$ rows $-\beta_j \le 0$ for $j \in \mathcal{A}^\perp$.

We verify that these $n + 1 + |\mathcal{A}^\perp|$ rows of $[A,\,B]$
are linearly independent. The $i$-th row in (i) has $B$-component
$-e_i^T$, so these $n$ rows in (i) are independent. The rows in (ii)-(iii) have $B$-component zero
and are therefore independent of (i). Within (ii)-(iii), their $A$-components are $\mathbf{1}_p^T$ and
$\{-e_j: j \in \mathcal{A}^\perp\}$;
the latter are independent, and
$\mathbf{1}_p \notin \mathrm{span}\{e_j: j \in \mathcal{A}^\perp\}$
because $\mathcal{A} \neq \varnothing$. Hence $\mathcal{I}(\mathbf{Y})$ is a valid maximal independent subset.

With this choice, $|\mathcal{I}(\mathbf{Y})|=n+1+|\mathcal{A}^{\perp}(\mathbf{Y})|$. Moreover, $A_{\mathcal{I}}=( \mathbf{X}^T, \mathbf{1}_p, -E_{\mathcal{A}^{\perp}}^T )^T$ where $E_{\mathcal{A}^\perp} \in \mathbb{R}^{|\mathcal{A}^\perp| \times p}$
selects rows indexed by $\mathcal{A}^\perp$ from $I_p$. By eliminating the $\mathcal{A}^{\perp}$-columns using the $-E_{\mathcal{A}^{\perp}}$ block, one obtains that $\mathrm{rank}(A_{\mathcal{I}(\mathbf{Y})})=|\mathcal{A}^{\perp}(\mathbf{Y})|+\mathrm{rank}(\tilde{\mathbf{X}}_{\mathcal{A}})$ where $\tilde{\mathbf{X}}=(\mathbf{X}^T, \mathbf{1})^T$. Finally we have that $\mathrm{div}(\mathbf{X}\hat{\beta}_{\mathrm{sc}})=\mathrm{rank}(\tilde{\mathbf{X}}_{\mathcal{A}})-1$ for a.e. $\mathbf{Y}$. Under the conditional homoskedastic Gaussian model for $\mathbf{Y}|\mathbf{X}$, it follows that $\mathrm{df}(\mathbf{X}\hat{\beta}_{\mathrm{sc}})=E[\mathrm{rank}(\tilde{\mathbf{X}}_{\mathcal{A}}) | \mathbf{X}]-1$.

\subsection{Consistency of degrees of freedom estimates}
\label{subsectionA5}

In this subsection, we show that the unbiased degrees of freedom estimate for the synthetic control estimator without covariates is consistent. The result and argument emulate that of \citeA{zou2007degrees}.

Let $\{Y_i,X_i\}_{i=1}^n$ be i.i.d. draws from a probability law $\mathcal{F}$.  Collect the entries in the n-tuple $\mathbf{Y}$ and $n\times p$ matrix $\mathbf{X}$, and define the $n$-tuple of residuals $\varepsilon = \mathbf{Y} -E[\mathbf{Y}|\mathbf{X}]$. We distinguish between the conditional expectation $E[\mathbf{Y}|\mathbf{X}]$, the best linear predictor $E^*[\mathbf{Y}|\mathbf{X}] = \mathbf{X}\beta^*$, and the simplex constrained best linear predictor $E^*_{\mathbb{S}^{p-1}}[\mathbf{Y}|\mathbf{X}] = \mathbf{X}\hat{\beta}^*$, where $\mathbb{S}^{p-1}$ is the probability simplex (\ref{SCwithoutCOVend}).
We refer to $\hat{\beta}$ as the synthetic control coefficient solving (\ref{eq:SCwithoutCOVbeginning})-(\ref{SCwithoutCOVend}), denote its active set by $\hat{\mathcal{B}}_n = \{j: \hat{\beta}_j > 0\}$  and its population analog by $\mathcal{B}^* = \{j: \hat{\beta}_j^* > 0\}$.

The Lagrangian for the sample problem (\ref{eq:SCwithoutCOVbeginning})-(\ref{SCwithoutCOVend}) is given by $\mathcal{L}_n(\beta;\lambda,\mu)=n^{-1}\|\mathbf{Y}-\mathbf{X}\beta\|_2^2+\lambda(\textbf{1}_p^T\beta-1)-\mu^T\beta$ with $\lambda\in\mathbb{R}$ and $\mu\in\mathbb{R}^p_{+}$. The population counterpart is $\mathcal{L}_0(\beta;\lambda,\mu)=E(Y_i-X_i'\beta)^2+\lambda(\textbf{1}_p^T\beta-1)-\mu^T\beta$. Let $(\hat{\beta}, \hat{\lambda},\hat{\mu})$ denote the saddle point of the sample Lagrangian, and $(\hat{\beta}^{*}, \hat{\lambda}^{*},\hat{\mu}^{*})$ the saddle point of the population Lagrangian.

\begin{assumption}
We assume that
\vspace{-1em}
\begin{itemize}
    \item[(i)] The probability simplex constraint is non-binding. Specifically, $\mathbf{X}\hat{\beta}^* \equiv E^*_{\mathbb{S}^{p-1}}[\mathbf{Y}|\mathbf{X}] = E^*[\mathbf{Y}|\mathbf{X}] \equiv \mathbf{X}\beta^*$.
    \item[(ii)] $\frac{1}{n}\mathbf{X}^T\mathbf{X} \rightarrow C$ as $n \to \infty$, and $C$ is nonsingular.
    \item[(iii)] $E[\|X_i\|^2]<\infty$ and $E[Y_i^2]<\infty$.
    \item[(iv)] The optimal solution is a nontransition point. Specifically, if $\hat{\beta}^*_j = 0$ then $\hat{\mu}^*_j > 0$, and if $\hat{\mu}^*_j = 0$ then $\hat{\beta}^*_j > 0$.
    \item[(v)] The errors $\varepsilon$ are normally distributed given $\mathbf{X}$, that is, $\varepsilon_i \ | \ \mathbf{X}\iid N(0,\sigma^2)$.
\end{itemize}
\label{assumptiona1}
\end{assumption}
\vspace{-1em}

\Cref{assumptiona1} (i) delivers asymptotic orthogonality of the fitted residuals with respect to the columns of $\mathbf{X}$. Since $\hat{\beta}^*$ typically exhibits sparsity, this assumption can be interpreted as a sparsity assumption on $\beta^{*}$. \Cref{assumptiona1} (ii) guarantees the uniqueness of the optimizer. \Cref{assumptiona1} (iii) is a regularity condition necessary to invoke the law of large numbers. \Cref{assumptiona1} (iv) corresponds to the ``nontransition point'' assumption in \citeA{zou2007degrees}. \Cref{assumptiona1} (v) is required to apply Stein's lemma and derive the closed-form expression for the degrees of freedom in \Cref{corollary:dof}.

\begin{theorem}[Consistency of the unbiased degrees of freedom estimate]
Suppose \Cref{assumptiona1} holds. Then, $\widehat{\textnormal{df}}(\mathbf{X}\hat{\beta}_{\text{sc}}) \overset{P}{\rightarrow}
\textnormal{df}(\mathbf{X}\hat{\beta}_{\text{sc}})$.
\label{theorem:dof_consistency}
\end{theorem}

\begin{proof}[Proof of \Cref{theorem:dof_consistency}]
We follow the proof strategy of Theorem 2 in \citeA{zou2007degrees} and proceed in two steps. First, we establish that $(\hat{\beta},\hat{\lambda},\hat{\mu})\overset{P}{\rightarrow} (\hat{\beta}^{*},\hat{\lambda}^{*},\hat{\mu}^{*})$. Second, we establish model selection consistency, which in turn implies the consistency of the degrees of freedom estimate.

The synthetic control coefficient estimate $\hat{\beta}$ and its population analog $\hat{\beta}^{*}$ are the minimizers of the following sample and population objective functions, respectively:
\begin{equation*}
Q_n(\beta)=
\begin{cases}
n^{-1}\|\mathbf{Y}-\mathbf{X}\beta\|_2^2 \ \ \ &\text{if} \ \beta\in\mathbb{S}^{p-1} \\[4pt]
+\infty \ \ \ \ \ \ \  \ \ \ \ \ \ &\text{otherwise}
\end{cases} \ \ \ \text{and} \ \ \
Q_0(\beta)=
\begin{cases}
E(Y_i-X_i'\beta)^2 \ \ \ &\text{if} \ \beta\in\mathbb{S}^{p-1} \\[4pt]
+\infty \ \ \ \ \ \ \  \ \ \ \ \ \ &\text{otherwise}
\end{cases}
\end{equation*}

To show $\hat{\beta}\overset{P}{\rightarrow} \hat{\beta}^{*}$, we verify the conditions of Theorem 2.1 of \citeA{newey1994large}. First, by \Cref{assumptiona1} (ii)-(iii), $Q_0(\beta)$ has a unique minimizer $\hat{\beta}^{*}$ on $\mathbb{S}^{p-1}$. Second, the parameter space $\mathbb{S}^{p-1}$ is compact, and $Q_0(\beta)$ is continuous on $\mathbb{S}^{p-1}$. Third, $Q_n(\beta)$ converges to $Q_0(\beta)$ uniformly in probability over $\mathbb{S}^{p-1}$. In particular, by \Cref{assumptiona1} (iii),  $Q_n(\beta)-Q_0(\beta)\overset{P}{\rightarrow} 0$ for each $\beta\in\mathbb{S}^{p-1}$ by the law of large numbers. Uniform convergence then follows from the compactness of $\mathbb{S}^{p-1}$ together with the continuity and quadratic-form structure of the objective functions.

Define the Lagrange dual functions
\begin{equation*}
D_n(\lambda,\mu)=
\begin{cases}
\underset{\beta}{\inf} \ \mathcal{L}_n(\beta;\lambda,\mu) \ \ \ &\text{if} \ \mu\geq 0 \\[4pt]
-\infty \ \ \ \ \ \ \  \ \ \ \ \ \  \ \ &\text{otherwise}
\end{cases}
 \ \ \ \text{and} \ \ \ D_0(\lambda,\mu)=
 \begin{cases}
\underset{\beta}{\inf} \ \mathcal{L}_0(\beta;\lambda,\mu) \ \ \ &\text{if} \ \mu\geq 0 \\[4pt]
-\infty \ \ \ \ \ \ \  \ \ \ \ \ \ \ \ \ \ \ &\text{otherwise}
\end{cases}
\end{equation*}
By standard convex duality arguments -- since Slater's condition holds for the probability simplex constraint -- strong duality holds and the dual optimum is attained. Because the primal optimizer is unique and the feasible set is the simplex, the KKT conditions imply that the associated dual multipliers are also unique. Hence $(\hat{\lambda},\hat{\mu})=\arg\max \ D_n(\lambda,\mu) $ and  $(\hat{\lambda}^{*},\hat{\mu}^{*})=\arg\max D_0(\lambda,\mu)$.

To prove the consistency of $(\hat{\lambda},\hat{\mu})$, we use a compactification argument and apply Theorem 2.1 of \citeA{newey1994large}. First, for each fixed $(\lambda,\mu)\in\mathbb{R}\times\mathbb{R}_{+}^p$, we have
\begin{align*}
\hat{\beta}_D(\lambda,\mu)&=\underset{\beta}{\arg\min} \ \mathcal{L}_n(\beta;\lambda,\mu)=\left(\frac{1}{n}\mathbf{X}^T\mathbf{X}\right)^{-1}\left(\frac{1}{n}\mathbf{X}^T\mathbf{Y}-\frac{1}{2}\left(\lambda\mathbf{1}_p-\mu\right)\right) \\[7pt]
\hat{\beta}_D^{*}(\lambda,\mu)&=\underset{\beta}{\arg\min} \ \mathcal{L}_0(\beta;\lambda,\mu)=\left(EX_iX_i'\right)^{-1}\left(EX_iY_i-\frac{1}{2}\left(\lambda\mathbf{1}_p-\mu\right)\right)
\end{align*}
and $\hat{\beta}_D(\lambda,\mu)\overset{P}{\rightarrow}  \hat{\beta}_D^{*}(\lambda,\mu)$ by \Cref{assumptiona1} (ii)-(iii) and the law of large numbers. Then $D_n(\lambda,\mu)-D_0(\lambda,\mu)=\mathcal{L}_n(\hat{\beta}_D(\lambda,\mu);\lambda,\mu)-\mathcal{L}_0(\hat{\beta}_D^{*}(\lambda,\mu);\lambda,\mu)\overset{P}{\rightarrow}  0$ follows by the continuous mapping theorem and the law of large numbers. Next, we show that $\hat{\lambda}=O_p(1)$, and $\hat{\mu}=O_p(1)$. This follows from the KKT condition $2n^{-1}\mathbf{X}^T(\mathbf{X}\hat{\beta}-\mathbf{Y})+\hat{\lambda}\mathbf{1}_p-\hat{\mu}=0$, combined with $\hat{\beta}\in\mathbb{S}^{p-1}, n^{-1}\mathbf{X}^T\mathbf{X}=O_p(1)$, and $n^{-1}\mathbf{X}^{T}\mathbf{Y}=O_p(1)$ from \Cref{assumptiona1} (iii). Because $(\hat{\lambda}^{*},\hat{\mu}^{*})$ is fixed and finite, there exists a deterministic compact set $K\subseteq \mathbb{R}\times\mathbb{R}_{+}^p$ containing both $(\hat{\lambda},\hat{\mu})$ and $(\hat{\lambda}^{*},\hat{\mu}^{*})$ with probability approaching one. By the law of large numbers and continuity of the quadratic-form map, we have $\sup_{(\lambda,\mu)\in K} |D_n(\lambda,\mu)-D_0(\lambda,\mu)|\overset{P}{\rightarrow} 0$. Because $D_0(\lambda,\mu)$ has a unique maximizer $(\hat{\lambda}^{*},\hat{\mu}^{*})$, these steps verify the conditions of \citeA[Theorem 2.1]{newey1994large}, yielding the desired consistency.

We now establish model selection consistency, i.e.,  $P(\hat{\mathcal{B}}_n=\mathcal{B}^{*})\rightarrow 1$. First, for $j\in\mathcal{B}^{*}$, the continuous mapping theorem implies that $\text{Sgn}(\hat{\beta}_j)\overset{P}{\rightarrow} \text{Sgn}(\hat{\beta}_j^{*})\neq 0$ since $\text{Sgn}$ is continuous at all $x$ other than zero. Thus $P(\hat{\mathcal{B}}_n\supseteq \mathcal{B}^{*})\rightarrow 1$.
Next, consider $j'\notin\mathcal{B}^{*}$. By \Cref{assumptiona1} (iv), $\hat{\mu}_{j'}^{*}>0$. Since $\hat{\mu}\overset{P}{\rightarrow} \hat{\mu}^{*}$, the event $\{\hat{\mu}_{j'}>0\}$ occurs with probability approaching 1. By complementary slackness for the sample problem, $\hat{\mu}_{j'}\hat{\beta}_{j'}=0$. Because $\hat{\mu}_{j'}>0$ implies $\hat{\beta}_{j'}=0$, it follows that $\hat{\beta}_{j'}=0$ with probability approaching 1. Therefore $P(\mathcal{B}^{*}\supseteq \hat{\mathcal{B}}_n)\rightarrow 1$ and $P(\hat{\mathcal{B}}_n=\mathcal{B}^{*})\rightarrow 1$. Consequently, we obtain $\widehat{\textnormal{df}}(\mathbf{X}\hat{\beta}_{\text{sc}})\overset{P}{\rightarrow} |\mathcal{B}^{*}|-1$. By the dominated convergence theorem, $\textnormal{df}(\mathbf{X}\hat{\beta}_{\text{sc}})=E[\widehat{\textnormal{df}}(\mathbf{X}\hat{\beta}_{\text{sc}})]\rightarrow |\mathcal{B}^{*}|-1$, and thus $\widehat{\textnormal{df}}(\mathbf{X}\hat{\beta}_{\text{sc}}) \overset{P}{\rightarrow}  \textnormal{df}(\mathbf{X}\hat{\beta}_{\text{sc}})$, as desired.
\end{proof}

\subsection{Consistency of the heteroskedasticity robust estimator}
\label{subsection:consistency_hc}

In this subsection, we establish the theoretical justification for the heteroskedasticity-robust information criterion (\ref{eq:robustIC}). Specifically, we show that its sample covariance penalty term is a consistent estimator of the population counterpart in (\ref{genSURE}). 

\begin{assumption}
We assume that
\begin{itemize} 
\item[(i)] The model is correctly specified as linear, i.e., $E[Y|X]=X\beta^{*}$.
\item[(ii)] The variance of $\varepsilon_i^2$ is bounded, i.e., $\var(\varepsilon_i^2)<\infty$.
\item[(iii)] The largest eigenvalue of $\frac{1}{n}\sum_{i=1}^n X_iX_i^T$ is stochastically bounded.
\end{itemize}
\label{assumptiona2}
\end{assumption}

\begin{theorem}
Suppose \Cref{assumptiona1} (i)-(iv) and \Cref{assumptiona2} hold. Then, the covariance penalty term of the heteroskedasticity-robust information criteria is consistent for its population counterpart:
\begin{equation*}
\frac{1}{n}\sum_{i=1}^n \hat{\varepsilon}_i^2\frac{\partial\hat{Y}_i}{\partial Y_i}-E\left[\frac{1}{n}\sum_{i=1}^{n}\sigma_{i}^{2}E_{Y|X}\left[\frac{\partial\hat{Y}_{i}}{\partial Y_{i}}\right]\right]:=S_n-E[T_n]\overset{P}{\rightarrow} 0.
\end{equation*}
\label{theorem:consistency_hc}
\end{theorem}
\begin{proof}[Proof of \Cref{theorem:consistency_hc}]


We write $\frac{\partial\hat{Y}_i}{\partial Y_i}=(H_{\hat{\mathcal{B}}_n})_{ii}$ where $H_{\hat{\mathcal{B}}_n}$ is the divergence matrix with the sample active set $\hat{\mathcal{B}}_n$. We decompose $S_n$ into three terms:
\begin{align*}
S_n&=\frac{1}{n}\sum_{i=1}^n \hat{\varepsilon}_i^2\left(H_{\hat{\mathcal{B}}_n}\right)_{ii}=\frac{1}{n}\sum_{i=1}^n \left(\varepsilon_i-X_i^T(\hat{\beta}-\beta^{*})\right)^2\left(H_{\hat{\mathcal{B}}_n}\right)_{ii} \\[7pt]
&=\underbrace{\frac{1}{n}\sum_{i=1}^n \varepsilon_i^2\left(H_{\hat{\mathcal{B}}_n}\right)_{ii}}_{S_{n1}}-\underbrace{\frac{2}{n}\sum_{i=1}^n \varepsilon_iX_i^T( \hat{\beta}-\beta^{*} ) \left(H_{\hat{\mathcal{B}}_n}\right)_{ii}}_{S_{n2}}
+\underbrace{\frac{1}{n}\sum_{i=1}^n ( X_i^T( \hat{\beta}-\beta^{*} ) )^2\left(H_{\hat{\mathcal{B}}_n}\right)_{ii}}_{S_{n3}}
\end{align*}



First, we show that $S_{n1}-E[T_n]\overset{P}{\rightarrow}  0$. We write that $S_{n1}=\frac{1}{n}\sum_{i=1}^n \varepsilon_i^2(H_{\mathcal{B}^{*}})_{ii}+\frac{1}{n}\sum_{i=1}^n \varepsilon_i^2[(H_{\hat{\mathcal{B}}_n})_{ii}-(H_{\mathcal{B}^{*}})_{ii}]$. Consider the second term:
\begin{equation*}
\left|\frac{1}{n}\sum_{i=1}^n \varepsilon_i^2\left[(H_{\hat{\mathcal{B}}_n})_{ii}-(H_{\mathcal{B}^{*}})_{ii}\right]\right|\leq \left| \frac{1}{n}\sum_{i=1}^n \varepsilon_i^2 \right| \times \max_{1\leq i\leq n}\left| (H_{\hat{\mathcal{B}}_n})_{ii}-(H_{\mathcal{B}^{*}})_{ii} \right|=\left(E[\varepsilon_i^2]+o_p(1)\right)\times o_p(1)=o_p(1),
\end{equation*}
where $\max_{1\leq i\leq n}| (H_{\hat{\mathcal{B}}_n})_{ii}-(H_{\mathcal{B}^{*}})_{ii} |\overset{P}{\rightarrow}  0$ since $P(| \max_{1\leq i\leq n}| (H_{\hat{\mathcal{B}}_n})_{ii}-(H_{\mathcal{B}^{*}})_{ii} | |>\eta )\leq P(\hat{\mathcal{B}}_n\neq \mathcal{B}^{*})\rightarrow 0$ for any $\eta>0$,
as shown in the proof of Theorem \ref{theorem:dof_consistency}. 
Therefore, the second term of $S_{n1}$ is $o_p(1)$. For the first term of $S_{n1}$, we apply Chebyshev's inequality. 
Since $H_{\mathcal{B}}$ is symmetric and idempotent, its diagonal values satisfy $0\leq (H_{\mathcal{B}})_{ii} \leq 1$, and we have that ${\var}(\frac{1}{n}\sum_{i=1}^n \varepsilon_i^2(H_{\mathcal{B}^{*}})_{ii})\leq{\var}(\frac{1}{n}\sum_{i=1}^n \varepsilon_i^2)=n^{-1}\var(\varepsilon_i^2)\rightarrow 0$ as $n\rightarrow\infty$ by \Cref{assumptiona2} (ii). Hence,
\begin{equation*}
\frac{1}{n}\sum_{i=1}^n \varepsilon_i^2(H_{\mathcal{B}^{*}})_{ii}- E\left[\frac{1}{n}\sum_{i=1}^n \varepsilon_i^2(H_{\mathcal{B}^{*}})_{ii}\right]\overset{P}{\rightarrow}  0.  
 \end{equation*}
Next, it follows that $E[T_n]-E\left[\frac{1}{n}\sum_{i=1}^n \sigma_i^2 (H_{\mathcal{B}^{*}})_{ii}\right]\rightarrow 0 $ by dominated convergence theorem and model selection consistency.  Since $E\left[\frac{1}{n}\sum_{i=1}^n \sigma_i^2 (H_{\mathcal{B}^{*}})_{ii}\right]=E\left[\frac{1}{n}\sum_{i=1}^n \varepsilon_i^2(H_{\mathcal{B}^{*}})_{ii}\right]$, we conclude that $S_{n1}-E[T_n]\overset{P}{\rightarrow}  0$, as desired.

Second, we show that $S_{n3}\overset{P}{\rightarrow}  0$. Recall that $0\leq (H_{\mathcal{B}})_{ii} \leq 1$. Hence,
\begin{equation*}
|S_{n3}|\leq \frac{1}{n}\sum_{i=1}^n ( X_i^T( \hat{\beta}-\beta^{*} ) )^2=(\hat{\beta}-\beta^{*})^T\left(\frac{1}{n}\sum_{i=1}^n X_iX_i^T\right)(\hat{\beta}-\beta^{*})\leq \left\|\frac{1}{n}\sum_{i=1}^n X_iX_i^T\right\|_{\text{op}}\|\hat{\beta}-\beta^{*}\|^2=o_p(1)
\end{equation*}
because $\hat{\beta}-\beta^{*}\overset{P}{\rightarrow}  0$ by \Cref{assumptiona2} (iii), as proved in \Cref{theorem:dof_consistency}.

Third, we show that $S_{n2}\overset{P}{\rightarrow}  0$. By Cauchy-Schwarz inequality and since $0\leq (H_{\mathcal{B}})_{ii} \leq 1$, we have that
\begin{align*}
|S_{n2}|&=\left| \frac{2}{n}\sum_{i=1}^n \varepsilon_iX_i^T( \hat{\beta}-\beta^{*} ) \left(H_{\hat{\mathcal{B}}_n}\right)_{ii} \right|\leq \frac{2}{n}\sum_{i=1}^n \left|\varepsilon_i\right|\left|X_i^T( \hat{\beta}-\beta^{*} ) \right| \\[7pt]
&\leq 2\left(\frac{1}{n}\sum_{i=1}^n \varepsilon_i^2\right)^{1/2}\left(\frac{1}{n}\sum_{i=1}^n ( X_i^T( \hat{\beta}-\beta^{*} ) )^2\right)^{1/2}=O_p(1)\times o_p(1)=o_p(1)
\end{align*}
by the results established in the previous steps.
\end{proof}

\section{Heteroskedasticity and Autocorrelation Robust Information Criteria}
\label{section:hac_ic}

In this section, we provide a novel SURE-based information criterion that is robust to both conditional heteroskedasticity and serial correlation. We maintain the Gaussian assumption
\begin{equation*}
\mathbf{Y}|\mathbf{X}\sim N\left(E\left[\mathbf{Y}|\mathbf{X}\right],\Sigma_{Y|X}\right),
\end{equation*}
where $\Sigma_{Y|X}\in\mathbb{R}^{n\times n}$ denotes the conditional variance-covariance matrix, which may be non-diagonal and exhibit heterogeneous variances along the diagonal.

We are interested in the risk $\mathcal{R}:=E\| \hat{\mathbf{Y}}-E[\mathbf{Y}|\mathbf{X}] \|_2^2$. 
Invoking a generalization of Stein's Lemma (\citeNP{liu1994siegel}), we can produce a generalization of the SURE, 
\begin{equation*}
\mathcal{R}:=E\left\Vert \mathbf{Y}-\hat{\mathbf{Y}}\right\Vert _{2}^{2}+2E\left[\mathrm{Tr}\left(\Sigma_{Y|X}E\left[\left.\nabla\hat{\mathbf{Y}}\right|\mathbf{X}\right]\right)\right]+\mathrm{const,}
\end{equation*}
where $E[\nabla\hat{\mathbf{Y}}|\mathbf{X}]$ is the expected divergence, i.e., the Jacobian of $\hat{\mathbf{Y}}$ with respect to $\mathbf{Y}$. The analytical expressions of the divergence $\nabla\hat{\mathbf{Y}}$ for a class of synthetic control methods are given in Appendix \ref{subsectionA4}.

However, estimating each individual entry of $\Sigma_{Y|X}$ raises a substantial challenge without a fully specified model of the dependence. Drawing on the classical heteroskedasticity-and-autocorrelation-robust (HAR) literature for estimating asymptotic covariances of regression estimators (see e.g. \citeA{muller2014hac,lazarus2018har} for surveys), we propose a HAR estimator for the information criteria.  Specifically, we apply a spectral decomposition to the penalty term in the generalized population information criteria; this reformulation naturally yields a tractable sample analog estimator.

\begin{proposition}
Suppose the expected divergence $E[\nabla\hat{\mathbf{Y}}|\mathbf{X}]$ is a real 
positive-definite matrix and admits a spectral decomposition of the form $E[\nabla\hat{\mathbf{Y}}|\mathbf{X}]=\sum_{k=1}^r d_kv_kv_k^T$ where $v_k\in\mathbb{R}^n$ is the $k$-th orthonormal eigenvector, $d_k\in\mathbb{R}$ is the $k$-th eigenvalue, and $r$ is the rank of the divergence matrix.

Then, the population generalized information criteria is
\begin{equation}
\mathrm{IC}=E\left\Vert \mathbf{Y}-\hat{\mathbf{Y}}\right\Vert _{2}^{2}+2E\left[\sum_{i=1}^n\sum_{j=1}^n E\left[\left.\mathbf{W}_i^T\mathbf{W}_j\right|\mathbf{X}\right]\right]
\label{equ:ic_hac}
\end{equation}
where $\mathbf{W}_i=( \sqrt{d_1}v_{1i},...,\sqrt{d_r}v_{ri} )\varepsilon_i\in\mathbb{R}^{r\times 1}$ for each $i=1,...,n$ and $\varepsilon_i=Y_i-E[Y_i|X]$.
\label{proposition_spectral}
\end{proposition}
\begin{proof}[Proof of \Cref{proposition_spectral}]
Note that the penalty term in the generalized population information criteria can be rewritten as follows
\begin{align*}
E\left[\mathrm{Tr}\left(\Sigma_{Y|X}E\left[\left.\nabla\hat{\mathbf{Y}}\right|\mathbf{X}\right]\right)\right]&= E\left[\sum_{i=1}^n\sum_{j=1}^nE[\varepsilon_i\varepsilon_j|\mathbf{X}]\left(\sum_{k=1}^r d_kv_{ki}v_{kj}\right)\right]=E\left[\sum_{i=1}^n\sum_{j=1}^n E\left[\left.\mathbf{W}_i^T\mathbf{W}_j\right|\mathbf{X}\right]\right].
\end{align*}
\end{proof}
Once \Cref{proposition_spectral} is established, the sample analog of the covariance penalty term in \Cref{equ:ic_hac} can be obtained by replacing each true error $\varepsilon_i$ with its empirical residual $\widehat{\varepsilon}_i$ and using kernel weights to ensure positive definiteness. Accordingly, our proposed heteroskedasticity and autocorrelation robust IC estimator is
\begin{equation}
\begin{aligned}
\widehat{\mathrm{IC}}_{\text{HAR}} &=\left\Vert \mathbf{Y}-\widehat{\mathbf{Y}}\right\Vert _{2}^{2}+\frac{n}{n-\widehat{\mathrm{df}}(\widehat{\mathbf{Y}})}\sum_{j=-n+1}^{n-1} \omega\left( \frac{j}{h_n}  \right)\widehat{\Gamma}(j), \ \ \ \text{with}\\[7pt]
\widehat{\Gamma}(j)&=\begin{cases}
\sum\limits_{\ell=j+1}^{n} \widehat{\boldsymbol{W}}_{\ell}^T \widehat{\boldsymbol{W}}_{\ell-j}   & \ \text{for} \ j\geq 0, \\[18pt]
\sum\limits_{\ell=-j+1}^{n} \widehat{\boldsymbol{W}}_{\ell+j}^T\widehat{\boldsymbol{W}}_{\ell}    & \ \text{for} \ j<0,
\end{cases}
\end{aligned}
\label{equ:haric}
\end{equation}
where $\widehat{\boldsymbol{W}}_i=\widehat{\varepsilon}_i( \sqrt{d_1}v_{1i},...,\sqrt{d_r}v_{ri} )$, $\widehat{\varepsilon}_i$ is the residual from the unpenalized synthetic control method, $\omega(\cdot)$ is a real-valued kernel, and $h_n$ is a bandwidth parameter. The factor $n/(n-\widehat{\mathrm{df}}(\widehat{\mathbf{Y}}))$ is a small-sample degrees of freedom adjustment.

The choices of the kernel functions $\omega(\cdot)$ and the bandwidth $h_n$ are discussed in \citeA{neweywest1987}, \citeA{andrews1991heteroskedasticity}  and the subsequent literature. Moreover, as suggested by \citeA{andrews1992improved}, one can prewhiten the ``transformed residuals'' $\widehat{\boldsymbol{W}}_i$ by fitting a vector autoregression (VAR) of order $r$ and then replacing $\widehat{\boldsymbol{W}}_i$ with the VAR residuals in \Cref{equ:haric}, which may lead to a smaller bias. Implementation details for computing the HAR covariance matrix are in the well-established R package \texttt{sandwich} (\citeNP{zeileis2004hac}), and these routines can be utilized to compute our HAR-information criteria estimator. However, the theoretical consequences of the kernel and bandwidth selection and the prewhitening step on our proposed HAR information criteria remain unknown and are left for future research.

\begin{figure}[h]
\begin{center}
\includegraphics[scale=0.25]{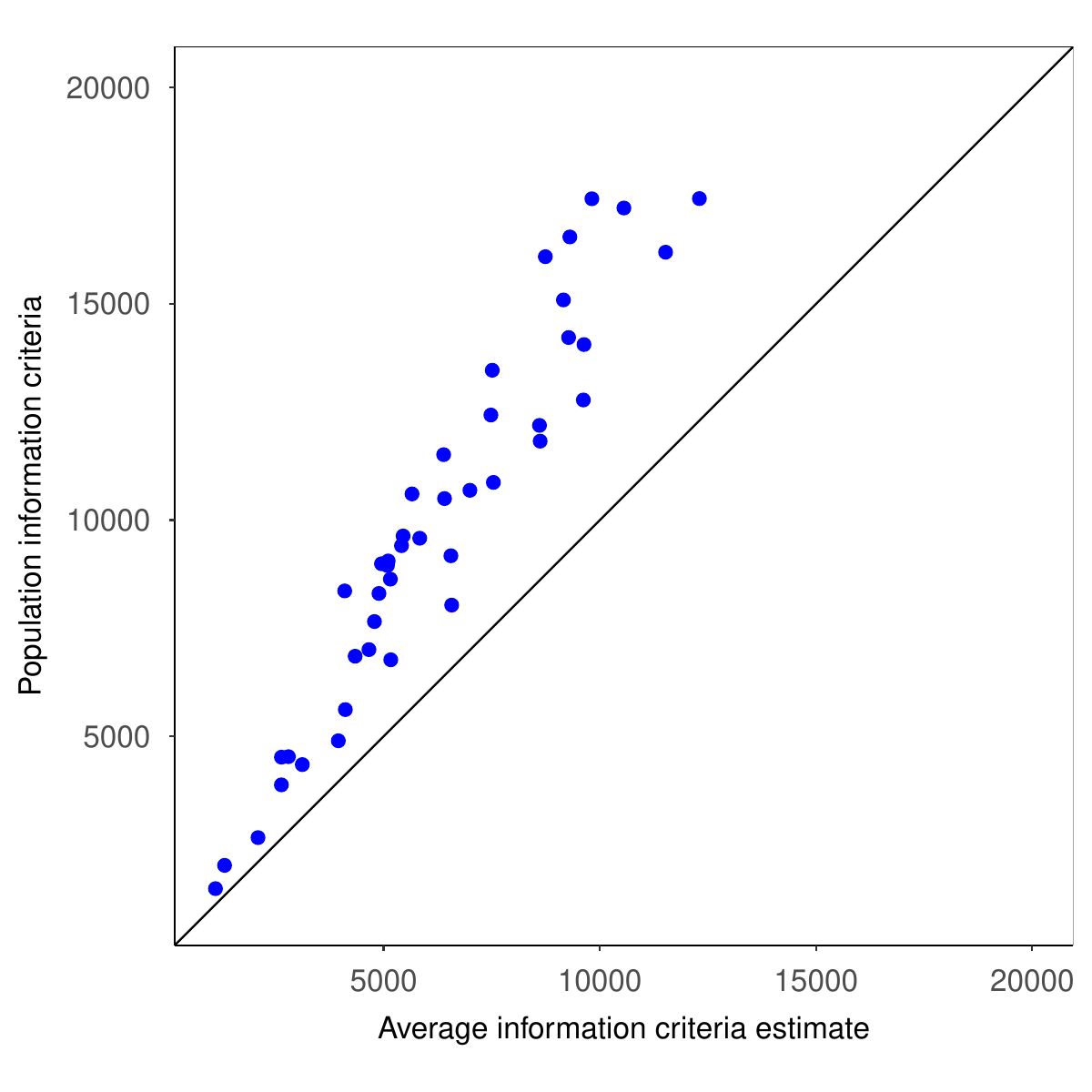} \
\includegraphics[scale=0.25]{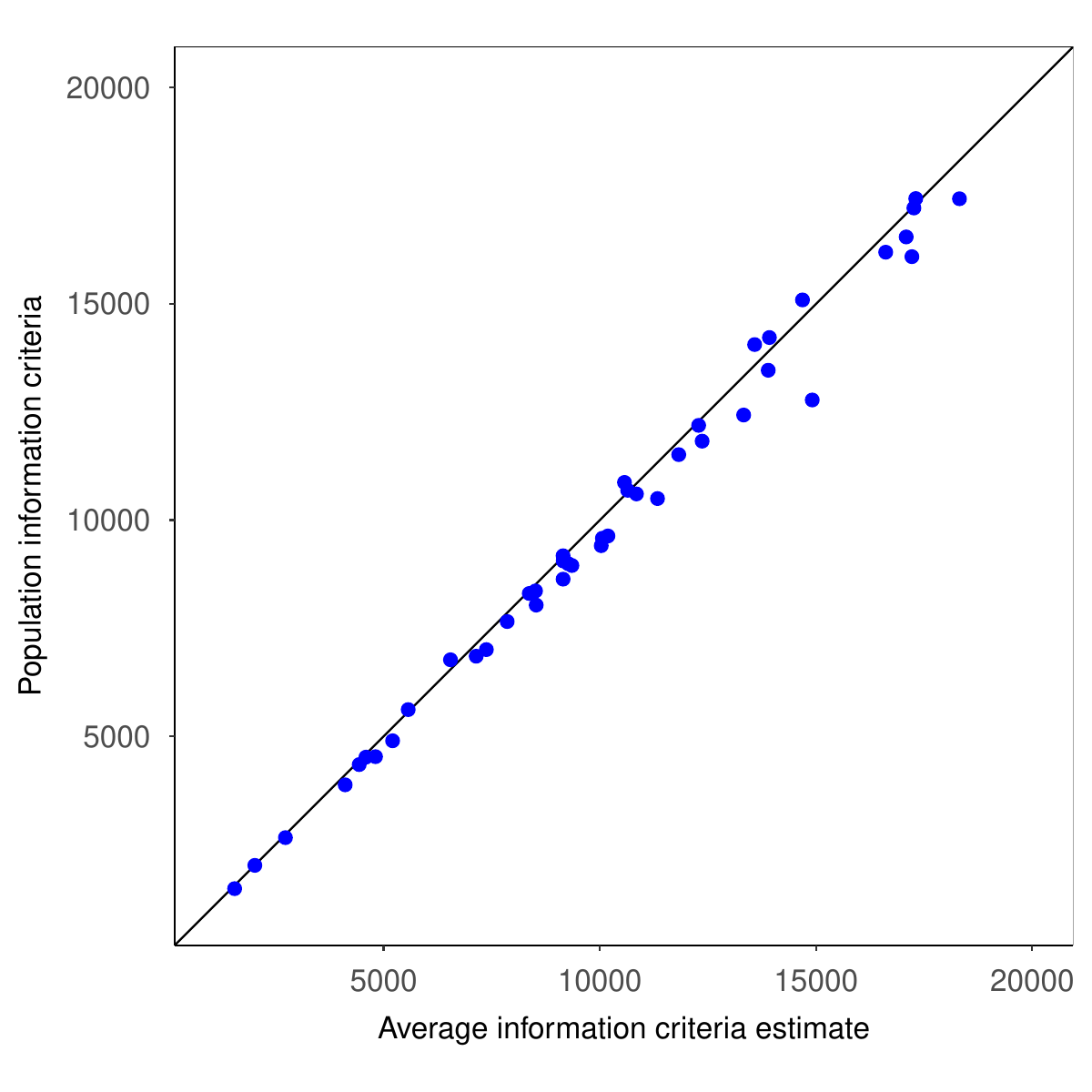} \
\includegraphics[scale=0.25]{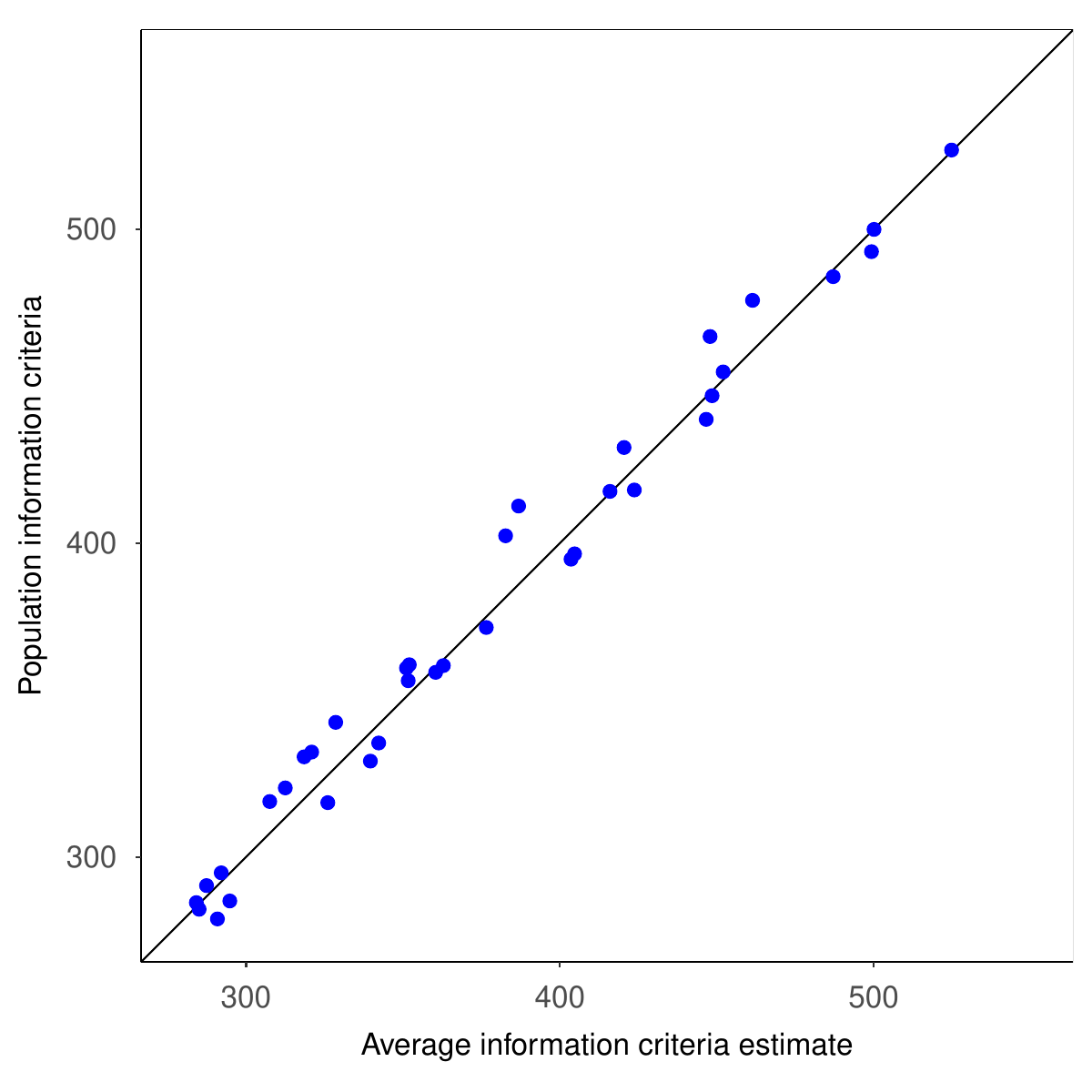}  
\end{center}

\caption{Numerical performance of the naive and HAR information criteria.}
\label{figure_hac_ic}
\end{figure}

We evaluate the finite sample performance of our proposed HAR-information criteria via Monte Carlo simulations. Specifically, we consider two data-generating processes
\begin{align*}
\text{HOMO}: \ \ \ \ \ \ & Y_i=X_i'\beta+\varepsilon_i, \ \text{where} \ \varepsilon_i\iid N(0,\sigma^2)  \\[7pt]
\text{HETE-AR(1)}: \ \ \ \ \ \ & Y_i=X_i'\beta+\varepsilon_i, \ \text{where} \ \varepsilon_i=\rho \varepsilon_{i-1}+u_i, \ u_i\iid N(0,(1-\rho)^2\exp(X_{i,1}/50)).
\end{align*}
For each design, we compute the average of the estimated information criteria over 300 replications and plot it against its population counterpart. The simulation results are presented in \Cref{figure_hac_ic}. The left panel corresponds to the naive information criteria estimate (\ref{eq:information-criteria-estimate}) in the HETE-AR(1) design. The center and right panels correspond to the HAR information criteria (\ref{equ:haric}) under HETE-AR(1) and HOMO designs respectively. As shown, the naive information criterion (\ref{eq:information-criteria-estimate}) tends to underestimate the population counterpart and is severely biased in the high-heteroskedasticity design, whereas the HAR information criterion (\ref{equ:haric}) is unbiased on average -- even with heteroskedasticity and autocorrelation -- and remains unbiased under homoskedasticity.

\section{Additional Figures}

\begin{figure}[H]
\includegraphics[scale=0.30]{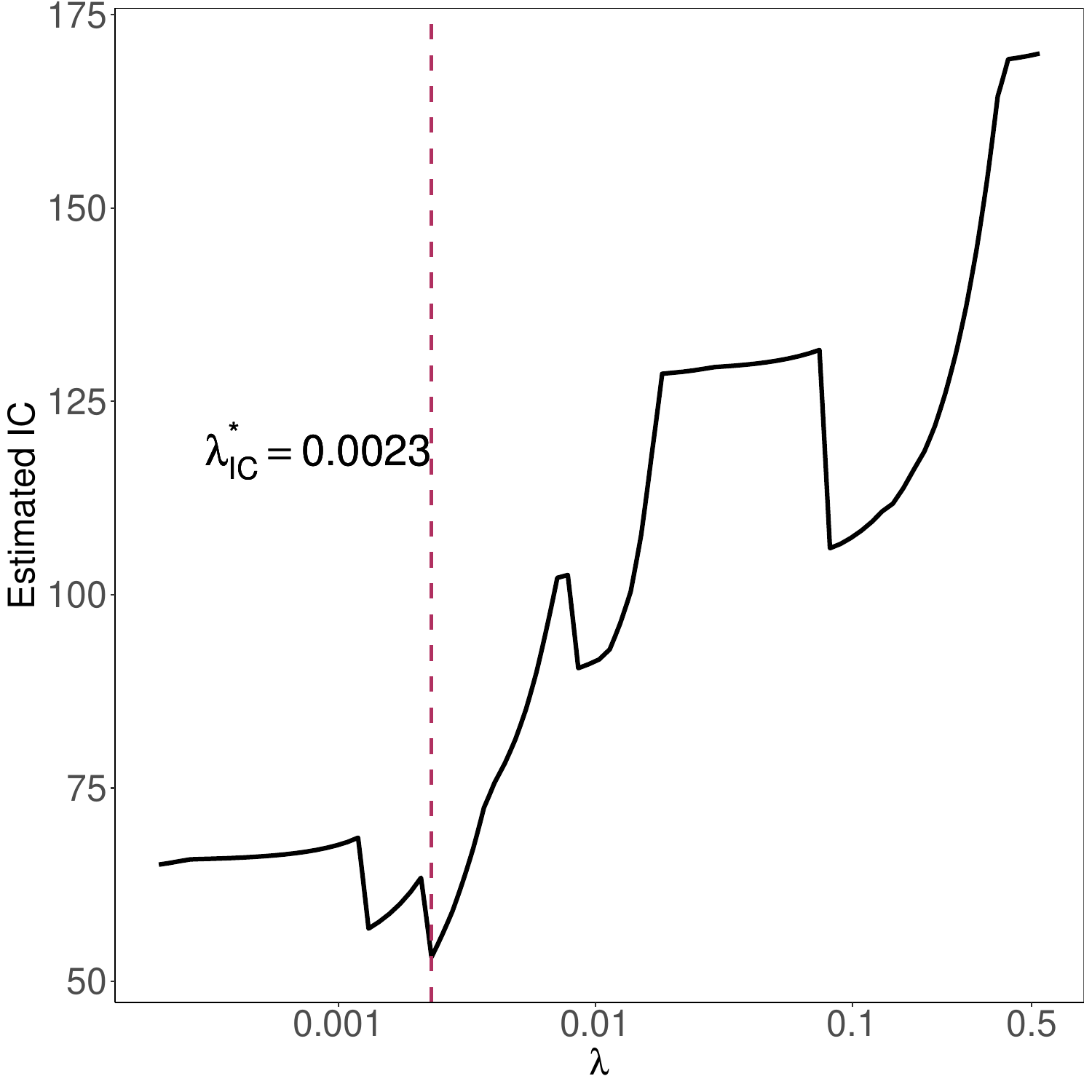}\hspace{1.5em}
\includegraphics[scale=0.30]{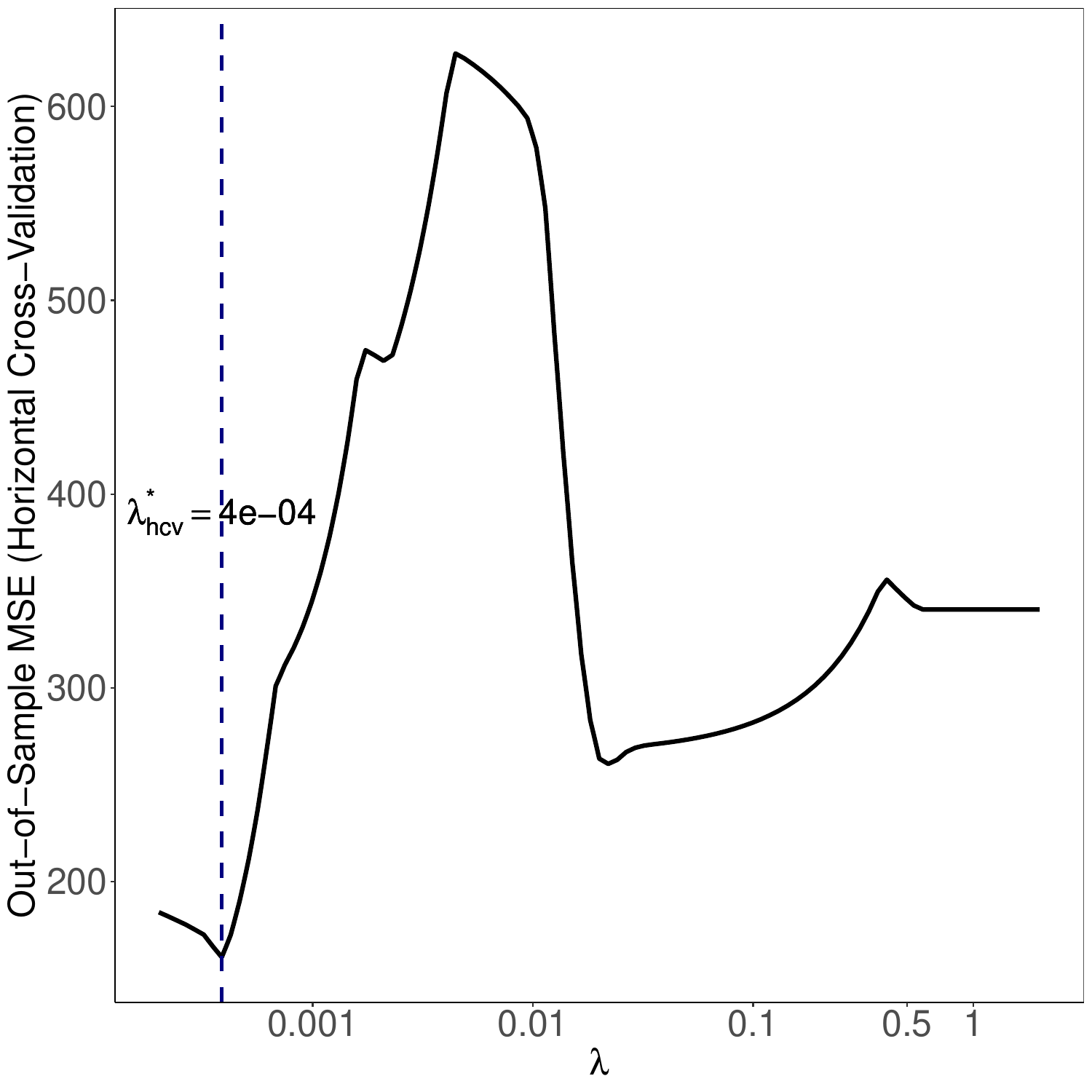}\\[1em]
\includegraphics[scale=0.30]{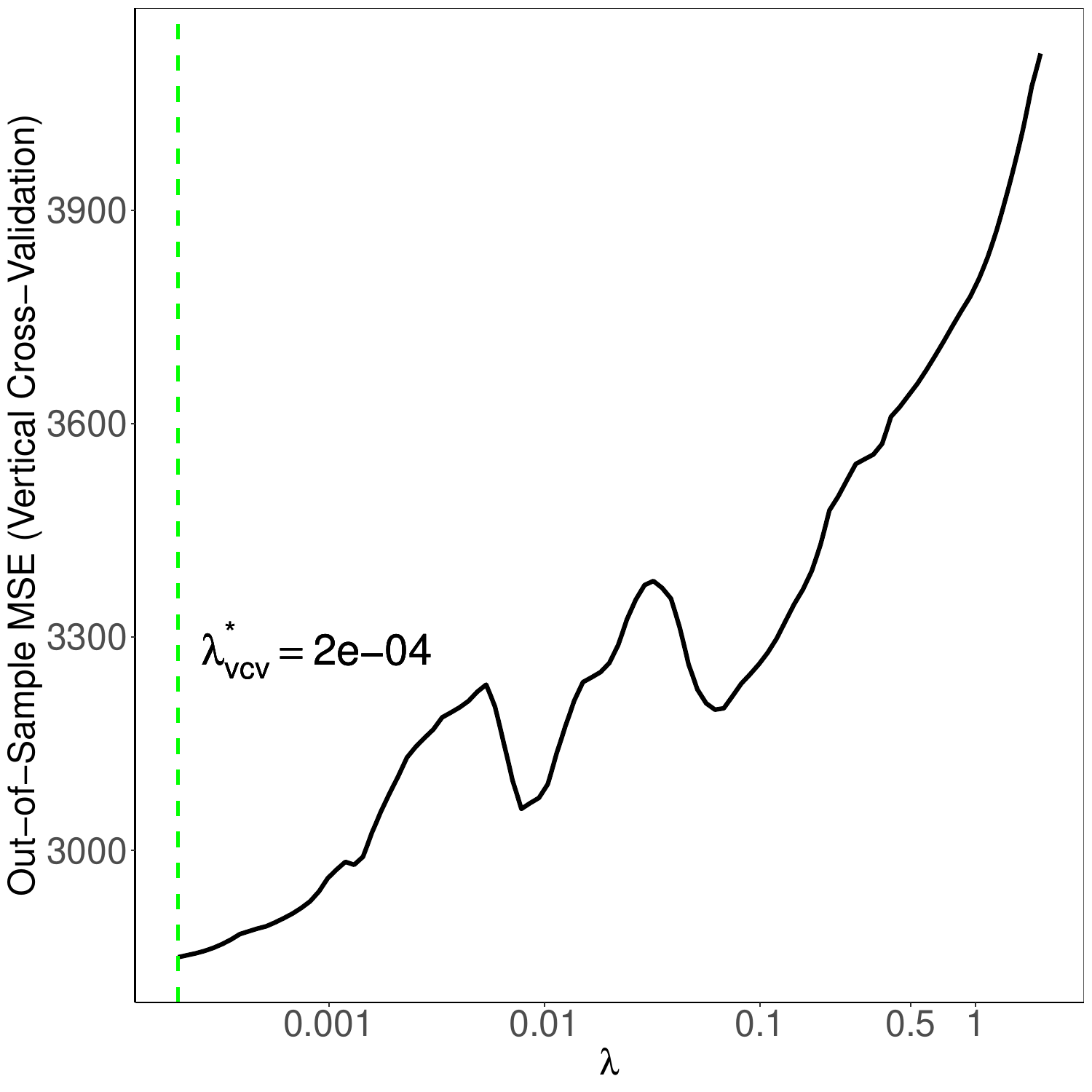}\hspace{1em}
\includegraphics[scale=0.30]{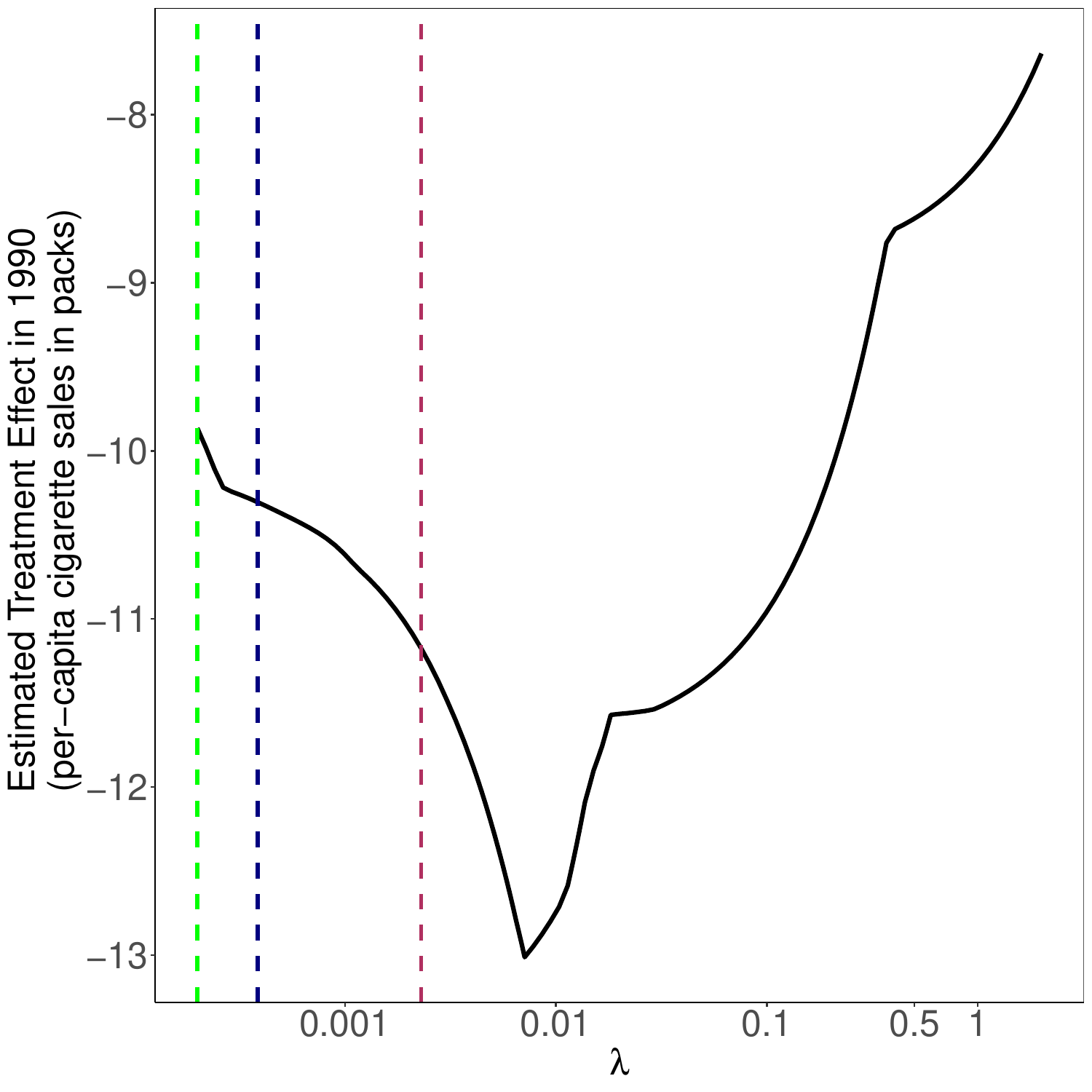} \\[1em]
\footnotesize Top-left: estimated information criterion; top-right: horizontal cross-validation loss; bottom-left: vertical cross-validation loss; bottom-right: estimated one-year treatment effect as a function of \(\lambda\).  
Vertical lines mark the minimizing penalties: \(\lambda^{\ast}_{VCV}\) (green), \(\lambda^{\ast}_{IC}\) (red), and \(\lambda^{\ast}_{HCV}\) (blue). There are no covariates.
\caption{\emph{Tuning parameter selection for PSCM and resulting treatment effect estimates in California Proposition 99 application.}}
\label{fig:California_tobacco_selection}
\end{figure}

\section{Details for Simulations}
\label{simdetails}

\subsection{Empirical simulation design}

In the first simulation design in \Cref{sec:simulation}, we assume that $Y_t$ and $\mathbf{X}_{t}=\left(X_{1t},...,X_{pt}\right)^{T}$ are jointly normal, which implies that $Y_t|\mathbf{X}_t$ is also normal. Under this specification, the Gaussian assumption underlying the theory is satisfied, and the conditional expectation is
\begin{equation}
E\!\left[Y_t \mid \mathbf{X}_t\right]
=
\left(L_0^{\top}L_{-0}\right)
\left(L_{-0}^{\top}L_{-0}+\Sigma_X\right)^{-1}
\mathbf{X}_t
\label{equ:expectation_expression}
\end{equation}
where $\Sigma_X$ is the $p\times p$ donor block of $\Sigma$. Consequently, both Stein's unbiased risk estimate and the exact true risk can be computed and directly compared.

To assess the robustness to the Gaussian assumption, the second simulation design keeps the Gaussian factor structure for donors, but generates the treated unit's idiosyncratic shocks from the empirical distribution of the fitted residuals. Specifically, let $Y_t^* = \psi_t^T L_0$ denote the factor component of the
treated unit. We simulate
\vspace{1.5em}
\begin{equation*}
\left(\begin{array}{c}
Y_{t}^{*}\\
\mathbf{X}_{t}
\end{array}\right)\sim N\left(\mathbf{0}_{p+1},\mathbf{L}\mathbf{L}^{T}+\left(\begin{array}{cc}
0 & 0\\
0 & \Sigma_{X}
\end{array}\right)\right)\\[7pt]
\end{equation*}
and construct $Y_t = Y_t^* + U_{0t}$, where the idiosyncratic shocks
$\{U_{0t}\}_{t=1}^T$ are drawn i.i.d.\ with replacement from the centered fitted treated
residuals $\{\hat U_{01}-\bar{\hat{U}}_0, \dots, \hat U_{0,T^{*}}-\bar{\hat{U}}_0\}$, obtained by
estimating the factor model on the Shijiazhuang panel. By construction, $U_{0t}$ is mean zero and independent of $\mathbf{X}_t$, so
\[
E\!\left[Y_t \mid \mathbf{X}_t\right]
=E\!\left[Y_t^{*} \mid \mathbf{X}_t\right]+E\!\left[U_{0t}\mid \mathbf{X}_t\right]
=E\!\left[Y_t^{*} \mid \mathbf{X}_t\right]
\]
and the same conditional mean formula \eqref{equ:expectation_expression} continues to apply.

\subsection{Details of alternative tuning parameter selection procedures }
\label{subsection:tuning_parameter_selection}

In \Cref{tab:selection_compare} and \Cref{fig:Risk vs lambda Gaussian sim}, the true risk can be computed exactly from the simulated DGP:
\begin{equation*}
\mathrm{Risk}(\lambda)=\sum_{t=1}^{T^{*}} \left( \hat{Y}_t(\lambda)- E\!\left[Y_t \mid \mathbf{X}_t\right]\right)^2,
\end{equation*}
where $T^{*}$ is the number of pre-treatment periods. Minimizing this criterion over $\lambda\in\Lambda$ yields the oracle penalty $\lambda_{\mathrm{risk}}$, which serves as the benchmark.

The oracle information criterion $\mathrm{IC}_{\mathrm{oracle}}$ is the infeasible Stein's unbiased risk estimate
\begin{equation*}
\mathrm{IC}_{\mathrm{oracle}}(\lambda)=\sum_{t=1}^{T^{*}} \left(Y_t-\hat{Y}_t(\lambda)\right)^2+2\sigma^2\hat{\mathrm{df}}(\lambda)
\end{equation*}
where $\sigma^2=\var(Y_t|\mathbf{X}_t)$ is the true conditional variance, which is available in closed form under the simulated DGP, and $\widehat{\mathrm{df}}(\lambda)=(1+\lambda)(|\mathcal{A}(\lambda)|-1)$ with $\mathcal{A}(\lambda)$ denoting the active donors.

The feasible information criterion replaces the true conditional variance with its empirical analog:
\begin{equation*}
\widehat{\mathrm{IC}}(\lambda)=\sum_{t=1}^{T^{*}} \left(Y_t-\hat{Y}_t(\lambda)\right)^2+2\widehat{\sigma}^2\hat{\mathrm{df}}(\lambda)
\end{equation*}
where \(\widehat{\sigma}^2 \) is a holdout-based estimate of the pre-treatment forecast error variance. Specifically, we fit an unpenalized synthetic control on the first two-thirds of the pre-treatment periods $\{1,...,\lfloor\tfrac{2}{3}T^{*}\rfloor\}$, use that fit to predict the remaining one-third $\{\lfloor\tfrac{2}{3}T^{*}\rfloor+1,...T^{*}\}$, and then take the sample variance of the resulting out-of-sample residuals.

We also consider three cross-validation methods for selecting the tuning parameters: (i) \emph{horizontal} cross-validation (\emph{pre-intervention hold-out on the treated unit}),
(ii) \emph{vertical} cross-validation (\emph{leave-one-out validation on the untreated units}), and
(iii) a \emph{rolling-window} cross-validation. Each method is described in detail below.

\textbf{(i) Horizontal cross-validation.} We divide the pre-treatment periods of the treated unit into a training and validation set: the first two thirds, $t=1,...,\lfloor\tfrac{2}{3}T^{*}\rfloor$, are used to estimate PSCM, and the remaining one-third $t=\lfloor\tfrac{2}{3}T^{*}\rfloor+1,\dots ,T^{*}$ are used for validation. The tuning parameter is selected to minimize the mean square error on the validation periods, we denote this choice by \(\lambda_{\text{HCV}}^*\).

\textbf{(ii) Vertical cross-validation.} Each donor is iteratively designated as a pseudo-treated unit. In each iteration, the selected donor is excluded from the donor pool, PSCM is re-estimated using the pre-treatment data of the remaining donors, and the fitted model is used to predict the pseudo-treated unit over its post-treatment periods. The resulting prediction mean-square errors are averaged across all iterations to compute the vertical validation loss. The tuning parameter that minimizes this average is denoted \(\lambda_{\text{VCV}}^*\).

\textbf{(iii) Rolling-window cross-validation.} Starting with the first half of the pre-treatment period, \(t=1,\dots ,\lfloor T^{*}/2\rfloor\), we estimate PSCM and compute the one-step-ahead prediction error at \(t=\lfloor T^{*}/2\rfloor+1\). The estimation window is then expanded by one period, the model re-estimated, and the next one-step-ahead error computed. Repeating this procedure until \(t=T^{*}\) produces a sequence of one-period-ahead prediction errors. This average defines the rolling-window loss, and the tuning parameter that minimizes this loss is denoted \(\lambda_{\text{RWCV}}^*\).

\end{appendices}

\clearpage
\bibliography{df_SC_appendix}

@article{abadie2003economic,
  title={The economic costs of conflict: A case study of the Basque Country},
  author={Abadie, Alberto and Gardeazabal, Javier},
  journal={American economic review},
  volume={93},
  number={1},
  pages={113--132},
  year={2003}
}

@article{abadie2021using,
  title={Using synthetic controls: Feasibility, data requirements, and methodological aspects},
  author={Abadie, Alberto},
  journal={Journal of Economic Literature},
  volume={59},
  number={2},
  pages={391--425},
  year={2021}
}

@article{athey2021matrix,
  title={Matrix completion methods for causal panel data models},
  author={Athey, Susan and Bayati, Mohsen and Doudchenko, Nikolay and Imbens, Guido and Khosravi, Khashayar},
  journal={Journal of the American Statistical Association},
  pages={1--15},
  year={2021},
  publisher={Taylor \& Francis}
}

@techreport{doudchenko2016balancing,
  title={Balancing, regression, difference-in-differences and synthetic control methods: A synthesis},
  author={Doudchenko, Nikolay and Imbens, Guido W},
  year={2016},
  institution={National Bureau of Economic Research}
}

@article{abadie2010synthetic,
  title={Synthetic control methods for comparative case studies: Estimating the effect of California's tobacco control program},
  author={Abadie, Alberto and Diamond, Alexis and Hainmueller, Jens},
  journal={Journal of the American statistical Association},
  volume={105},
  number={490},
  pages={493--505},
  year={2010},
  publisher={Taylor \& Francis}
}

@article{abadie2015comparative,
  title={Comparative politics and the synthetic control method},
  author={Abadie, Alberto and Diamond, Alexis and Hainmueller, Jens},
  journal={American Journal of Political Science},
  volume={59},
  number={2},
  pages={495--510},
  year={2015},
  publisher={Wiley Online Library}
}

@article{chernozhukov2018t,
  title={A $ t $-test for synthetic controls},
  author={Chernozhukov, Victor and Wuthrich, Kaspar and Zhu, Yinchu},
  journal={arXiv preprint arXiv:1812.10820},
  year={2018}
}

@article{xu2017generalized,
  title={Generalized synthetic control method: Causal inference with interactive fixed effects models},
  author={Xu, Yiqing},
  journal={Political Analysis},
  volume={25},
  number={1},
  pages={57--76},
  year={2017},
  publisher={Cambridge University Press}
}

@article{bai2002determining,
  title={Determining the number of factors in approximate factor models},
  author={Bai, Jushan and Ng, Serena},
  journal={Econometrica},
  volume={70},
  number={1},
  pages={191--221},
  year={2002},
  publisher={Wiley Online Library}
}

@article{liu1994siegel,
  title={Siegel's formula via Stein's identities},
  author={Liu, Jun S},
  journal={Statistics \& Probability Letters},
  volume={21},
  number={3},
  pages={247--251},
  year={1994},
  publisher={Elsevier}
}

@article{cavallo2013catastrophic,
  title={Catastrophic natural disasters and economic growth},
  author={Cavallo, Eduardo and Galiani, Sebastian and Noy, Ilan and Pantano, Juan},
  journal={Review of Economics and Statistics},
  volume={95},
  number={5},
  pages={1549--1561},
  year={2013},
  publisher={The MIT Press}
}

@article{bifulco2017using,
  title={Using synthetic controls to evaluate the effect of unique interventions: The case of Say Yes to Education},
  author={Bifulco, Robert and Rubenstein, Ross and Sohn, Hosung},
  journal={Evaluation review},
  volume={41},
  number={6},
  pages={593--619},
  year={2017},
  publisher={SAGE Publications Sage CA: Los Angeles, CA}
}

@article{bohn2014did,
  title={Did the 2007 Legal Arizona Workers Act reduce the state's unauthorized immigrant population?},
  author={Bohn, Sarah and Lofstrom, Magnus and Raphael, Steven},
  journal={Review of Economics and Statistics},
  volume={96},
  number={2},
  pages={258--269},
  year={2014},
  publisher={The MIT Press}
}

@article{pieters2016effect,
  title={Effect of democratic reforms on child mortality: a synthetic control analysis},
  author={Pieters, Hannah and Curzi, Daniele and Olper, Alessandro and Swinnen, Johan},
  journal={The Lancet Global Health},
  volume={4},
  number={9},
  pages={e627--e632},
  year={2016},
  publisher={Elsevier}
}

@article{heersink2017disasters,
  title={Disasters and elections: Estimating the net effect of damage and relief in historical perspective},
  author={Heersink, Boris and Peterson, Brenton D and Jenkins, Jeffery A},
  journal={Political Analysis},
  volume={25},
  number={2},
  pages={260--268},
  year={2017},
  publisher={Cambridge University Press}
}

@article{peri2019labor,
  title={The labor market effects of a refugee wave synthetic control method meets the mariel boatlift},
  author={Peri, Giovanni and Yasenov, Vasil},
  journal={Journal of Human Resources},
  volume={54},
  number={2},
  pages={267--309},
  year={2019},
  publisher={University of Wisconsin Press}
}

@article{billmeier2013assessing,
  title={Assessing economic liberalization episodes: A synthetic control approach},
  author={Billmeier, Andreas and Nannicini, Tommaso},
  journal={Review of Economics and Statistics},
  volume={95},
  number={3},
  pages={983--1001},
  year={2013},
  publisher={The MIT Press}
}

@article{abadie2021penalized,
  title={A penalized synthetic control estimator for disaggregated data},
  author={Abadie, Alberto and L{'}Hour, J{\'e}r{\'e}my},
  journal={Journal of the American Statistical Association},
  number={just-accepted},
  pages={1--34},
  year={2021},
  publisher={Taylor \& Francis}
}

@article{kellogg2020combining,
  title={Combining matching and synthetic control to trade off biases from extrapolation and interpolation},
  author={Kellogg, Maxwell and Mogstad, Magne and Pouliot, Guillaume A and Torgovitsky, Alexander},
  journal={Journal of the American Statistical Association},
  number={just-accepted},
  pages={1--30},
  year={2021},
  publisher={Taylor \& Francis}
}

@article{ben2021augmented,
  title={The augmented synthetic control method},
  author={Ben-Michael, Eli and Feller, Avi and Rothstein, Jesse},
  journal={Journal of the American Statistical Association},
  number={just-accepted},
  pages={1--34},
  year={2021},
  publisher={Taylor \& Francis}
}

@article{cattaneo2021prediction,
  title={Prediction intervals for synthetic control methods},
  author={Cattaneo, Matias D and Feng, Yingjie and Titiunik, Rocio},
  journal={Journal of the American Statistical Association},
  volume={116},
  number={536},
  pages={1865--1880},
  year={2021},
  publisher={Taylor \& Francis}
}

@article{claeskens2008model,
  title={Model selection and model averaging},
  author={Claeskens, Gerda and Hjort, Nils Lid},
  journal={Cambridge Books},
  year={2008},
  publisher={Cambridge University Press}
}

@article{meyer2000degrees,
  title={On the degrees of freedom in shape-restricted regression},
  author={Meyer, Mary and Woodroofe, Michael},
  journal={The annals of Statistics},
  volume={28},
  number={4},
  pages={1083--1104},
  year={2000},
  publisher={Institute of Mathematical Statistics}
}

@article{zou2007degrees,
  title={On the ``degrees of freedom'' of the lasso},
  author={Zou, Hui and Hastie, Trevor and Tibshirani, Robert},
  journal={The Annals of Statistics},
  volume={35},
  number={5},
  pages={2173--2192},
  year={2007},
  publisher={Institute of Mathematical Statistics}
}

@article{stein1981estimation,
  title={Estimation of the mean of a multivariate normal distribution},
  author={Stein, Charles M},
  journal={The annals of Statistics},
  pages={1135--1151},
  year={1981},
  publisher={JSTOR}
}

@article{mukherjee2015degrees,
  title={On the degrees of freedom of reduced-rank estimators in multivariate regression},
  author={Mukherjee, Ashin and Chen, Kun and Wang, Naisyin and Zhu, Ji},
  journal={Biometrika},
  volume={102},
  number={2},
  pages={457--477},
  year={2015},
  publisher={Oxford University Press}
}

@article{candes2013unbiased,
  title={Unbiased risk estimates for singular value thresholding and spectral estimators},
  author={Candes, Emmanuel J and Sing-Long, Carlos A and Trzasko, Joshua D},
  journal={IEEE transactions on signal processing},
  volume={61},
  number={19},
  pages={4643--4657},
  year={2013},
  publisher={IEEE}
}

@article{deledalle2012risk,
  title={Risk estimation for matrix recovery with spectral regularization},
  author={Deledalle, Charles-Alban and Vaiter, Samuel and Peyr{\'e}, Gabriel and Fadili, Jalal and Dossal, Charles},
  journal={arXiv preprint arXiv:1205.1482},
  year={2012}
}

@article{mazumder2020computing,
  title={Computing the degrees of freedom of rank-regularized estimators and cousins},
  author={Mazumder, Rahul and Weng, Haolei},
  journal={Electronic Journal of Statistics},
  volume={14},
  number={1},
  pages={1348--1385},
  year={2020},
  publisher={Institute of Mathematical Statistics and Bernoulli Society}
}

@article{minami2020degrees,
  title={Degrees of freedom in submodular regularization: A computational perspective of Stein's unbiased risk estimate},
  author={Minami, Kentaro},
  journal={Journal of Multivariate Analysis},
  volume={175},
  pages={104546},
  year={2020},
  publisher={Elsevier}
}

@article{chen2020degrees,
  title={On degrees of freedom of projection estimators with applications to multivariate nonparametric regression},
  author={Chen, Xi and Lin, Qihang and Sen, Bodhisattva},
  journal={Journal of the American Statistical Association},
  volume={115},
  number={529},
  pages={173--186},
  year={2020},
  publisher={Taylor \& Francis}
}

@article{tibshirani2015stein,
  title={Stein's unbiased risk estimate},
  author={Tibshirani, Ryan and Wasserman, L},
  journal={Course notes from Statistical Machine Learning},
  pages={1--12},
  year={2015}
}

@book{friedman2001elements,
  title={The elements of statistical learning},
  author={Friedman, Jerome and Hastie, Trevor and Tibshirani, Robert},
  volume={1},
  number={10},
  year={2001},
  publisher={Springer series in statistics New York}
}

@article{li2018better,
  title={Better lucky than rich? Welfare analysis of automobile licence allocations in Beijing and Shanghai},
  author={Li, Shanjun},
  journal={The Review of Economic Studies},
  volume={85},
  number={4},
  pages={2389--2428},
  year={2018},
  publisher={Oxford University Press}
}

@article{daljord2021black,
  title={The Black Market for Beijing License Plates},
  author={Daljord, {\O}ystein and Pouliot, Guillaume and Xiao, Junji and Hu, Mandy},
  journal={arXiv preprint arXiv:2105.00517},
  year={2021}
}

@article{white1980heteroskedasticity,
  title={A heteroskedasticity-consistent covariance matrix estimator and a direct test for heteroskedasticity},
  author={White, Halbert},
  journal={Econometrica: journal of the Econometric Society},
  pages={817--838},
  year={1980},
  publisher={JSTOR}
}

@article{breusch1979simple,
  title={A simple test for heteroscedasticity and random coefficient variation},
  author={Breusch, Trevor S and Pagan, Adrian R},
  journal={Econometrica: Journal of the econometric society},
  pages={1287--1294},
  year={1979},
  publisher={JSTOR}
}

@article{tibshirani2012degrees,
  title={Degrees of freedom in lasso problems},
  author={Tibshirani, Ryan and Taylor, Jonathan},
  journal={The Annals of Statistics},
  volume={40},
  number={2},
  pages={1198--1232},
  year={2012},
  publisher={Institute of Mathematical Statistics}
}

@article{kato2009degrees,
  title={On the degrees of freedom in shrinkage estimation},
  author={Kato, Kengo},
  journal={Journal of Multivariate Analysis},
  volume={100},
  number={7},
  pages={1338--1352},
  year={2009},
  publisher={Elsevier}
}

@article{efron2004estimation,
  title={The estimation of prediction error: covariance penalties and cross-validation},
  author={Efron, Bradley},
  journal={Journal of the American Statistical Association},
  volume={99},
  number={467},
  pages={619--632},
  year={2004},
  publisher={Taylor \& Francis}
}

@book{hastie1990generalized,
  title={Generalized Additive Models},
  author={Hastie, TJ and Tibshirani, RJ},
  volume={43},
  year={1990},
  publisher={CRC Press}
}

@article{arkhangelsky2021synthetic,
  title={Synthetic difference-in-differences},
  author={Arkhangelsky, Dmitry and Athey, Susan and Hirshberg, David A and Imbens, Guido W and Wager, Stefan},
  journal={American Economic Review},
  volume={111},
  number={12},
  pages={4088--4118},
  year={2021}
}

@article{ferman2021synthetic,
  title={Synthetic controls with imperfect pretreatment fit},
  author={Ferman, Bruno and Pinto, Cristine},
  journal={Quantitative Economics},
  volume={12},
  number={4},
  pages={1197--1221},
  year={2021},
  publisher={Wiley Online Library}
}

@book{hastie2009elements,
  title={The elements of statistical learning: data mining, inference, and prediction},
  author={Hastie, Trevor and Tibshirani, Robert and Friedman, Jerome H},
  volume={2},
  year={2009},
  publisher={Springer}
}

@article{fathi2022relaxing,
  title={Relaxing the Gaussian assumption in shrinkage and SURE in high dimension},
  author={Fathi, Max and Goldstein, Larry and Reinert, Gesine and Saumard, Adrien},
  journal={The Annals of Statistics},
  volume={50},
  number={5},
  pages={2737--2766},
  year={2022},
  publisher={Institute of Mathematical Statistics}
}

@article{andrews1991heteroskedasticity,
  title={Heteroskedasticity and autocorrelation consistent covariance matrix estimation},
  author={Andrews, Donald WK},
  journal={Econometrica: Journal of the Econometric Society},
  pages={817--858},
  year={1991},
  publisher={JSTOR}
}

@article{andrews1992improved,
  title={An improved heteroskedasticity and autocorrelation consistent covariance matrix estimator},
  author={Andrews, Donald WK and Monahan, J Christopher},
  journal={Econometrica: Journal of the Econometric Society},
  pages={953--966},
  year={1992},
  publisher={JSTOR}
}

@article{zeileis2004hac,
 title={Econometric Computing with HC and HAC Covariance Matrix Estimators},
 volume={11},
number={10},
 journal={Journal of Statistical Software},
 author={Zeileis, Achim},
 year={2004},
 pages = {1--17}
}

@article{newey1994large,
  title={Large sample estimation and hypothesis testing},
  author={Newey, Whitney K and McFadden, Daniel},
  journal={Handbook of econometrics},
  volume={4},
  pages={2111--2245},
  year={1994},
  publisher={Elsevier}
}

@article{muller2014hac,
  title={HAC corrections for strongly autocorrelated time series},
  author={M{\"u}ller, Ulrich K},
  journal={Journal of Business \& Economic Statistics},
  volume={32},
  number={3},
  pages={311--322},
  year={2014},
  publisher={Taylor \& Francis}
}

@article{lazarus2018har,
  title={HAR inference: Recommendations for practice},
  author={Lazarus, Eben and Lewis, Daniel J and Stock, James H and Watson, Mark W},
  journal={Journal of Business \& Economic Statistics},
  volume={36},
  number={4},
  pages={541--559},
  year={2018},
  publisher={Taylor \& Francis}
}

@article{neweywest1987,
 author = {Whitney K. Newey and Kenneth D. West},
 journal = {Econometrica},
 number = {3},
 pages = {703--708},
 publisher = {[Wiley, Econometric Society]},
 title = {A Simple, Positive Semi-Definite, Heteroskedasticity and Autocorrelation Consistent Covariance Matrix},
 urldate = {2025-06-01},
 volume = {55},
 year = {1987}
}

@article{shen2002adaptive,
  title={Adaptive model selection},
  author={Shen, Xiaotong and Ye, Jianming},
  journal={Journal of the American Statistical Association},
  volume={97},
  number={457},
  pages={210--221},
  year={2002},
  publisher={Taylor \& Francis}
}

@article{shen2006optimal,
  title={Optimal model assessment, selection, and combination},
  author={Shen, Xiaotong and Huang, Hsin-Cheng},
  journal={Journal of the American Statistical Association},
  volume={101},
  number={474},
  pages={554--568},
  year={2006},
  publisher={Taylor \& Francis}
}

@article{zou2005regularization,
  title={Regularization and variable selection via the elastic net},
  author={Zou, Hui and Hastie, Trevor},
  journal={Journal of the Royal Statistical Society Series B: Statistical Methodology},
  volume={67},
  number={2},
  pages={301--320},
  year={2005},
  publisher={Oxford University Press}
}

@article{tibshirani2013uniqueness,
author = {Ryan J. Tibshirani},
title = {{The lasso problem and uniqueness}},
volume = {7},
journal = {Electronic Journal of Statistics},
number = {none},
publisher = {Institute of Mathematical Statistics and Bernoulli Society},
pages = {1456 -- 1490},
year = {2013},
doi = {10.1214/13-EJS815},
URL = {https://doi.org/10.1214/13-EJS815}
}

@book{schrijver1998theory,
  title={Theory of linear and integer programming},
  author={Schrijver, Alexander},
  year={1998},
  publisher={John Wiley \& Sons}
}

@book{rockafellar1997convex,
  title={Convex analysis},
  author={Rockafellar, R Tyrrell},
  volume={28},
  year={1997},
  publisher={Princeton university press}
}

@book{meyer2023matrix,
  title={Matrix analysis and applied linear algebra},
  author={Meyer, Carl D},
  year={2023},
  publisher={SIAM}
}

@article{dossal2013degrees,
  title={The degrees of freedom of the lasso for general design matrix},
  author={Dossal, Charles and Kachour, Maher and Fadili, MJ and Peyr{\'e}, Gabriel and Chesneau, Christophe},
  journal={Statistica Sinica},
  pages={809--828},
  year={2013},
  publisher={JSTOR}
}
\end{document}